\documentclass[pdflatex,sn-mathphys-num]{sn-jnl}

\usepackage{placeins}
\usepackage{graphicx}%
\usepackage{multirow}%
\usepackage{amsmath,amssymb,amsfonts}%
\usepackage{amsthm}%
\usepackage{mathrsfs}%
\usepackage[title]{appendix}%
\usepackage{xcolor}%
\usepackage{textcomp}%
\usepackage{manyfoot}%
\usepackage{booktabs}%
\usepackage{algorithm}%
\usepackage{algorithmicx}%
\usepackage{algpseudocode}%
\usepackage{listings}%
\usepackage{amsmath,amssymb,amsthm,graphicx,a4wide,enumerate,url,tikz}
\usepackage[small,bf]{caption}
\usepackage{subcaption}
\usepackage{colortbl}
\usepackage{color}
\usepackage{physics}
\usepackage{bm}
\usepackage{tikz}
\usetikzlibrary{matrix}
\usepackage{algorithm,algpseudocode}
\usepackage[title]{appendix}
\usepackage{booktabs}
\usepackage{nomencl}

\usepackage[markup=underlined]{changes}

\theoremstyle{thmstyleone}%

\theoremstyle{thmstyletwo}%

\theoremstyle{thmstylethree}%

\begin{document}

\title[Quantifying topological features and irregularities in zebrafish patterns]{Quantifying topological features and irregularities in zebrafish patterns using the sweeping-plane filtration}

\author*[1]{\fnm{Nour} \sur{Khoudari}}\email{nkhoudar@purdue.edu}

\author[2]{\fnm{John} \sur{Nardini}}\email{nardinij@tcnj.edu}

\author[1]{\fnm{Alexandria} \sur{Volkening}}\email{avolkening@purdue.edu}

\affil*[1]{\orgdiv{Department of Mathematics}, \orgname{Purdue University}, \orgaddress{\street{150 N.\ University St.}, \city{West Lafayette}, \postcode{47907}, \state{Indiana}, \country{USA}}}

\affil[2]{\orgdiv{Department of Mathematics and Statistics}, \orgname{The College of New Jersey}, \orgaddress{\street{2000 Pennington Rd.}, \city{Ewing}, \postcode{08628}, \state{New Jersey}, \country{USA}}}

\abstract{Complex patterns emerge across a wide range of biological systems. While such patterns often exhibit remarkable robustness, variation and irregularity exist at multiple scales and can carry important information about the underlying agent interactions driving collective dynamics. Many methods for quantifying biological patterns focus on large-scale, characteristic features (such as stripe width or spot number), but questions remain on how to characterize messy patterns. In the case of cellular patterns that emerge during development or regeneration, understanding where patterns are most susceptible to variability may help shed light on cell behavior and the tissue environment. Motivated by these challenges, we introduce methods based on topological data analysis to classify and quantify messy patterns arising from agent-based interactions, by extracting meaningful biological interpretations from persistence barcode summaries. To compute persistent homology, our methods rely on a sweeping-plane filtration which, in comparison to the Vietoris--Rips filtration, is more rarely applied to biological systems. We demonstrate how results from the sweeping-plane filtration can be interpreted to quantify stripe patterns---with and without interruptions---by analyzing \textit{in silico} zebrafish skin patterns, and we generate new quantitative predictions about which pattern features may be most robust or variable. Our work provides an automated framework for quantifying features and irregularities in spot and stripe patterns and highlights how different approaches to persistent homology can provide complementary insight into biological systems.}





\maketitle

\section{Introduction}

Spatial pattern formation is present in biological systems at many scales, with examples including cells organizing during tissue development or regeneration \cite{Giniunaite2020,Buttenschon2020,VolkeningRev,Kondo2021rev}, as well as flocking, herding, and swarming of animal or insect populations \cite{Mogilner1999,Bernoff2011,Huepe2008,Dorsogna2006,VicsekReview,Katz2011fish,Lukeman2010keshet}. Complementing experiments, mathematical models can shed light on the agent behaviors that govern the formation of these patterns. Often models focus on capturing large-scale, characteristic features (e.g., the presence of stripes or number of spots), but biological patterns are messy and imperfect. For example, wild-type zebrafish feature gold and blue stripes in their skin \cite{ParichyRev2019,IrionRev2019,Kondo2021rev}, and these stripes are remarkably robust in comparison to mutant patterns \cite{Frohnhofer}. At closer inspection, however, blue stripes are considered more likely to develop interruptions than gold stripes \cite{volkening2018}, and subtle differences may arise in different anatomical regions of the fish body \cite{McCluskey2021}. Determining when and where patterns are most susceptible to non-characteristic elements and variability has the potential to provide new insight into the mechanisms driving self-organization and the effects of tissue growth on cell behavior. With this motivation, we develop a methodology for quantifying features and irregularities in stripe and spot patterns based on topological data analysis. Our approach relies on persistent homology, and we show how to interpret topological summaries from the sweeping-plane filtration \cite{Bendich2016,Nardini2021} in terms of biologically meaningful characterizations of zebrafish patterns \cite{volkening2018}.

\begin{figure}[t!]
\includegraphics[width=\textwidth]{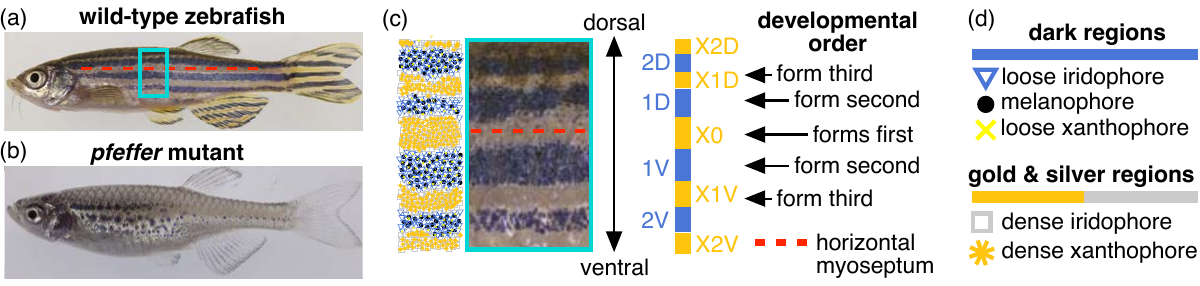}\vspace{0.5\baselineskip}
\caption{Introduction to zebrafish skin patterns. (a) Wild-type zebrafish are characterized by dark stripes and light interstripes, while (b) \textit{\textit{pfeffer}} \cite{Maderspacher2003,ParTur130,PatDev127} mutants feature more variable patterns made up of dark blue spots on a silver background \cite{Frohnhofer} (c) During wild-type development over the course of several weeks, stripes and interstripes form sequentially in the dorsal (upward) and ventral (downward) directions, starting from the center of the fish \cite{Jan,Quigley2002}. The first interstripe to form---X$0$---appears at the horizontal myoseptum (red dashed line), which helps align stripes horizontally \cite{Frohnhofer,Yamaguchi,Quigley2002}. As the fish continues to grow, stripes $1$D and $1$V form, followed by interstripes X$1$D and X$1$V. (d) Three main types of pigment cells self-organize to produce the pigmentation patterns in zebrafish. Gold and silver regions in the skin contain dense xanthophores and dense iridophores, whereas dark regions consist of loose xanthophores, loose iridophores, and melanophores \cite{Cleveland2023,volkening2018}. Images (a), (b), and (c, \textit{in vivo} image) are adapted and cropped from Fadeev et al.\ \cite{fadeev2015} and licensed under CC-BY $4.0$ (\url{https://creativecommons.org/licenses/by/4.0/}); we added the cyan box, red line, text, information about gold and blue stripes, and dorsal--ventral guide. Images (c, left panel) and (d) are adapted from Volkening et al. \cite{volkening2018} under CC-BY $4.0$.}
\label{fig:biology}
\end{figure}

As we show in Fig.~\ref{fig:biology}, the blue stripes and gold ``interstripes" \cite{IrionRev2019,Kondo2021rev,VolkeningRev,ParichyRev2019} in zebrafish form due to the interactions of pigment cells. Stripes and interstripes appear sequentially during development, starting from a gold interstripe along a central morphological feature on the body. Over several weeks as the fish doubles in size, pattern formation then progresses dorsally and ventrally; see Fig.~\ref{fig:biology}(c).
Studying the mechanisms driving this process in zebrafish---a model system for understanding animal pigmentation---can provide insight into developmental biology and genetics. For example, altering the tissue environment can influence the behavior of cells and disrupt patterns \cite{eskova2017gain}, highlighting the role of factors beyond cell behavior in patterning \cite{PatPLos}. Other mutations change cell--cell interactions in unknown ways \cite{McGuirl2020}, leading to patterns with spots or wider or more frequently interrupted stripes (e.g., \cite{Maderspacher2003,Watanabe2006,2012changing,IrionGap,Iwashita}). These mutations offer scientists the opportunity to uncover the functional impact of genes on cell behavior and organism phenotype. 
Because many zebrafish genes have counterparts in the human genome \cite{genome}, research on zebrafish has the potential for broader impact as well.

Motivated by the variety of skin patterns in wild-type and mutant zebrafish, mathematical models have been developed to describe this patterning process. These include macroscopic models of cell density in the form of partial differential or integro-differential equations \cite{Bullara,Konow2021,Gaffney,Nakamasu,painter,Yamaguchi,Woolley2017}, and microscopic models in the form of on- or off-lattice agent-based models \cite{volkening2015,volkening2018,volkening2020,VolkeningRev,Bullara,Owen2020,MorDeutsch,Konow2021,Cops,Shinbrot}. Because agent-based models treat cells as individual entities, they are a natural means of studying self-organization during tissue development, including pattern formation in zebrafish. However, extracting analytical insight into agent-based models is challenging, making it difficult to broadly characterize model behavior. Moreover, as detailed, stochastic agent-based models get more realistic, they face many of the same challenges as empirical data. Whether \textit{in vivo} or \textit{in silico}, assessing messy, variable spatial data often relies on qualitative observation, which limits the perspective that one can take and may necessitate a focus on characteristic---e.g., commonly occurring, large-scale, or defining---features in patterns, such as the presence of stripes in Fig.~\ref{fig:biology}(a).

Beyond characteristic features, we are interested in understanding variability in patterns across organisms at the population scale, and in characterizing irregularities and biological messiness across a single pattern at the organism or tissue scale. In order to unpack variability and messiness in complex biological systems, it is necessary to take a quantitative perspective. There are many approaches to quantifying qualitative data \cite{VolkeningQuant}, including order parameters \cite{VicsekReview,Topaz2015}, pattern simplicity scores \cite{Miyazawa2010Simplicity,Djur2019Trout}, pair-correlation functions \cite{Gavagnin2018yates,Bull2024,Binder2013}, and techniques from topological data analysis \cite{Edelsbrunner2008,Carlsson2009, Otter2017mason,Ghrist2014,Munch2020Shape,Feng2021tda,Chazal2021}. Topological data analysis, in particular, has recently been used to extract information 
from data for many biological systems (e.g., \cite{Coral,PelegTDA2023,Ulmer2019topaz,Lawson2019cancer,Hartsock2025,Bendich2016,Thorne, Nardini2021,Stolz2022,Topaz2015,Bhaskar2019,Bhaskar2023,Ciocanel2021,McGuirl2020,Cleveland2023}). One of the main tools in topological data analysis is persistent homology \cite{Edelsbrunner2008, Edelsbrunner2002,Otter2017mason,Chazal2021}, which involves filtering through data and identifying features such as connected components and loops that are present as some scale $r$ is increased.

Computing persistent homology involves choosing a method for filtering through data, and some methods naturally lend themselves to specific types of data. For data in the form of point clouds, for example, one means of studying shape is to place a ball of radius $r$ around each point and let these balls grow; as the scale $r$ increases, balls intersect and one tracks the topological features present (see Fig.~\ref{fig:TDA}(a)). This approach is related to building Vietoris--Rips simplicial complexes \cite{Topaz2015}. Filtering based on sublevel sets, on the other hand, is particularly useful for data that can be represented as images with one channel of color intensity \cite{Turke2021,Lawson2019cancer,Kramar2016}; in this case, one can sequentially threshold the color intensity at different heights $r$ and study the shape of the sublevel sets, building a filtered cubical complex. As other examples, the edge weight can be used as the filtration parameter $r$ in weighted networks \cite{Otter2017mason}, or, for binary images, one can sweep a plane across the image, characterizing shape as more data are uncovered \cite{Nardini2021}. These filtration methods offer different perspectives \cite{Otter2017mason,StolzThesis}, highlighting the value of applying multiple filtrations to data.

In the case of agent-based models (e.g., \cite{Topaz2015,Ulmer2019topaz,Bhaskar2019,Bhaskar2023,Ciocanel2021}), including for zebrafish \cite{McGuirl2020,Cleveland2023}, many studies consider persistent homology based on the Vietoris--Rips filtration. Notably, McGuirl \textit{et al.} \cite{McGuirl2020} and Cleveland \cite{Cleveland2023} applied persistent homology to the positions of pigment cells in cropped zebrafish patterns from the agent-based model \cite{volkening2018}. They \cite{McGuirl2020,Cleveland2023} interpreted the results of the Vietoris--Rips filtration in terms of the number of stripes and spots in \textit{in silico} patterns, providing a valuable analysis of variability across stochastic simulations. However, questions remain that persistent homology with the Vietoris--Rips filtration may be less amenable to answering. First, if one considers an individual zebrafish, there is variability in the pattern across the body, but we do not expect the Vietoris--Rips filtration to provide this information. Moreover, the quantification pipelines \cite{McGuirl2020,Cleveland2023} crop the domain, focusing on the central portion of the pattern that is most well-formed. These methods also require two important pieces of \textit{a priori} information: first, patterns must be pre-sorted as striped or spotted. And second, rather than directly identifying interruptions or breaks in stripes, these pipelines \cite{McGuirl2020,Cleveland2023} flag stripe patterns as irregular based on whether the stripe count is less than the target number for the model \cite{volkening2018}. This makes it difficult to generalize the approach \cite{McGuirl2020,Cleveland2023} to patterns simulated under new parameters. Questions also remain about where and when stripes are most susceptible to developing interruptions or irregularities.

Motivated by these questions and challenges, here we take a new persistent-homology perspective on messy spot and stripe patterns. Our work centers on the sweeping-plane filtration, and we develop a methodology to automatically classify agent-based patterns as spots, perfect stripes, or irregular (e.g., broken) stripes, with no prior assumptions on the expected number of stripes. While our work centers on wild-type and \textit{pfeffer} mutant zebrafish patterns generated by the model \cite{volkening2018}, we expect our methodology to be more widely applicable, and our code is available on GitHub \cite{2025repo}. As our main contribution, we show how to interpret topological summaries from the sweeping-plane filtration in terms of biologically meaningful information about the location and width of irregularities in stripe patterns. We also automatically characterize stripe width and spot size, and count spots and stripes, even in the presence of irregularities. We discuss how the Vietoris--Rips and sweeping-plane filtrations provide complementary insight into zebrafish patterns. Our work further highlights the value of considering multiple filtrations when computing persistent homology for complex biological systems, and it provides new experimentally-testable predictions about which features may be most robust in wild-type zebrafish and where patterns may be most susceptible to irregularities.

\section{Background and methods}
\label{sec:background}

Here we give a brief overview of zebrafish skin patterns (Sect.\ \ref{sec:biobackground}) and describe the agent-based model \cite{volkening2018} which generated the simulated patterns that we quantify (Sect.\ \ref{sec:modelbackground}). In Sect.\ \ref{sec:TDAbackground}, we introduce topological data analysis and discuss two techniques for computing persistent homology: the Vietoris--Rips filtration, which has been widely used to quantify biological patterns (Sect.~\ref{sec:VRbackground}), and the sweeping-plane filtration \cite{Bendich2016,Nardini2021}, which is comparatively less common in agent-based modeling studies and serves as the basis of our work (Sect.\ \ref{sec:SPbackground}).

\subsection{Biological background on zebrafish skin patterns}\label{sec:biobackground}

Zebrafish (\textit{Danio rerio}) are known for their skin patterns consisting of dark blue stripes and gold interstripes, which form over several weeks through the self-organizing interactions of pigment cells \cite{delta,Inaba,Patterson2014,IrionRev2019,ParichyRev2019,VolkeningRev,Jan}. As we show in Fig.~\ref{fig:biology}(d), black melanophores, loose blueish iridophores, and loose yellow xanthophores make up blue stripes and spots in the skin, whereas dense silver iridophores and gold xanthophores occupy light regions. At the cellular level, neighboring cells are separated by approximately $50$ micrometers ($\mu\text{m}$) \cite{2016heterotypic,ParTur256,TakahashiMelDisperse}, and, on the pattern scale, the widths of stripes and interstripes in adult zebrafish are around $500~\mu\text{m}$ \cite{fadeev2016,delta,volkening2018}. The formation of these skin patterns involves cell migration, differentiation, division, and competition \cite{TakahashiMelDisperse,Mahalwar,Dooley,Mcmen2014,Budi,Gur2020,Nakamasu,Yamaguchi}, and the dynamics underlying cell interactions may include signal diffusion \cite{Patterson2014} or extensions of various lengths \cite{eom2017macrophage,delta,Inaba}. Notably, melanophores differentiate from precursors, appearing largely \textit{in situ} in the skin, while xanthophores mainly divide from existing cells \cite{Mahalwar,Budi,Dooley,Singh,PatPLos,Mcmen2014,walderich2016homotypic}. Empirical understanding of iridophores continues to evolve and grow, with research suggesting both division and precursor differentiation are important \cite{Gur2020,Singh}.

As we show in Fig.~\ref{fig:biology}(a,c), an adult wild-type zebrafish typically has four to five dark stripes, and four light interstripes \cite{Jan}. Researchers conventionally name these features according to their position along the dorsal--ventral axis based on developmental order (e.g., \cite{Frohnhofer,volkening2018}). The central interstripe, which begins appearing around three weeks post fertilization, is labeled ``X0" \cite{Jan}. This interstripe forms in conjunction with the horizontal myoseptum, an anatomical structure along the center of the body that provides horizontal directionality \cite{Frohnhofer}. As the fish continues to grow, the first two dark stripes form flanking the central interstripe; these are denoted ``1D" and ``1V" \cite{Jan}. At around six weeks post fertilization, though growth rates differ across experimental conditions \cite{McMen2016,Parichy,volkening2018}, the next two gold interstripes appear (``X1D" and ``X1V"). Later on the last two blue stripes, ``2V" and ``2D", develop, eventually followed by interstripes ``X2D" and ``X2V". There are also many altered skin patterns that form in zebrafish due to genetic mutations that affect cell interactions or cause one or more pigment cell types to be absent. Even when mutant patterns contain spots rather than stripes, such as in the \textit{pfeffer} mutant zebrafish \cite{Maderspacher2003,ParTur130,PatDev127} in Fig.~\ref{fig:biology}(b), it is common to refer to the ``X0 region" and to discuss whether dark spots appear in the positions of stripes 1D and 1V \cite{Frohnhofer}. For wild-type and mutant patterns, many of the cell interactions involved remain to be fully understood, and epithelial growth during self-organization may also play a role in patterning \cite{Cleveland2023,volkening2020,ParTur256}.

\subsection{Focal agent-based model and simulated data}\label{sec:modelbackground}

We apply our methodology for quantifying features and irregularities in spot and stripe patterns to an agent-based model of cell behavior in zebrafish skin \cite{volkening2018}. Developed by Volkening and Sandstede, this detailed, stochastic off-lattice model describes the sequential appearance of wild-type stripes and interstripes on a growing domain and reproduces the formation of various mutant patterns. The model \cite{volkening2018} is a useful basis for our work because it was previously considered \cite{McGuirl2020,Cleveland2023} from the perspective of topological data analysis. Analyzing \textit{in silico} patterns that were cropped to remove the messier dorsal and ventral stripes and interstripes (i.e., X2D, X1D, X1V, and X2V in Fig.~\ref{fig:biology}(c)), these studies \cite{McGuirl2020,Cleveland2023} computed persistent homology based on the Vietoris--Rips filtration and interpreted the results. Motivating our methodology using the same model sets up a natural case study for us to determine what alternative filtrations can tell us about the same patterns \cite{mcguirl2020zebrafish}. Because off-lattice agent-based models naturally produce data in the form of point clouds, the model \cite{volkening2018} also motivates our pipeline in Sect.~\ref{sec:step2} for transforming these data into a format more appropriate for the sweeping-plane filtration, as we expect our methodology to be applicable to other cellular patterns as well.

We briefly discuss the agent-based model \cite{volkening2018} here, and refer to the original reference for full details. Broadly, this model \cite{volkening2018} tracks the interactions of five types of pigment cells; see Fig.~\ref{fig:biology}(d). Cell agents are represented as particles in continuous space, with ($x$,$y$)-coordinates marking their positions on growing two-dimensional domains. For instance, $\textbf{M}_i(t) \in \mathbb{R}^2$ is the position of the $i$th melanophore, and $\textbf{I}^\text{l}_j(t) \in \mathbb{R}^2$ is the position of the $j$th loose iridophore at time~$t$. Each simulated day, these positions are scaled deterministically to reflect epithelial growth as uniform spatial expansion. Cell movement is implemented through coupled ordinary differential equations, with each cell following an equation describing the forces exerted on it by neighboring cells. Based in the biological literature, the model also incorporates cell differentiation, division, competition, and transitions in agent type, which follow stochastic discrete-time rules \cite{volkening2018}. As an example, to model melanophores differentiating from precursors that are randomly distributed in the skin, Volkening and Sandstede \cite{volkening2018} uniformly at random select $N$ candidate ($x$,$y$)-coordinates in the domain per simulated day; these positions are then evaluated for possible differentiation (e.g., appearance of a new melanophore agent) based on noisy rules that depend on the types of cells in their local or long-range neighborhoods. 

Through the cell interactions in the model \cite{volkening2018}, patterns emerge autonomously on growing domains that capture the full fish height and roughly a third of the fish body length. The initial condition at $21$ days post fertilization features a single strip of dense iridophores at the center of the domain, representing the future location of the X0 interstripe guided by the horizontal myoseptum \cite{volkening2018}. We consider the patterns that form after simulating the model from $21$ to $66$ days post fertilization, at which point the domain corresponds to a third of the patterned body length of a juvenile zebrafish\footnote{In the empirical community, fish age is commonly described using developmental stage or standardized standard length (i.e., the length from the snout to the base of the tail fin, for a reference zebrafish), because there is some variability in growth rates across experimental conditions \cite{Parichy}. The time point of $66$ days post fertilization in the model \cite{volkening2018} corresponds to a zebrafish with standard length of $12.63$ millimeters (mm). The domain sizes in the model \cite{volkening2018} account for about one third of the standard length, after removing a small un-patterned region around the fish eye.}. The domain has periodic boundary conditions in $x$, and wall-like boundaries at its top and bottom edges. As an important point for our study, we highlight that noise effectively builds on noise as pattern formation occurs. In particular, the model \cite{volkening2018} suggests that the initial condition at three weeks post fertilization provides the first instructions for stochastic cell division and differentiation, guiding the formation of stripes 1V and 1D. However, if a cell appears out of place due to stochasticity, this influences where cells appear at the next time step. We thus expect stripes and interstripes to be messier and more irregular as pattern formation proceeds dorsally and ventrally, and our methodology allows us to quantitatively test this in Sect.~\ref{sec:results}.

\begin{figure}[t!]
    \centering 
    \includegraphics[width=1\linewidth]{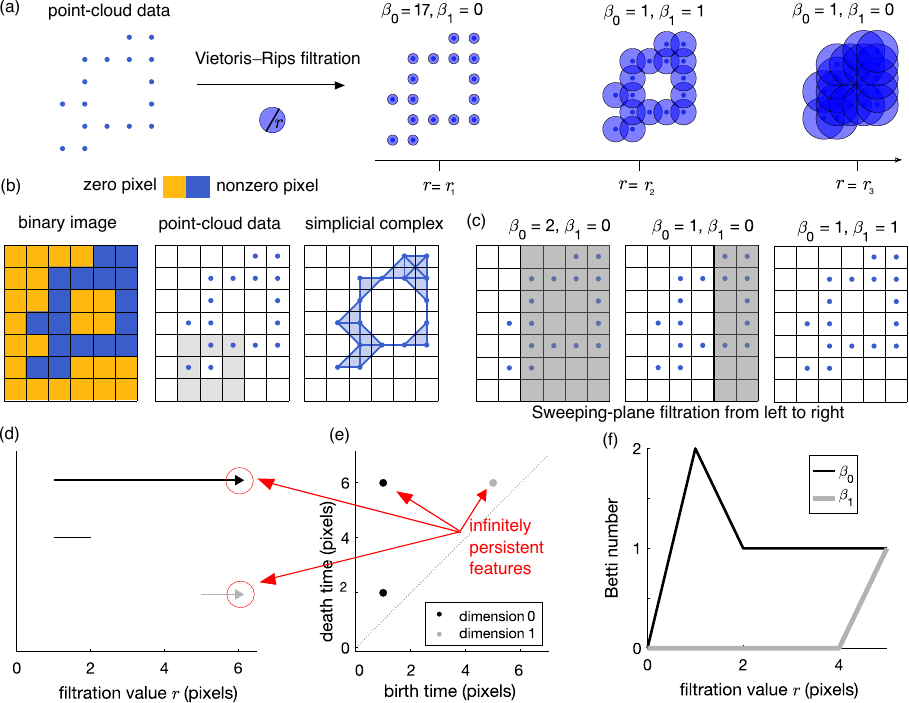}\vspace{0.5\baselineskip}
    \caption{Computing persistent homology using the Vietoris--Rips filtration and the sweeping-plane filtration. (a) The Vietoris--Rips filtration starts with point-cloud data, and it can be thought of roughly as a means of studying shape as balls of diameter $r$ around each point grow in size \cite{Topaz2015}. We report the $0$th and $1$st Betti numbers, indicating the number of connected components and loops, for a few $r$ values. (b) The sweeping-plane filtration offers a different perspective on the shape of data. It starts with a binary image---here blue (nonzero-valued) pixels and gold (zero-valued) pixels---and the centroids of the nonzero pixels form a point cloud \cite{Nardini2021}. A simplicial complex can be constructed from that point cloud by placing an edge ($1$-simplex) between two points if they are within the Moore neighborhood of each other and placing a triangle ($2$-simplex) between three pairwise-connected points \cite{Nardini2021}. The Moore neighborhood for an example point is highlighted in light gray. (c) In the sweeping-plane filtration, we slide a plane across the image, uncovering sequentially more points and constructing a sequence of simplicial complexes during this process; we highlight the corresponding $0$th and $1$st Betti numbers at a few steps. The (d) barcode, (e) persistence diagram, and (f) Betti curves associated with the sweeping-plane filtration provide summaries of the topological shape of our example data in (b). We note that bars ending with arrows in the barcode---and points with death time greater than or equal to the maximum filtration value in the persistence diagram---represent infinitely persistent features.}
    \label{fig:TDA}
\end{figure}

\subsection{Topological data analysis}\label{sec:TDAbackground}

Our approach to quantitatively describing messy patterns relies on techniques from topological data analysis (TDA), particularly persistent homology \cite{Munch2020Shape,Otter2017mason,Carlsson2009,Edelsbrunner2008,Chazal2021}. (We overview persistent homology informally, and refer to \cite{Carlsson2009,chazal,Edelsbrunner2008,Ghrist2014,Otter2017mason,Topaz2015} for more technical definitions.) Broadly, persistent homology is a means of characterizing shape in data, and it has been widely applied to quantify empirical and \textit{in silico} patterns and complex systems \cite{Coral,PelegTDA2023,Ulmer2019topaz}, providing insight into cancer histology \cite{Lawson2019cancer}, microscopy images \cite{Hartsock2025}, brain artery trees \cite{Bendich2016}, vascular networks \cite{Thorne, Nardini2021,Stolz2022}, flocking \cite{Topaz2015,Bhaskar2019}, cell sorting \cite{Bhaskar2023}, intracellular transport \cite{Ciocanel2021}, and zebrafish patterns \cite{McGuirl2020,Cleveland2023}.  There are many approaches to computing persistent homology, but they all involve defining a filtration---e.g., a growing, nested sequence of topological subspaces based on some parameter $r$ that allows one to filter through data at different scales \cite{Chazal2021,Heiss2021}. (In the case of multi-parameter persistence, more complicated filtrations can involve more than one scaling parameter \cite{Hu2021,CarlssonMulti,CarlssonMulti2,Harrington2019}.) During the filtration process, one tracks the presence of connected components, loops, and higher-dimensional topological features in the data.

More specifically, persistent homology often involves describing data in terms of $n$-simplices that connect $n + 1$ data points, for $n = 0$, $1$, $2$, and so on. For example, a point is a $0$-simplex, an edge between two points is a $1$-simplex, and a filled-in triangle associated with three points is a $2$-simplex \cite{Otter2017mason,Topaz2015}. A finite collection of points, edges, filled-in triangles, and other $n$-simplices that satisfies certain properties is a ``simplicial complex" $K_r$; see \cite{Otter2017mason} for definitions. 
For a given simplicial complex $K_{r}$, one can define vector spaces with the $0$-, $1$-, and $n$-simplices as their bases. For instance, the boundary map $\partial_2$ operates on $2$-simplices and outputs the edges of triangles, $\partial_1$ applies to $1$-simplices and outputs the endpoints of edges, and $\partial_0$ operates on $0$-simplices and outputs zero \cite{Nardini2021}. By computing the kernel and image of these boundary maps, one can define the homology groups associated with the simplicial complex $K_r$ as $H_0(K_r)=\text{Kernel}(\partial_0)/\text{Image}(\partial_1)$ and $H_1(K)=\text{Kernel}(\partial_1)/\text{Image}(\partial_2)$ \cite{Nardini2021}. The dimension of each of these vector spaces or homology groups---where $\beta_n(K_r) = \text{dim}(H_n(K_r))$---gives the number of $n$-dimensional holes present \cite{Topaz2015,Chazal2021}. These dimensions are called ``Betti numbers". For example, $\beta_0(K_r)$ and $\beta_1(K_r)$ are the Betti numbers that quantify the numbers of connected components and loops associated with the simplicial complex $K_r$. A filtration process, in turn, provides a means of building a ``filtered simplicial complex" $K = \{K_{r_0} \subset K_{r_1} \subset ... \subset K_{r_\text{max}}\}$, consisting of nested simplicial complexes, each of which is constructed from the filtration values $r_0<r_1<...<r_\text{max}$ \cite{Otter2017mason}.

To study the shape of data across scales, the first step in computing persistent homology is thus choosing a filter. 
Each filtering method offers a different perspective on the data, and part of our motivation for this study is to highlight how the choice of filtration affects biological interpretations and insights. With this in mind, we overview two filtrations: the Vietoris--Rips filtration, which has previously been applied to cropped \textit{in silico} zebrafish patterns \cite{McGuirl2020,Cleveland2023}, and the sweeping-plane filtration, which lends itself to branching and vascular data \cite{Bendich2016,Nardini2023,Nardini2021}. First, because it is perhaps the most widely used filtration in agent-based modeling, we discuss the Vietoris--Rips filtration in Sect.~\ref{sec:VRbackground} as background. 
Second, in Sect.~\ref{sec:SPbackground}, we overview the sweeping-plane filtration, which is comparatively less common in studies of complex biological systems. 
This filtration method---which has not previously been applied to spot and stripe patterns to our knowledge---provides new insight into the shape of these data and serves as the basis for our methodology for quantifying features and irregularities in zebrafish skin patterns. Lastly, we discuss barcodes, persistence diagrams, and Betti curves, three common ways of visualizing persistent homology, in Sect.~\ref{sec:barcode}.

\subsubsection{Persistent homology with the Vietoris--Rips filtration}\label{sec:VRbackground}

To compute persistent homology using the Vietoris--Rips filtration, we start with point-cloud data, such as the ($x$,$y$)-coordinates of cells. Given these data and some measurement of distance, this filtration connects points whose pairwise distance are all less than the parameter $r$. This can be thought of as placing a ball of diameter $r$ around each point and drawing an $n$-simplex between any $n + 1$ points whose balls all pairwise intersect.
By increasing the scaling parameter $r$, one then builds a filtered simplicial complex $K = \{K_{r_0} \subset K_{r_1} \subset ... \subset K_{r_\text{max}}\}$ that captures underlying structures that persist across scales; see Fig.~\ref{fig:TDA}(a). Each data point is a $0$-simplex, edges connecting points that are within a distance $r$ away from one another are $1$-simplices, and collections of three points with pairwise distances less than $r$ are $2$-simplices \cite{Topaz2015,Bhaskar2019}. As $r$ grows from $0$ to a sufficiently large value, the shape of our manifold evolves from isolated points to a single connected component.

\subsubsection{Persistent homology with the sweeping-plane filtration}\label{sec:SPbackground}

While the Vietoris--Rips filtration (Sect.~\ref{sec:VRbackground}) is well-suited to point-cloud data, the sweeping-plane filtration is most appropriate for binary images representing systems with a branching nature, such as vascular or neuronal networks \cite{Bendich2016,Nardini2021}. (Motivated by the angiogenesis study of Nardini \textit{et al.} \cite{Nardini2021}, our methodology in Sect.~\ref{sec:methodology} considers binary images, so we discuss the sweeping-plane filtration in terms of this data format.) To compute persistent homology using the sweeping-plane filtration on a binary image, we slide a line across the image in a given direction, slowly uncovering more of the pattern; see Fig.~\ref{fig:TDA}(a)-(b). 
Compared to Vietoris--Rips, where the filtration parameter $r$ is the ball diameter, the filtration parameter $r$ in the sweeping-plane approach denotes the distance---in pixels---that we have moved the line in the sweeping direction from the boundary.
At each filtration step, one moves this line by a fixed number of pixels, and a point cloud is generated from the centroids of all of the nonzero-valued pixels that have been uncovered. These pixel centroids are the $0$-simplices. Any two points within the Moore neighborhood (the $8$ pixels surrounding a single pixel) of each other are connected by an edge (a $1$-simplex), and any triplet of points connected pairwise with an edge are connected with a filled triangle (a $2$-simplex) \cite{Nardini2021}, as we show in Fig.~\ref{fig:TDA}(b). The result is a simplicial complex $K_r$ at the filtration value $r$. This process is repeated for all considered $r$ values to create the sweeping-plane filtration in the chosen direction.

\subsubsection{Visualizing and interpreting persistent homology}\label{sec:barcode}

Whether considering the Vietoris--Rips filtration on point-cloud data, the sweeping-plane filtration \cite{Bendich2016,Nardini2021} on binary images, or another filtration, persistent homology allows us to track when topological features such as connected components or loops appear or disappear as a function of the filtration parameter $r$. If a topological feature appears at $r=r_{\text{birth}}$ and disappears at $r=r_{\text{death}}>r_{\text{birth}}$, we say that the ``birth time" of this feature is $r_{\text{birth}}$, its ``death time" is $r_{\text{death}}$, and its ``persistence" is $r_{\text{death}}-r_{\text{birth}}$ \cite{Topaz2015}. Persistence is often visualized using barcodes and persistence diagrams \cite{Topaz2015}. As we show in Fig.~\ref{fig:TDA}(d) for the sweeping-plane filtration, a barcode consists of vertically stacked horizontal bars, each representing the lifespan of a topological feature. With the filtration parameter $r$ along the horizontal axis, the left endpoint of a bar corresponds to the birth time $r_{\text{birth}}$ of the topological feature, and its right endpoint is at $r_{\text{death}}$. Closely related, a persistence diagram is a scatter plot with birth times as the horizontal axis and death times as the vertical axis. Each point $(r_{\text{birth}},r_{\text{death}})$ refers to the lifespan of one topological feature; see Fig.~\ref{fig:TDA}(e). Another way to visualize persistent homology is by graphing the total number of connected components ($0$-dimensional holes) and loops ($1$-dimensional holes) present as a function of the filtration step $r$. To visualize how the dimension-$0$ and dimension-$1$ topological features evolve as one filters through data, researchers often plot Betti curves, e.g., graphs of $\beta_0(K_r)$ and $\beta_1(K_r)$ as a function of the filtration value $r$, see Fig.~\ref{fig:TDA}(f).

After visualizing persistent homology, the next step is interpreting the results in terms of the specific data under consideration. Relating topological features to biologically meaningful quantities is a challenging task. In the case of simulated zebrafish patterns, however, prior studies \cite{McGuirl2020,Cleveland2023} have interpreted barcodes from the Vietoris--Rips filtration in terms of the numbers of spots and uninterrupted stripes in patterns. Specifically, by applying persistent homology to the positions of melanophores on a domain that is periodic in $x$, the methods of McGuirl \textit{et al.} \cite{McGuirl2020} and Cleveland \textit{et al.} \cite{Cleveland2023} compute the number of stripes as the number of highly persistent loops with sufficiently low birth times. For spotted patterns, the number of dark spots is related to the number of highly persistent connected components \cite{McGuirl2020}. The sweeping-plane filtration, in contrast, has been applied primarily to patterns that resemble networks and its results are generally summarized using barcodes or Betti curves. For example, Nardini \textit{et al.} \cite{Nardini2021} swept across binary images from simulations of tumor-induced angiogenesis. In this previous study, the authors performed plane sweeping in four directions (top-to-bottom, bottom-to-top, left-to-right, and right-to-left) and presented the results using Betti numbers and persistence images. Because each filtration offers a different perspective on data, here we are interested in developing a methodology to interpret barcodes from the sweeping-plane filtration as direct characterizations of stripe and spot patterns in binary image data, as we discuss in Sect.~\ref{sec:methodology}.

\section{Results: Our methodology for quantifying messy patterns using the sweeping-plane filtration}
\label{sec:methodology}

\begin{figure}[t!]
\includegraphics[width=\textwidth]{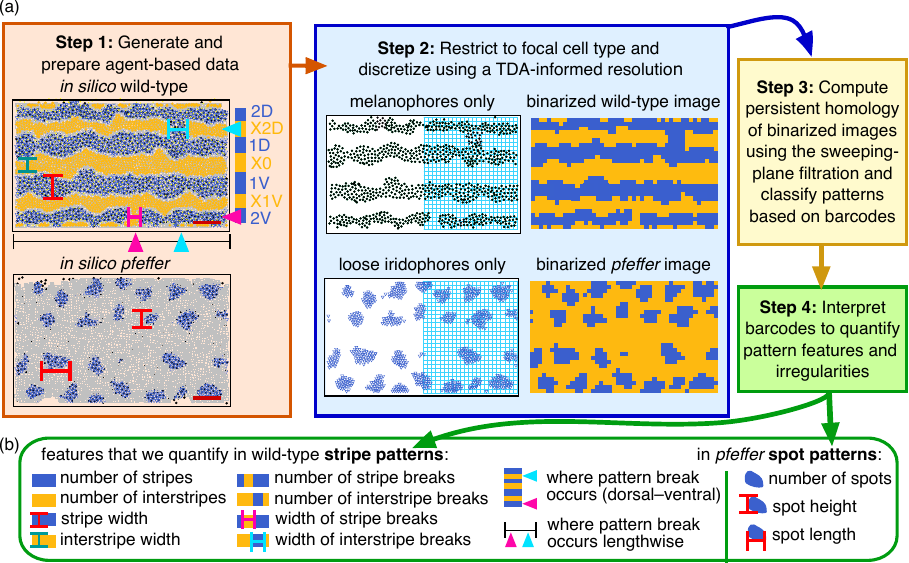}\vspace{0.5\baselineskip}
\caption{Summary of our quantification pipeline, including input patterns and output features. (a) Our methodology involves four main steps: (1) generating point-cloud pattern data; (2) discretizing these data to produce binary images; (3) computing persistent homology using the sweeping-plane filtration to classify patterns as spots, unbroken stripes, or stripes with various types of breaks; and (4) interpreting barcodes to quantify pattern features and irregularities. Our study focuses on wild-type and \textit{pfeffer} mutant zebrafish-skin patterns generated by the agent-based model \cite{volkening2018} for concreteness, and we highlight examples of these patterns in Step $1$ with red scale bars indicating $500$~$\mu$m. (b) By interpreting the results of persistent homology, we provide quantitative summaries of biologically meaningful features, including the number of stripes, number of spots, stripe width, and spot size. To better understand irregularities that emerge during development, we also identify where breaks or interruptions occur in stripe patterns, and we estimate interruption width.}
\label{fig:summary}
\end{figure}

We now describe our methodology for quantifying features and irregularities in stripe and spot patterns arising from the self-organization of individual agents. Our associated code is publicly available on GitHub \cite{2025repo}, and, while we frame our approach around \textit{in silico} zebrafish patterns \cite{volkening2018} for concreteness, we expect our methods to be more generally applicable to other biological systems. As we summarize in Fig.~\ref{fig:summary}, our pipeline consists of four main steps: first, we generate and prepare pattern data from the agent-based model \cite{volkening2018} (Sect.\ \ref{sec:step1}). Second, we transform these agent-based (point-cloud) data to binary, pixelated images, which are the natural input for TDA computations with the sweeping-plane filtration (Sect.\ \ref{sec:step2}). This step involves selecting an image resolution or pixel size, and we develop a method to select this resolution based on persistent homology. Third, we use topological summaries from the sweeping-plane filtration \cite{Nardini2021,Bendich2016} to classify patterns as stripes, interrupted stripes, or spots (Sect.\ \ref{sec:step3}). Fourth, we further interpret persistent homology visualizations to quantify biologically meaningful features and irregularities in messy stripe or spot patterns (Sect.\ \ref{sec:step4}). In particular, we determine the number of stripes, the number of spots, stripe and interstripe width, and spot size. Most importantly, we show how to interpret barcodes from the sweeping-plane filtration as information not only about pattern features, but also about defects, characterizing when and where interruptions appear in stripes.

\begin{figure}[t!]
    \centering 
    \includegraphics[width=1\linewidth]{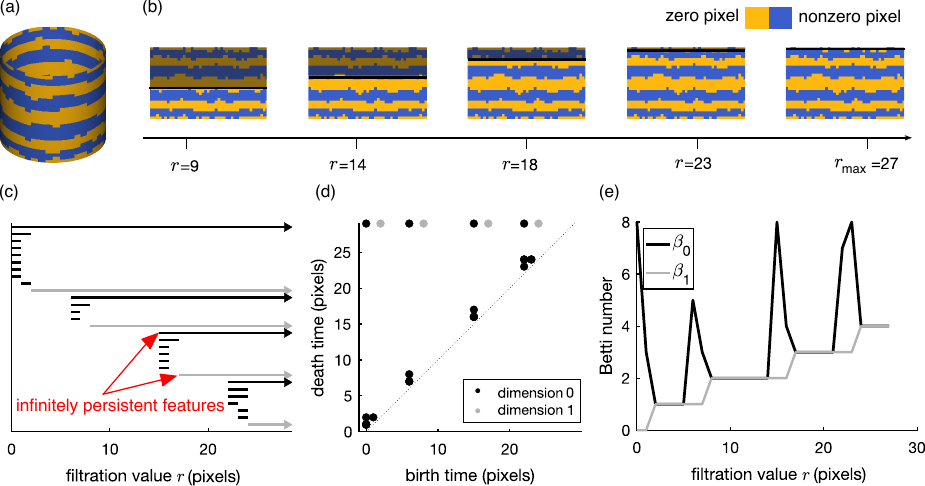}\vspace{0.5\baselineskip}
    \caption{Example of computing and visualizing persistent homology for a binary image of an \textit{in silico} zebrafish pattern. (a) The agent-based model \cite{volkening2018} has periodic boundary conditions in $x$, so we associate the left and right boundaries when sweeping from top to bottom or bottom to top. (b) As an illustrative example, we filter from bottom to top on a binary image with a voxel width of $\varepsilon=80$ $\mu$m here. At each filtration step, we ignore the grayed-out region of the image above the sweeping line. (c) This barcode describing the dimension-$0$ and dimension-$1$ topological features associated with the filtration in (c). Each bar ending in an arrow corresponds to a persistent topological feature. The results of persistent homology can also be represented by (d) a persistence diagram, where each point denotes the birth and death times of a topological feature; or by (e) plotting the Betti numbers $\beta_0$ (in black) and $\beta_1$ (in gray) as curves over different filtration values $r$. The maximum $r$ value is determined based on the length of the image in pixels in the sweeping direction. Because we consider the bottom-to-top sweeping-plane filtration in this example, $r_{\text{max}}=27$~pixels, and any features with a death time of $r_\text{death} = 27$~pixels are infinitely persistent.}
    \label{fig:persistent_homology}
\end{figure}

When computing persistent homology using the sweeping-plane filtration, one could choose to sweep across an image in any direction, but we focus on four directions: top-to-bottom (TB), bottom-to-top (BT), left-to-right (LR), and right-to-left (RL). These directions are natural for our application because horizontal stripes form sequentially in zebrafish skin in the dorsal and ventral directions during development; see Fig.~\ref{fig:biology} and Sect.~\ref{sec:biobackground}. Following the approach \cite{Nardini2021}, we thus consider $\beta_i(K^{\nu}_{r})$ for dimension $i\in\{0,1\}$ and sweeping direction $\nu\in\{$TB, BT, LR, RL$\}$. Here $K^\nu$ refers to the filtered simplicial complex associated with the sweeping direction $\nu$. Because the agent-based model \cite{volkening2018}, has periodic boundary conditions in the horizontal direction and wall-like boundary conditions at the top and bottom of the domain, we take this into account when we implement the TB or BT 
filtrations. In particular, we effectively wrap the pattern into a cylinder by gluing together the left and right boundaries when sweeping up or down; see Fig.~\ref{fig:persistent_homology}(a)--(b). When we compute persistent homology with the LR or RL 
filtrations, on the other hand, we do not include periodicity.

For reference, we define some notation 
that we use throughout Sections~\ref{sec:step2}--\ref{sec:step4} below:
\begin{itemize}
    \item \textit{Persistent feature} (P): Given a sweeping direction $\nu\in\{$TB, BT, LR, RL$\}$, we refer to a persistent feature as a connected component or loop that has a death time $r_\text{death} \ge r_\text{max}$, where $r_\text{max}$ represents the maximum filtration value (in pixels) that we consider in sweeping direction $\nu$. This maximum value is based on the dimensions of the binary image that we are considering. For example, $P^{\text{dim}0,\nu}_j$ denotes the $j$th persistent connected component (dimension-$0$ hole) in the barcode for the filtered simplicial complex $K^\nu$. In all cases, we assign features an order based first on their birth times and then on their persistence, so that the feature with the earliest birth time and longest persistence appears first. 
    \item \textit{Non-persistent feature} (NP): A non-persistent feature is a connected component or loop that dies for some filtration value $r_{\text{death}}<r_{\text{max}}$, where $r_\text{max}$ is the maximum filtration value (in pixels) in the sweeping direction $\nu$ under consideration.
    \item \textit{Zero-born feature} (ZB): A zero-born feature is a connected component or loop that is born at filtration value $r_{\text{birth}}=0$ pixels in the sweeping direction  $\nu$ under consideration.
    \item \textit{Nonzero-born feature} (NZB): A nonzero-born feature is a connected component or loop that is born at a filtration value $r_{\text{birth}}>0$ pixels in the sweeping direction $\nu$ under consideration.
\end{itemize}
We layer these terms when building our methodology in Sections~\ref{sec:step2}--\ref{sec:step4}. For example,
$\text{PNZB}^{\text{dim}0,\nu}_j$ denotes the $j$th persistent, nonzero-born connected component---when ordered based first on increasing birth time and second on decreasing persistence---in the filtered simplicial complex associated with the sweeping-plane filtration in the direction $\nu$. We are also often interested in $\beta_0(K^{\nu}_{r_\text{max}})$ and $\beta_1(K^{\nu}_{r_\text{max}})$, the numbers of connected components and holes, respectively, that are present at the maximum filtration value $r=r_{\text{max}}$ for each direction $\nu\in\{$TB, BT, LR, RL$\}$.

\subsection{Step 1: Generating and preparing agent-based data}\label{sec:step1}

As we discuss in Sect.~\ref{sec:biobackground}, we use the agent-based model \cite{volkening2018} to generate \textit{in silico} zebrafish patterns that feature variable and messy stripes and spots. This model \cite{volkening2018} is the focus of two studies \cite{McGuirl2020, Cleveland2023} that quantify patterns based on the Vietoris--Rips filtration, so it sets up an excellent case study for us to determine what information alternative approaches to persistent homology can provide about the same patterns. Specifically, we use a public dataset \cite{mcguirl2020zebrafish} containing simulations of the model \cite{volkening2018}. We focus on two conditions and consider $1000$ stochastic simulations for each: wild-type and \textit{pfeffer} mutant patterns\footnote{The \textit{pfeffer} mutant (encoding csf1rA) lacks xanthophores \cite{Maderspacher2003,ParTur130,PatDev127}, so the model \cite{volkening2018} simulates \textit{pfeffer} simply by turning off loose and dense xanthophore differentiation, with all other parameters the same as in the wild-type condition. Since the model \cite{volkening2018} features stochastic cell interactions, repeated simulations lead to slightly different \textit{in silico} patterns, offering predictions about biological variability.}; see examples of wild-type and \textit{pfeffer} zebrafish in Fig.~\ref{fig:biology}(a)--(b) and example simulations in Fig.~\ref{fig:summary}(a) under Step $1$. 
We choose these two conditions because they allow us to illustrate our methodology on stripe (wild-type) and spot (\textit{pfeffer}) patterns; future work could also consider other mutants, many of which are spotted. Following the TDA studies \cite{McGuirl2020,Cleveland2023}, we focus on simulated pattens representing juvenile zebrafish. Based on the model growth rates and initial conditions \cite{volkening2018}, this means that we consider patterns at the simulation time corresponding to $66$ days post fertilization\footnote{The initial condition for the model \cite{volkening2018} is $21$ days post fertilization, so quantifying patterns at $66$ days means that we apply our methods to simulated data \cite{mcguirl2020zebrafish} at time point $45$.}, which is the final time point for the wild-type simulation data \cite{mcguirl2020zebrafish}. We note that McGuirl \textit{et al.} considered \textit{pfeffer} patterns at $76$ days post fertilization, but we consider the same stochastic simulations at the earlier time of $66$ days post fertilization.

At $66$ days post fertilization, model domains \cite{volkening2018} are $3.71$ mm long and $2.215$ mm high; as we discuss in Sect.~\ref{sec:modelbackground}, these domains capture the full fish height and roughly a third of its patterned body length. 
For \textit{pfeffer} mutant patterns \cite{Maderspacher2003,ParTur130,PatDev127,Frohnhofer}, as in Fig.~\ref{fig:summary}, we expect to see blue spots of various sizes roughly aligned in stripes. For wild-type patterns, we expect to see three gold interstripes (X1D, X0, and X1V) and four blue stripes (2D, 1D, 1V, and 2V) at this time point, and it is also common to have some gold cells making up partially formed interstripes at the dorsal and ventral domain boundaries. One of the measurements that Volkening and Sandstede \cite{volkening2018} used to judge the success of their model was a low number of wild-type patterns with interruptions in interstripes X1D, X0, and X1V at $66$ days post fertilization. Notably, they did not consider interruptions in blue stripes, as breaks occur more frequently for these patterns in real fish \cite{volkening2018}, and this means the modeling focus was largely on the portion of each pattern spanned between X1D and X1V. Because the patterns are generally messier near their dorsal and ventral boundaries and may not be fully formed there, prior TDA studies \cite{McGuirl2020,Cleveland2023} cropped the top and bottom of the domain before applying persistent homology, removing stripes 2D and 2V. Importantly, here we are interested in both pattern features and pattern irregularities, and, unlike prior approaches, we do not crop the simulated patterns \cite{mcguirl2020zebrafish} before quantifying them.

\subsection{Step 2: Choosing a TDA-informed resolution and transforming point-cloud data to images}
\label{sec:step2}

\begin{figure}[t!]
    \centering 
    \includegraphics[width=1\linewidth]{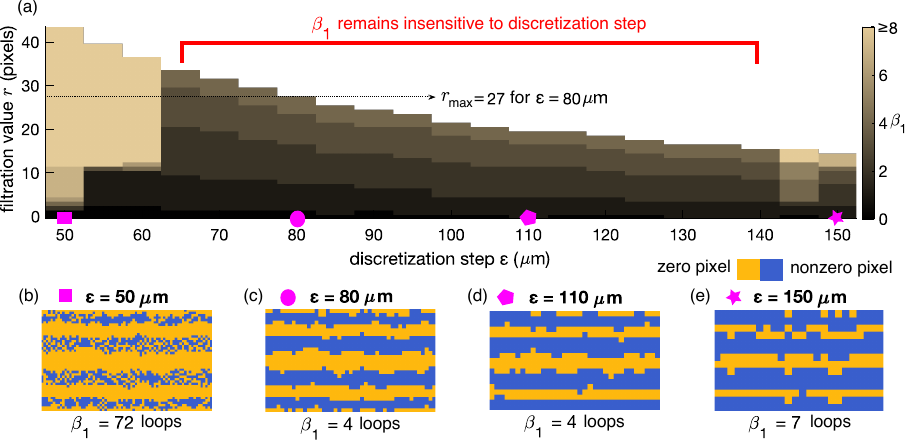}\vspace{0.5\baselineskip}
    \caption{Our TDA-based method for choosing a voxel width to construct binary images from point-cloud data. (a) In order to make agent-based patterns \cite{volkening2018} suitable for the sweeping-plane filtration, we transform them into binary images in Step~$2$ of our pipeline (see Fig.~\ref{fig:summary}), and this necessitates the choice of a discretization step $\varepsilon$. To determine $\varepsilon$, we apply the bottom-to-top sweeping-plane filtration to binary images based on the same patterns discretized using different resolutions. We then construct a heatmap of the Betti number $\beta_1$ versus filtration value ($r$ in pixels) and voxel width ($\varepsilon$ in $\mu\text{m}$), across $25$ sample unbroken striped patterns. (b)--(e) We show some binary images corresponding to the same pattern at different resolutions. Our goal is to choose a voxel width that neither introduces noise (e.g., the gold pixels in blue stripes in (b)) nor creates artifacts that are not present in the pattern (e.g., the gold-stripe break in (e)). Voxel widths in the range $65$--$140$ $\mu$m lead to $\beta_1(r)$ values that are fairly insensitive (up to a scaling by the image size in pixels), and we use this observation to choose $\varepsilon$.}
    \label{fig:voxel_width}
\end{figure}

Given the positions of pigment cells in simulated patterns \cite{mcguirl2020zebrafish}, the next step is to represent these point-cloud data as binary images so that we can compute persistent homology with the sweeping-plane filtration. To do so, we first choose a focal cell type for wild-type and \textit{pfeffer} patterns, and we base our binary images on the presence or absence of this cell population. We select the ($x$, $y$)-coordinates of black melanophores
for wild-type patterns and the ($x$, $y$)-coordinates of loose blue iridophores
for \textit{pfeffer} patterns. These cell types both occupy the dark stripes in wild-type patterns and the dark spots in \textit{pfeffer} mutants\footnote{We could alternatively use melanophores to produce binary images for wild-type and \textit{pfeffer} patterns. However, sparse melanophores are randomly distributed in \textit{pfeffer} patterns, so we follow the same approach as McGuirl \textit{et al.} \cite{McGuirl2020} in using loose iridophores for TDA with \textit{pfeffer}.}. Binary images are composed of two pixel intensities, $0$ or $1$. Given some spatial discretization step $\varepsilon$, we transform cell coordinates to binary images by binning cells in voxels or ``pixels". The value of each pixel is set to $1$ if it contains one or more cells of the focal type, or to $0$ if it does not. We represent $0$-intensity pixels as gold and $1$-intensity pixels as blue to relate these images to zebrafish patterns; see Fig.~\ref{fig:voxel_width}(a)-(e). For the developmental time that we consider, the full domain is patterned, so it is not a strong simplification to assume that any regions that are not dark blue, as signaled by the presence of black melanophores or loose blue iridophores, are gold (i.e. containing interstripe cells).

An important step in converting a point cloud into a binary image is choosing a spatial discretization step or voxel width $\varepsilon$, which controls the image resolution. Our goal is to identify a discretization step that preserves essential pattern details while minimizing noise resulting from the discretization that may generate artifacts during topological data analysis. In order to help make our quantification pipeline more widely applicable and reduce subjectivity in hyperparameter selection, we develop an automated approach to select the discretization step $\varepsilon$. Our methodology for choosing $\varepsilon$ is motivated in part by the concept of ``Betti CROCKER plots" introduced by Topaz \textit{et al.} \cite{Topaz2015}. Developed to analyze time-dependent data on flocking and swarming, CROCKER plots, which can be represented as heatmaps, are a means of visualizing how Betti numbers vary as a function of the filtration step $r$ and time \cite{Topaz2015}. With time as the horizontal axis and the filtration step $r$ as the vertical axis, CROCKER plots summarize the Betti curves in a given dimension for multiple snapshots of data; each vertical slice of the heatmap, corresponding to a given time, is a Betti curve in the traditional sense. With this as motivation, we instead consider heatmaps of Betti numbers across filtration values $r$ (vertical axis) for binary images that we generate based on various discretization steps $\varepsilon$ (horizontal axis); see Fig.~\ref{fig:voxel_width}.

Specifically, we consider a sample of $25$ unbroken striped patterns, and for each of these patterns we generate a collection of binary-image representations for a range of candidate voxel widths $\varepsilon$. For each binary image, we compute persistent homology based on the BT 
filtration with periodic boundary conditions in the $x$-direction; see Fig.~\ref{fig:persistent_homology} for an example. We then plot heatmaps of the Betti numbers $\beta_1(K^\text{BT}_r)$ as a function of the filtration step $r$ for binary images based on different voxel widths $\varepsilon$. As we show in Fig.~\ref{fig:voxel_width}, there is a range of voxel widths over which the number of loops $\beta_1(K^\text{BT}_r)$ behaves the same regardless of the image resolution. (In Fig.~\ref{fig:voxel_width}, this insensitivity range, in terms of the first Betti number, is about $65$--$140$ $\mu$m.) We choose the voxel width for all patterns in this study to be the mean (over the $25$ samples) of the $10^{\text{th}}$ percentiles of these ranges and then round to the nearest integer, leading to $\varepsilon=80$ $\mu\text{m}$, a value that is slightly larger than the average distance between most neighboring pigment cells \cite{2016heterotypic,ParTur256,TakahashiMelDisperse}.  The endpoints of the insensitivity range vary slightly across stochastic patterns, so we average over approximately the $10^{\text{th}}$ percentile to help ensure that our discretization step is large enough to capture all patterns. We stop the discretization at the largest multiple of $\varepsilon=80$ $\mu\text{m}$ that is less than or equal to the width/height of the point-cloud data, so any partial intervals beyond that are dropped.

We note that the voxel width $\varepsilon$ could alternatively be determined using heatmaps of the Betti numbers $\beta_1(K^\text{TB}_r)$ associated with the TB 
filtration. However, we find that the Betti numbers $\beta_0(K^\text{TB}_r)$ and $\beta_0(K^\text{BT}_r)$, corresponding to the number of connected components under the TB and BT 
filtrations respectively, are less useful for selecting the voxel width. 
Sweeping orthogonal to the stripe orientation captures variations in the spacing, width, and alignment of stripes, as well as deviations from perfectly straight stripe boundaries, and this manifests as oscillations in the number of connected components $\beta_0$. Thus, heatmaps of Betti numbers $\beta_0(K^\text{TB}_r)$ and $\beta_0(K^\text{BT}_r)$ are more variable. 
In addition, we do not consider the LR and RL 
filtrations when selecting $\varepsilon$ because they do not fully capture the domain's periodicity. 
Another approach would be to consider multi-parameter persistent homology \cite{Hu2021,CarlssonMulti,CarlssonMulti2,Harrington2019} as a means of filtering across both $\varepsilon$ and $r$, and we suggest this is a valuable direction for future work.

\subsection{Step 3: Interpreting topological summaries to classify patterns as stripes, interrupted stripes, or spots} \label{sec:step3}

\begin{figure}[t!]
    \centering 
    \includegraphics[width=1\linewidth]{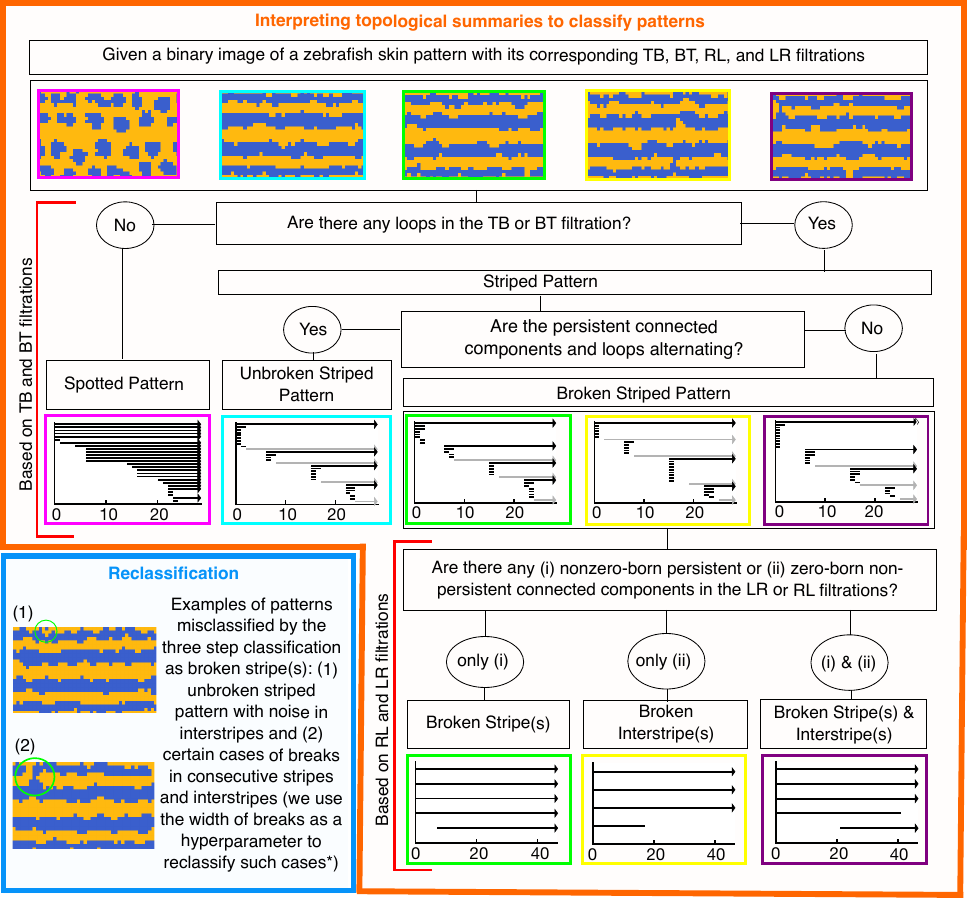}\vspace{0.5\baselineskip}
    \caption{Summary of our algorithm for classifying messy patterns by type based on persistent homology with the sweeping-plane filtration. We first use signatures of the barcodes associated with the TB and BT filtrations to distinguish between spotted, perfectly striped, or irregularly striped patterns. (Perfectly striped patterns are those with no interruptions or breaks; see Fig.~\ref{fig:summary}.) We next interpret the results of persistent homology with the RL and LR filtrations to further separate irregularly striped patterns into those with breaks in stripes, breaks in interstripes, or breaks in both stripes and interstripes. This process relies on assigning dimension-$0$ and dimension-$1$ features an order based first on their birth time and then on their persistence, so that features with low $r_\text{birth}$ values and high persistence values appear first. (The first row of example barcodes correspond to the BT filtration and the second row of example barcodes corresponds to the RL filtration.) As we discuss in Sect.~\ref{sec:step3}, our algorithm leads to misclassifications in rare cases, and we provide more details on misclassified patterns in Figures~\ref{fig:misclassified_patterns}--\ref{fig:common_misclassifications} in the appendix.}
    \label{fig:algorihm_diagram}
\end{figure}

Before we can interpret persistent homology as detailed information about stripes and spots in Step $4$ (Sect.~\ref{sec:step4}) of our pipeline, we must first classify patterns by type. With our $1000$ wild-type and $1000$ \textit{pfeffer} binary images in hand, all with a pixel width of $80$~$\mu$m, our goal is to automatically and blindly distinguish between patterns that are (1) spotted{, (2) ``perfectly" striped, or (3) ``irregularly" striped. As we show in Fig.~\ref{fig:summary}, we define ``perfect" stripe patterns as images in which there are no breaks or interruptions in gold or blue stripes. For ``irregular stripes", we consider three sub-categories: (3a) patterns with \textit{broken stripe(s)} feature a gold interruption breaking at least one blue stripe into more than one piece; (3b) patterns with \textit{broken interstripe(s)} have a blue break dividing at least one gold interstripe into multiple sections; and (3c) patterns with \textit{broken stripe(s) and interstripe(s)} feature both gold and blue interruptions. (We do not use ``irregular" to indicate that broken wild-type stripes are uncommon \textit{in vivo}, as we do not know of large-scale quantitative studies investigating this. It is also worth noting that, because the majority of mutant zebrafish patterns feature spots or stripes with more frequent interruptions \cite{Maderspacher2003,Watanabe2006,2012changing,IrionGap,Iwashita}, we expect that being able to blindly sort patterns into these categories is directly valuable for a much wider selection of patterns than we consider in this study.)

Our methodology for automatically sorting patterns by type relies on persistent homology. Broadly, for each blue and gold image in our dataset, we compute persistent homology using the sweeping-plane filtration (Sect.~\ref{sec:SPbackground}) for dimensions $i=\{0,1\}$ and directions $\nu=\{$TB, BT, LR, RL$\}$. We identify signatures in the barcodes associated with each image that allow us to first distinguish between stripe or spot patterns, and then further classify stripe patterns depending on whether or not they feature interruptions. As we discuss in Sect.~\ref{sec:step2}, the blue pixels are the signal in our patterns, and gold pixels play the role of the background from the perspective of the sweeping-plane filtration. Since we enforce periodic boundary conditions in the $x$-direction as in the agent-based model \cite{volkening2018} when we sweep from top to bottom or from bottom to top, a blue stripe wrapping around the domain is a persistent (dimension-$1$) loop; see Fig.~\ref{fig:persistent_homology}. A blue spot, in turn, is a (dimension-$0$) connected component, similar to studies \cite{McGuirl2020,Cleveland2023} with the Vietoris--Rips filtration. Because we do not enforce periodic boundary conditions when we sweep from left to right or from right to left, we do not expect loops for the associated filtered simplicial complexes.

As an example to build intuition for our classification pipeline, we discuss persistent homology for one wild-type zebrafish pattern under the BT  
filtration in Fig.~\ref{fig:persistent_homology}. The four uninterrupted blue stripes give rise to four dimension-$1$ holes, visible as the gray bars in the barcode in Fig.~\ref{fig:persistent_homology}(c) and the gray points in the persistence diagram in Fig.~\ref{fig:persistent_homology}(d). The arrows in the barcode indicate infinite persistence, and we refer to the associated infinitely persistent features as ``persistent". We also highlight the four steps in the Betti-$1$ curve in Fig.~\ref{fig:persistent_homology}(e). Each step in the Betti-$1$ curve, corresponding to the birth of a new loop or uncovering of enough of another blue stripe to span the full domain length, is proceeded by a jump in $\beta_0(K^\text{BT}_r)$. These jumps capture the roughness or non-uniformity of stripe boundaries, and they are also visible as the short black bars proceeding the birth of each new loop (gray) in the barcode in Fig.~\ref{fig:persistent_homology}(c).

The diagram in Fig.~\ref{fig:algorihm_diagram} summarizes our classification methodology, which consists of three main steps followed by an additional process to handle some rare cases. The first step of our algorithm focuses on the barcodes from the TB and BT  
filtrations. Given a matrix of ones and zeros representing a blue and gold image, we start by classifying the pattern broadly as striped or spotted. Since our periodic boundary conditions in $x$ mean that blue stripes are loops in the TB and BT 
filtrations, we define stripe patterns as those whose barcodes have at least one persistent loop. We define spotted patterns, on the other hand, as those with no loops. Critically, if a wild-type pattern has breaks in all of its blue ``stripes" $2$V, $1$V, $2$D, and $1$D, we consider it spotted. We suggest it is a subjective, qualitative choice whether such patterns should be called ``striped" or ``spotted".  Notably, all $1000$ \textit{pfeffer} patterns that we quantify have no loops and are classified as spotted (as expected), and only $3$ out of $1000$ wild-type patterns have breaks in all of their blue stripes and are classified as spotted; see Sect.~\ref{sec:results}.

Still focusing on the barcodes for the TB and BT 
filtrations, our next step is to further classify the patterns that we have identified as striped into two sub-categories: unbroken (e.g., perfect) or broken (e.g., irregular). As we show in Fig.~\ref{fig:algorihm_diagram}, the presence of alternating persistent connected components and loops in the TB and BT barcodes is a signature of \textit{unbroken striped patterns} in our dataset. Here we mean alternating in the sense of the birth times $\{r_\text{birth}\}$ of these features, and we focus on the birth times of persistent dimension-$0$ and dimension-$1$ holes. If there are any consecutive persistent connected components or loops in the TB or BT barcodes, we classify the striped pattern as broken.

The third step of our classification pipeline in Fig.~\ref{fig:algorihm_diagram} focuses on the dimension-$0$ barcodes from the RL and LR 
filtrations and further sorts broken striped patterns into three sub-categories: \textit{broken stripe(s)}, \textit{broken interstripe(s)}, or \textit{broken stripe(s) and interstripe(s)}.  
We find that a signature of the LR and RL barcodes for a striped pattern with broken stripe(s) but uninterrupted interstripes is the presence of at least one nonzero-born persistent connected component in either of these filtrations. Similarly, we identify patterns with broken interstripe(s) but uninterrupted stripes as those with at least one zero-born, non-persistent connected component in either the LR or RL barcode. This is because the death of a zero-born dimension-$0$ feature marks the merging of two connected components, each initially corresponding to a blue stripe.  
Finally, we classify a striped pattern as having breaks in both stripe(s) and interstripe(s) if there are both nonzero-born, persistent connected component(s) and zero-born, non-persistent connected component(s) in the LR or RL filtrations.

Our three-step process above can handle the vast majority of our $2000$ patterns, but there are a few special cases that we address by introducing additional checks to prevent misclassifications.
First, some cases of patterns with breaks in two consecutive stripes and interstripes can satisfy the condition of alternating persistent connected components and loops in the TB and BT barcodes. To avoid misclassifying such patterns as \textit{unbroken}, we add an extra step of checking that every pattern with alternating persistent connected components and loops in the TB and BT barcodes does not also satisfy the RL and LR barcode signatures of patterns with breaks in both stripe(s) and interstripe(s). We note that $6$ out of the $1000$ wild-type patterns in our dataset are patterns with broken stripe(s) and interstripe(s) that would be misclassified as \textit{unbroken striped patterns} without this additional check.  

Second, in rare cases, stochastic cell differentiation in the model \cite{volkening2020} can lead to one or more stray melanophores at the dorsal or ventral domain boundaries\footnote{Because the model \cite{volkening2018} includes local competition between cells in the gold and blue regions and limits differentiation in dense areas to prevent overcrowding, it is very unlikely to observe stray blue pixels anywhere except near the dorsal or ventral domain boundaries, where the pattern has formed most recently.}. The result is a pattern like the first one in the misclassification box in Fig.~\ref{fig:algorihm_diagram}, where a stray blue pixel is near the top of the domain.  Qualitatively, this is an unbroken striped pattern, but our three-step method would classify it as \textit{broken stripe(s)}. To gain intuition for why, notice that an isolated blue pixel is topologically the same as a long horizontal strip of blue pixels that does not fully span the domain; also 
see Fig.~\ref{fig:misclassified_patterns}(a) in the appendix.
We introduce a hyperparameter-based condition using measurements of the widths of the breaks in ``stripe(s)" to resolve this. (See Sect.~\ref{sec:defects} for how we count stripes and measure the width of each interruption.) This additional condition uses that a common signature of patterns with noisy blue pixels in interstripe regions is that our methods in Sect.~\ref{sec:defects} output a stripe break with negative width. If this happens, we flag the break as incorrectly identified, reduce our count of the number of breaks in stripes by one, and reclassify the pattern as an \textit{unbroken striped pattern} if the new break count is zero. Such cases of patterns flagged by our algorithm due to negative widths form $0.1$\% of our data\footnote{Encountering the opposite issue---having a stray gold pixel in a blue stripe---is very rare in our dataset because of the nature of the agent-based model \cite{volkening2018} and our choice of voxel width in Step~$2$ (Sect.~\ref{sec:step2}) of our pipeline. However, if a stray gold pixel is present in a striped pattern, our three-step classification algorithm could misclassify it. For an unbroken striped pattern with a stray gold pixel in the interior of a blue stripe, the birth times of the persistent connected components and loops for the TB and BT  
filtrations do not alternate, and our algorithm would consider the pattern as broken. Thus, as an added fail-safe, we introduce another check for unbroken patterns of whether all connected components in the RL and LR barcodes are zero-born persistent. This helps ensure that unbroken striped patterns with stray gold pixels in the interior of blue stripes are correctly classified. Notably, in all the stripe patterns that we quantified, none had stray gold pixels in the interior of blue stripes that would trigger the fail-safe check. However, some striped patterns have stray gold pixels on the boundaries of some stripes. Such edge cases are not addressed by the fail-safe logic, and this can lead to misclassifications, as we discuss in the main text body.}.

The last special case that we address is rare misclassifications of particularly unlucky patterns with breaks in stripes and interstripes that occur next to each other 
($x$-position-wise) in a consecutive stripe and interstripe; see the bottom pattern in the misclassification box in Fig.~\ref{fig:algorihm_diagram} for an example. In such cases, the breaks are not clear from the barcode signature and the pattern is classified with a single break type---usually as \textit{broken stripe(s)}. 
To resolve that, we introduce a hyperparameter-based condition to check if the width of any break(s) is greater than $40$\% of the image length (this usually happens when there are multiple breaks of different types in a consecutive stripe and interstripe) and reclassify the pattern with \textit{breaks in both stripe(s) and interstripe(s)}; see Fig.~\ref{fig:misclassified_patterns}(b) in the appendix. Once this happens, we increase the count of breaks in stripes and interstripes by one, and reclassify the pattern as \textit{broken stripe(s) and interstripe(s)}. Such cases of patterns flagged by our algorithm due to unreasonably large interruption widths form $1.2$\% of the data.

To help support the application of our methods to other datasets, Fig.~\ref{fig:common_misclassifications} in the appendix highlights a few examples of rare patterns where are classification algorithm---even after including the additional checks that we discuss above---fails. For example, broken stripe(s) patterns can be misclassified as \textit{broken stripe(s) and interstripe(s)} due to the presence of stray gold pixels on the left or right boundary of the domain. In particular, a broken stripe(s) pattern with one stray gold pixel at the right boundary of a stripe will have a RL barcode with a zero-born, non-persistent connected component of very low persistence (exactly one pixel).  
Notably, this RL barcode is similar to that of a true \textit{broken stripe(s) and interstripe(s)} pattern with an interstripe break located exactly one pixel away from the right boundary. This makes it challenging to distinguish between these two cases based solely on their barcodes; see Fig.~\ref{fig:common_misclassifications}(a) in the appendix. These misclassifications, which we estimate to occur less than $1\%$ of the time based on manually checking $100$ randomly selected images, represent a limitation of our method.

In summary, whereas prior zebrafish studies \cite{McGuirl2020,Cleveland2023} with the Vietoris--Rips filtration input known spot patterns into a separate quantification algorithm than known stripe patterns, our approach in Step~$3$ blindly and automatically classifies patterns. The three main steps of our classification process are hyperparameter-free, and we introduce hyperparameters only to address rare misclassification events associated with stray cells or consecutive breaks.
As an alternative approach to the additional process that we introduce, we could instead clean patterns before applying persistent homology as in \cite{Cleveland2023}. For example, we could swap the color of any pixel surrounded by pixels of the opposite color, and this would prevent issues related to stray pixels. Notably, we find that only $32$ of the $2000$ stripe and spot patterns that we quantify enter the reclassification process in our pipeline; the other approximately $98$\% of patterns are classified based on our main three-step algorithm. For more details on misclassified patterns and how we resolve some of them using hyperparameters, see Figures~\ref{fig:misclassified_patterns}--\ref{fig:common_misclassifications} in the appendix.

\subsection{Step 4: Characterizing pattern features and irregularities} \label{sec:step4}

After blindly classifying patterns by type in Step $3$ (Sect.~\ref{sec:step3}) of our pipeline, we now turn to quantifying the appropriate features and irregularities in each pattern using topological summaries based on the sweeping-plane filtration. For spot patterns, we characterize the number of blue spots and their size. For perfect striped patterns without interruptions, we interpret barcode information in terms of the number of blue stripes, and we quantify stripe and interstripe width based on topological data analysis. As our main result, for stripe patterns with interruptions, we show how results from the sweeping-plane filtration can naturally be interpreted in terms of the numbers of stripes and interstripes, the numbers of interruptions in stripes and interstripes, the widths of these interruptions, and the locations of these defects along the dorsal--ventral axis of the domain. Our methodology for counting stripes in unbroken striped patterns and spots in spot patterns is in Sect.\ \ref{sec:countStripesSpots}, for quantifying stripe width and spot size is in Sect.\ \ref{sec:sizeStripesSpots}, and for characterizing broken striped patterns in detail is in Sections \ref{sec:defects}--\ref{sec:defects2}.

\subsubsection{Counting spots and stripes in unbroken patterns}\label{sec:countStripesSpots}

For patterns classified as \textit{spotted} in Step~$3$ of our pipeline (see Fig.~\ref{fig:summary}), we estimate the number of blue spots as the number of persistent connected components based on the TB and BT 
filtrations. Specifically, we define:
\begin{align} \text{number of spots} ~&=~ \max \left\{ \beta_0\left(K^{\text{TB}}_{r_{\text{max}}}\right), \, \beta_0\left(K^{\text{BT}}_{r_{\text{max}}}\right) \right\},
\label{eq:spots}
\end{align}
where $r_\text{max}$ is the maximum filtration parameter that we consider when sweeping from left to right or from right to left, and $\beta_0(K^\nu_{r_\text{max}})$ for $\nu \in \{\text{TB},\text{BT}\}$ denotes the number of connected components at filtration step $r = r_\text{max}$. Our domains have height $2215$~$\mu$m and our voxel width is $80$~$\mu$m, so $r_\text{max} = 27$~pixels here. Because we enforce the model's \cite{volkening2018} use of periodic boundary conditions in $x$ under the TB and BT 
filtrations, a spot that spans the left and right boundaries of the domain is considered one spot.

As we discuss in Sect.~\ref{sec:step3} and Fig.~\ref{fig:persistent_homology}, complete, uninterrupted blue stripes form loops when we compute persistent homology based on the TB or BT 
filtrations. This observation suggests that we could interpret the number of persistent dimension-$1$ features, i.e.,  $\beta_1(K^{\text{TB}}_{r_{\text{max}}})$ or $\beta_1(K^{\text{BT}}_{r_{\text{max}}})$, as the stripe count. However, this approach only works in the case of unbroken striped patterns, and we are interested in counting stripes and interstripes even in the presence of breaks. Thus, to align our methodology for unbroken striped patterns with our methodology for striped patterns with various breaks in Sections~\ref{sec:defects}--\ref{sec:defects2}, we instead estimate the number of stripes based on the number of persistent zero-born connected components according to the RL and LR 
filtrations. Specifically, for patterns that we classify as \textit{unbroken striped patterns} in Step~$3$ of our pipeline, we define:
\begin{align}
\text{number of blue stripes} ~&=~ \max_{\substack{\text{zero-born} \\ \text{dim.\ 0}}} \left\{ \beta_0\left(K^{\text{RL}}_{r_{\text{max}}}\right), \, \beta_0\left(K^{\text{LR}}_{r_{\text{max}}}\right) \right\},\label{eqn:stripe}
\end{align}
where the domain length $r_\text{max} =45$ pixels, and we use the notation:
\begin{align}
  \max_{\substack{\text{conditions}}}\{\beta_a, \beta_b\}  = \text{max}~(&\beta_a - \text{\#bars in associated barcode $a$ that do not satisfy conditions},\\
  &\beta_a - \text{\#bars in associated barcode $b$ that do not satisfy conditions} )\label{eq:notation}
\end{align}
throughout this manuscript. While we choose to compute persistent homology based on the positions of blue pixels in this study, one could count the number of interstripes by instead taking the gold pixels as the signal and repeating our process in Steps~$1$--$4$ of Fig.~\ref{fig:summary} with dense iridophores, rather than melanophores or loose iridophores, as the focal cell type.

\subsubsection{Quantifying stripe width and spot size}\label{sec:sizeStripesSpots}

Here we develop a method to interpret topological summaries from the sweeping-plane filtration in terms of stripe width. Our approach relies on assigning topological features an order, as is common in barcodes, based first on their birth time, and then---for any features with the same birth time---based on their persistence, with features that persist for a longer time taking precedence. For example, in Fig.~\ref{fig:stripe_width}(a)--(b), the first persistent feature that we encounter when sweeping from top to bottom is stripe 2D. It gives rise to a zero-born persistent connected component, a persistent loop with $r_\text{birth} = 2$~pixels, and several dimension-$0$ features with low persistence that capture the roughness of the stripe boundary. With the terminology from the start of Sect.~\ref{sec:methodology} in hand, we refer to the persistent dimension-$0$ feature $\text{P}^{\text{dim}0,\text{TB}}_1$ and the persistent dimension-$1$ feature $\text{P}^{\text{dim}1,\text{TB}}_1$ as associated with the first stripe that we encounter when sweeping from top to bottom (i.e., stripe $2$D here). On the other hand, if we sweep from bottom to top, stripe $2$D is the last of four stripes uncovered in the example in Fig.~\ref{fig:stripe_width}. We thus refer to the persistent dimension-$0$ feature $\text{P}^{\text{dim}0,\text{BT}}_4$ and the persistent dimension-$1$ feature $\text{P}^{\text{dim}1,\text{BT}}_4$ as also associated with stripe 2D in this example.

Using this terminology, we exploit the fact that persistent connected components identify the ``outer limits” of a stripe, i.e., the first blue pixels of a stripe when sweeping vertically, and persistent loops identify the ``inner limits” of the stripe, i.e., the first strip of blue pixels in a stripe that spans the entire horizontal axis, to estimate stripe width. We estimate the width of each unbroken stripe using the birth times of its corresponding persistent connected component and loop for the TB and BT 
filtrations, before transforming from units of pixels to physical units ($\mu$m here). Specifically, for patterns classified as \textit{unbroken stripe(s)}, we define the
minimum and maximum width of the $j\text{th}$ blue stripe as:

\begin{align}
\text{max width of stripe $j$} ~&=~ \varepsilon \left( r_{\text{max}} - \hat{r}_{\text{birth}}\left(\text{P}^{\text{dim}0,\text{TB}}_j  \right) - \hat{r}_{\text{birth}}\left(\text{P}^{\text{dim}0,\text{BT}}_{N_\text{stripes}-j+1} \right) \right), \label{eq:width1}\\
\text{min width of stripe $j$} ~&=~ \varepsilon \left( r_{\text{max}} - \hat{r}_{\text{birth}}\left(\text{P}^{\text{dim}1,\text{TB}}_j  \right) - \hat{r}_{\text{birth}}\left(\text{P}^{\text{dim}1,\text{BT}}_{N_\text{stripes}-j+1} \right) \right),\label{eq:width2}
\end{align}

where $j$ corresponds to the $j\text{th}$ blue stripe as we sweep from top to bottom, $N_\text{stripes}$ is the number of blue stripes that we find based on Eqn.~\eqref{eqn:stripe}, $\varepsilon = 80$~$\mu$m is the voxel width, $r_\text{max} = 27$ pixels is the domain height, and the function $\hat{r}_\text{birth}(\text{P})$ outputs the birth time for feature P.

For example, for the pattern in Fig.~\ref{fig:stripe_width}, the birth times of $\text{P}^{\text{dim}0,\text{TB}}_3$ and $\text{P}^{\text{dim}0,\text{BT}}_2$ act as bounds on the vertical range of stripe 1D. Considering dimension-$1$ features, pairing the birth times of $\text{P}^{\text{dim}1,\text{TB}}_3$ and $\text{P}^{\text{dim}1,\text{BT}}_2$, in turn, provides an estimate of the minimum width of stripe 1D. And more generally, for unbroken striped patterns, pairing the birth times for topological features $j$ and $N_\text{stripes}-j+1$ when sweeping from top to bottom or from bottom to top, respectively, provides information on the width of the $j$th stripe from the dorsal boundary. While we consider persistent homology based only on blue pixels in this study, we note that one could 
estimate interstripe width in a similar way by selecting a different focal cell type in Step~$1$ (Sect.~\ref{sec:step1}). For striped patterns classified with breaks of any kind, however, Equations~\eqref{eq:width1}--\eqref{eq:width2} do not suffice because some corresponding persistent features will be ``missing". As an example, for the \textit{broken stripe(s)} pattern in Fig.~\ref{fig:algorihm_diagram} (green box), the first persistent feature that we encounter when sweeping bottom-to-top is stripe $2$V. It corresponds to a zero-born persistent connected component and several dimension-0 features, but no persistent loop due to the break in stripe $2$V. The first persistent loop that we encounter while sweeping bottom-to-top in this example corresponds to stripe $1$V, which is the first unbroken stripe from the bottom in this pattern. Using the terminology in Equations~\eqref{eq:width1}--\eqref{eq:width2} to define the width of stripe $2$V, the persistent dimension-$0$ feature $\text{P}^{\text{dim}0,\text{BT}}_1$ is associated with stripe $2$V, while the persistent dimension-$1$ feature $\text{P}^{\text{dim}1,\text{BT}}_1$ is associated with stripe $1$V.  
For this reason, we focus on reporting measurements of stripe widths for unbroken patterns only.
As future work, we expect that a rough estimate of the minimum or maximum width of broken stripes could instead be calculated using the birth and/or death time of the non-persistent feature closest to the supposedly ``missing" persistent feature in the barcode. For example, if a persistent connected component is missing, we can use the birth time of the first born non-persistent connected component corresponding to that stripe. Whereas, if a persistent loop is missing, we can use the death time of the last born non-persistent connected component corresponding to that stripe.

\begin{figure}[t!]
    \centering 
    \includegraphics[width=1\linewidth]{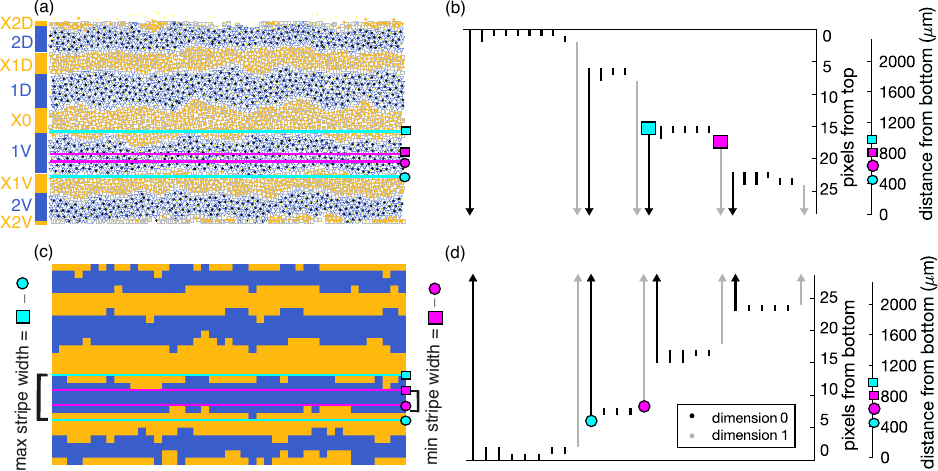}\vspace{0.5\baselineskip}
    \caption{Interpreting topological summaries from the TB and BT filtrations to quantify stripe width. We show (a) an unbroken striped pattern \cite{volkening2018} and (c) its corresponding binary image, as well as (b)--(d) barcodes describing connected components and loops in these data under the TB and BT filtrations, respectively. We use dimension-$0$ features to estimate the maximum width (spanned by the cyan lines) of stripe $1$V, the second stripe from the bottom of the pattern; to estimate its minimum width (spanned by magenta lines), we use dimension-$1$ features.
    Specifically, in (b)--(d), we indicate the birth times of the connected components and loops that we use to estimate the width of stripe $1$V in cyan and magenta, respectively. Square symbols correspond to birth times based on the TB filtration, and circles indicate the BT filtration. We emphasize that the additional axis (distance from bottom) in (b)--(d) serves as a visual guide, indicating the position from the bottom in $\mu$m of the square and circle symbols. In (b), the values on this axis are computed as the equivalent in $\mu$m of $r_{\max}$ (the image height in pixels) minus the number of pixels from the top, as illustrated in Equations~\eqref{eq:width1}--\eqref{eq:width2}.}
    \label{fig:stripe_width}
\end{figure}

We use a similar idea to quantify spot height and width; see 
Fig.~\ref{fig:spot_interpretation} for an example. Specifically, for patterns classified as \textit{spotted}, we define the height estimates of the $j$th spot as:
\begin{align}
\text{max height of spot $j$} ~&=~ \varepsilon \left( r_{\text{max}}  - \hat{r}_{\text{birth}}\left(\text{P}^{\text{dim}0,\text{TB}}_j  \right) - \hat{r}_{\text{birth}}\left(\text{P}^{\text{dim}0,\text{BT}}_{N_\text{spots}-j+1} \right) \right), \label{eq:spot_height}
\end{align}
where $j$ corresponds to the order of spots in terms of birth time (and then decreasing persistence) as we sweep from top to bottom, $N_\text{spots}$ is the number of spots that we find based on Eqn.~\eqref{eq:spots}, $\varepsilon = 80$~$\mu$m is the voxel width, $r_\text{max} = 27$ pixels is the domain height, and $\hat{r}_\text{birth}(\text{P})$ outputs the birth time for feature P.

To estimate spot length, we apply the same technique in Eqn.~\eqref{eq:spot_height}, but using LR and RL filtrations instead, as below:
\begin{align}
\text{max length of spot $j$} ~&=~ \varepsilon \left( r_{\text{max}}  - \hat{r}_{\text{birth}}\left(\text{P}^{\text{dim}0,\text{LR}}_j  \right) - \hat{r}_{\text{birth}}\left(\text{P}^{\text{dim}0,\text{RL}}_{N^{'}_\text{spots}-j+1} \right) \right),\label{eq:spot_length}
\end{align}
where $j$ corresponds to the order of spots in terms of birth time (and then decreasing persistence) as we sweep from left to right, $r_\text{max} = 45$~pixels is the domain length, $N^{'}_\text{spots} =\max \left\{ \beta_0\left(K^{\text{RL}}_{r_{\text{max}}}\right), \, \beta_0\left(K^{\text{LR}}_{r_{\text{max}}}\right) \right\}$ is the number of spots in the pattern assuming no periodicity when sweeping horizontally ($N^{'}_\text{spots}\geq N_\text{spots}$), and the function $\hat{r}_\text{birth}(\text{P})$ outputs the birth time for feature P. 

Due to the assumption of periodicity in the 
$x$-direction when sweeping vertically (but not horizontally), a spot that intersects both the left and right boundaries of the domain is assigned two length values and a single height value. The equations above give a rough estimate of the dimensions since it is challenging to track the exact order in which the spots are born in each direction just by looking at the barcodes. A manual check of $100$ spots shows that around $20$\% had inaccurate height estimates, while $34$\% had inaccurate length estimates due to mismatches in the corresponding connected components. We observe that length estimates exhibit higher error than height estimates, and we expect that this may be related to the nature of \textit{pfeffer} patterns---spots appear to be roughly organized in horizontal stripes, with no consistent vertical alignment. As we discuss in Sect.~\ref{sec:conclusion}, we suggest that the Vietoris--Rips filtration is more readily amenable to interpretation in terms of spot size \cite{Cleveland2023,McGuirl2020}, while we find that the sweeping-plane filtration naturally provides information about stripe width.

\begin{figure}[t!]
    \centering 
    \includegraphics[width=1\linewidth]{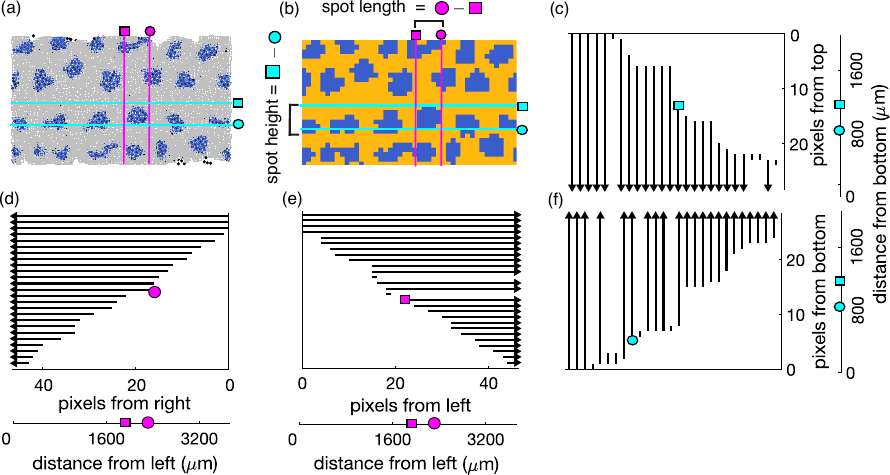}\vspace{0.5\baselineskip}
    \caption{Combining  and interpreting topological summaries from the TB, BT, LR, and RL filtrations to quantify spot size. As an illustrative example, we show (a) a \textit{pfeffer} pattern \cite{volkening2018}; (b) its corresponding binary image, with the length (magenta) and height (cyan) of an example spot indicated; and its associated dimension-$0$ barcodes based on the (c) TB, (d) RL, (e) LR, and (f) BT filtrations. We estimate the length of the $k$th spot using the birth times of the $k$th bar (cyan square in (c)) in our TB filtration and the ($N_\text{spots} - k +1)$th bar (cyan circle in (f)) in our BT filtration. (We order spot features based first on increasing birth time and then on decreasing persistence.) Similarly, we highlight the birth times of the connected components in the LR and RL filtrations that we use to estimate stripe length in magenta in (d)--(e). See Equations~\eqref{eq:spot_height}--\eqref{eq:spot_length}.}
    \label{fig:spot_interpretation}
\end{figure}

\subsubsection{Characterizing irregular striped patterns: number and width of breaks}\label{sec:defects}

Whereas stripe width and spot size have  \cite{Cleveland2023,McGuirl2020} been characterized based on the Vietoris--Rips filtration (Sect.~\ref{sec:VRbackground}), we now turn to a new perspective on irregular and broken striped patterns that interpreting topological summaries from the sweeping-plane filtration offers. We suggest that characterizing messy striped patterns is a particular strength of this approach. Our methodology in this section focuses on three main characteristics of irregular striped patterns: (1) detecting and counting stripe and interstripe interruptions; (2) counting stripes even in the presence of breaks; and (3) quantifying the horizontal width of interruptions. (See Sect.~\ref{sec:defects2} for our approach to identifying the locations of breaks.)
Importantly, if, for example, a gold bridge breaks a blue stripe into two pieces (as in the example pattern in the third column of Fig.~\ref{fig:algorihm_diagram}), our method counts those two pieces as one stripe. This mirrors what we would do ourselves if we looked at the pattern qualitatively. 

First, for patterns that we classify as \textit{broken stripe(s)}, \textit{broken interstripe(s)}, or \textit{broken stripe(s) and interstripe(s)} in Step~$3$ (Sect.~\ref{sec:step3}) of our pipeline, we detect and count the number of breaks in stripes and interstripes by interpreting the dimension-$0$ barcodes for the RL and LR 
filtrations. In particular, we consider nonzero-born persistent connected components and zero-born non-persistent connected components, and count stripe breaks as below:
\begin{align*}
\text{number of breaks in stripes} &= \max_{\substack{\text{nonzero-born} \\ \text{dim.\ 0}}} \left\{ \beta_0\left(K^{\text{RL}}_{r_{\text{max}}}\right), \, \beta_0\left(K^{\text{LR}}_{r_{\text{max}}}\right) \right\},
\end{align*}
where $r_\text{max} = 45$ pixels is the domain length, and we use the definition of $\text{max}_\text{condition}$ in Eqn.~\eqref{eq:notation}. We estimate the number of breaks in interstripes, in turn, as the maximum number of zero-born dimension-$0$ features with $r_\text{death} < r_\text{max}$ for the RL or LR 
filtrations. For example, whether sweeping leftward or rightward in Fig.~\ref{fig:gap_width}, notice the bridge in stripe $2$D is indicated by the presence of a connected component with birth time $r_\text{birth} >0$, while the break in interstripe X1V is captured by a connected component with death time $r_\text{death} < r_\text{max}$.

Once we know the number of interruptions in each striped pattern, we are ready to estimate the number of stripes---regardless of if they are unbroken (i.e., spanning the full length of the domain) or broken---according to:
\begin{align}
\text{number of stripes} &= \text{number of interstripe breaks} ~+\min_{\substack{\text{zero-born} \\ \text{dim.\ 0}}} \left\{ \beta_0\left(K^{\text{RL}}_{r_{\text{max}}}\right), \, \beta_0\left(K^{\text{LR}}_{r_{\text{max}}}\right) \right\},\label{eqn:stripe2}
\end{align}
where the domain length $r_\text{max} =45$ pixels, and we make the simplifying assumption that
\begin{align*}
    \text{number of broken interstripes} &= \text{number of interstripe breaks}.
\end{align*}
Importantly, breaks in interstripes are very uncommon in wild-type zebrafish patterns, and our methods suggest that, when present, they generally occur at most once per interstripe; see later in this section for our approach to identifying break locations. We thus assume that the number of breaks in interstripes is the same as the number of broken interstripes. We note that Eqn.~\eqref{eqn:stripe2} contains an additional term relative to Eqn.~\eqref{eqn:stripe} because we compute persistent homology based only on blue pixels and treat gold pixels as background throughout our pipeline. We expect that one could also count the number of interstripes and the number of breaks per interstripe by instead applying persistent homology based on the gold pixels in our pattern images.

\begin{figure}[t!]
    \centering 
    \includegraphics[width=1\linewidth]{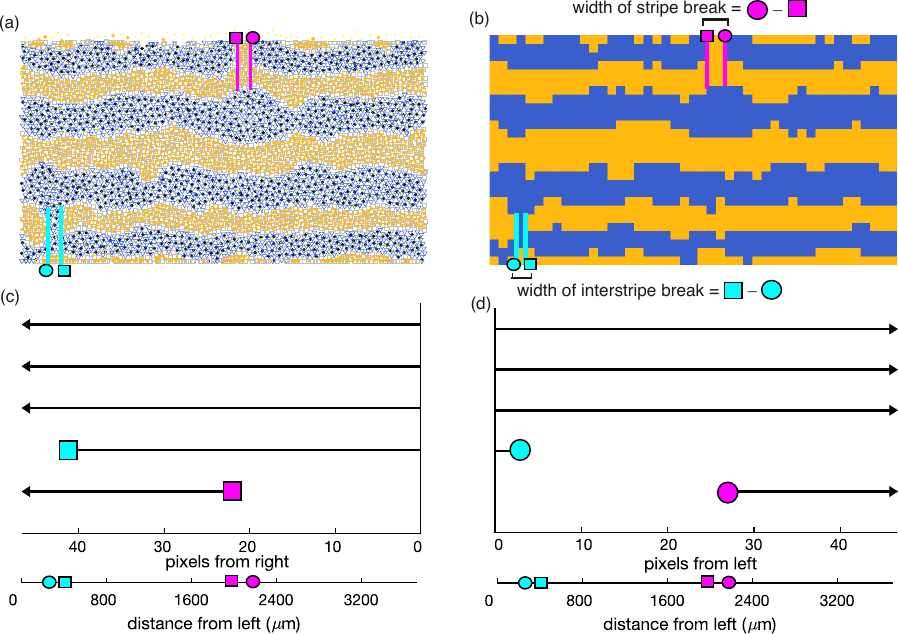}\vspace{0.5\baselineskip}
    \caption{Interpreting persistent homology based on the LR and RL filtrations as information about irregularities in striped patterns. While birth times of topological features provide insight into stripe width (see Fig.~\ref{fig:stripe_width}), death times lend themselves to estimating the widths of bridges that interrupt stripes. To illustrate our method for estimating the width of a stripe or interstripe interruption, we consider (a) an example agent-based pattern; (b) its corresponding binary image; (c) the dimension-$0$ barcode based on sweeping from left to right; and (d) the dimension-$0$ barcode associated with sweeping from right to left. The width of the interstripe break (spanned by cyan lines in (a)--(b)) is captured by the difference in the death time of the zero-born, non-persistent feature (cyan square) in (c) and the death time of the zero-born, non-persistent feature (cyan circle) in (d). Similarly, the death times of features that we use to estimate the width of the stripe break (spanned by magenta lines in (a)--(b)) are highlighted with magenta symbols in (c)--(d). Our measurements of gap width should be considered lower bounds on gap width, given the pattern discretization; also see Fig.~\ref{fig:multiple_breaks_patterns} and the appendix for pathological examples that can conceivably arise.}
    \label{fig:gap_width}
\end{figure}

With the number of interruptions in hand, we then estimate the width of each break $j$ as:

\begin{align}
\text{width of stripe break $j$} = \varepsilon \left( \hat{r}_{\text{birth}}\left( \text{PNZB}^{\text{dim}0,\text{LR}}_j \right) + \hat{r}_{\text{birth}}\left( \text{PNZB}^{\text{dim}0, \text{RL}}_{N_\text{sb}-j+1} \right) - r_{\text{max}} \right),\label{eq:gapwidth} \\
\text{width of interstripe break $j$} = \varepsilon \left( r_{\text{max}} - \hat{r}_{\text{death}}\left( \text{NPZB}^{\text{dim}0,\text{LR}}_j \right) - \hat{r}_{\text{death}}\left( \text{NPZB}^{\text{dim}0, \text{RL}}_{N_\text{ib}-j+1} \right) \right),\label{eq:gapwidth2}
\end{align} 
where $j$ in Eqn.~\eqref{eq:gapwidth} corresponds to the order of appearance of the right edge of a stripe break as we sweep left-to-right; $j$ in Eqn.~\eqref{eq:gapwidth2} corresponds to the order of appearance of the left edge of an interstripe break as we sweep left-to-right; $N_\text{sb}$ is the number of stripe breaks; $N_\text{ib}$ is the number of interstripe breaks; $\varepsilon = 80$ $\mu$m is the voxel width; the function $\hat{r}_\text{birth}(\text{P})$ outputs the birth time for feature P; the function $\hat{r}_\text{death}(\text{P})$ outputs the death time of feature P; $\text{PNZB}^{\text{dim}0,\nu}_j$ indicates the $j$th persistent, nonzero-born connected component for sweeping direction $\nu \in \{\text{LR}, \text{RL}\}$; and $\text{NPZB}^{\text{dim}0,\nu}_j$ denotes the $j$th non-persistent, zero-born connected component. 

As we discuss in Sect.~\ref{sec:results}, the model \cite{volkening2018} tends to produce wild-type patterns that have at most one stripe break and at most one interstripe break, so Eqn.~\eqref{eq:gapwidth} applies directly for the majority of the striped patterns in our dataset \cite{mcguirl2020zebrafish}; see Fig.~\ref{fig:gap_width}. Equations~\eqref{eq:gapwidth}--\eqref{eq:gapwidth2} also apply to cases of multiple breaks as long as the numbers of relevant topological features (persistent, nonzero-born connected components for stripe breaks; and non-persistent, zero-born connected components for interstripe breaks) are the same in both the LR and RL barcodes. However, if these numbers are different, difficulties arise. This situation is often a signature of a break occurring at the right or left domain boundaries. For example, if a pattern has one stripe break in the interior of the domain and another along the right domain boundary, we expect two persistent, nonzero-born connected components in the RL barcode and one in the LR barcode. One of each of these features is associated with the stripe break in the domain interior. The remaining PNZB feature in the RL barcode captures the left edge of the break at the domain boundary, so the domain boundary is its right edge.
We thus estimate here that the width of the stripe break at the boundary is $\varepsilon \min_{j\in\{1,2\}}\left( \hat{r}_{\text{birth}}\left(\text{PNZB}^{\text{dim}0, \text{RL}}_j \right) - 0\right)$, and then use the remaining PNZB features 
in Eqn.~\eqref{eq:gapwidth} to estimate the width of the interior stripe break; see Fig.~\ref{fig:multiple_breaks_patterns}(a) in the appendix for details..

There are a few limitations of our method for estimating break width that are important to discuss. First, 
we may estimate break width incorrectly when several breaks of the same type align at the same location across multiple stripes or interstripes. This makes it difficult to associate the topological features in the barcodes with the correct break, and we end up estimating the distance between the left edge of one break and the right edge of another rather than estimating actual break widths; see Fig.~\ref{fig:multiple_breaks_patterns}(b) in the appendix. We also encounter difficulties in some cases when patterns classified as \textit{broken stripe(s) and interstripe(s)} have breaks in consecutive stripes and interstripes. Specifically, our methods can fail when a non-persistent nonzero-born connected component is present; see the second pattern in the misclassification box in Fig.~\ref{fig:algorihm_diagram} and Fig.~\ref{fig:misclassified_patterns}(b) in the appendix for details. 
We note that, for patterns 
classified as \textit{broken stripe(s) and interstripe(s)}---specifically when the number of PNZB and NPZB connected components is not the same in the RL and LR barcodes or when NPNZB connected components appear in the RL or LR barcodes---we do not estimate break width because we do not expect to correctly match the connected components with their corresponding breaks; see Fig.~\ref{fig:misclassified_patterns}(b).

\subsubsection{Characterizing irregular striped patterns: break position}\label{sec:defects2}

Continuing our focus on irregular striped patterns from Sect.~\ref{sec:defects}, we now develop our method for interpreting persistent homology based on the sweeping-plane filtration to characterize the spatial position of stripe and interstripe breaks. Specifically, we (1) capture the position of each break along the horizontal length (anterior--posterior axis) of the pattern domain; and (2) identify the position of each interruption along the dorsal--ventral axis.

\begin{figure}[t!]
    \centering 
    \includegraphics[width=1\linewidth]{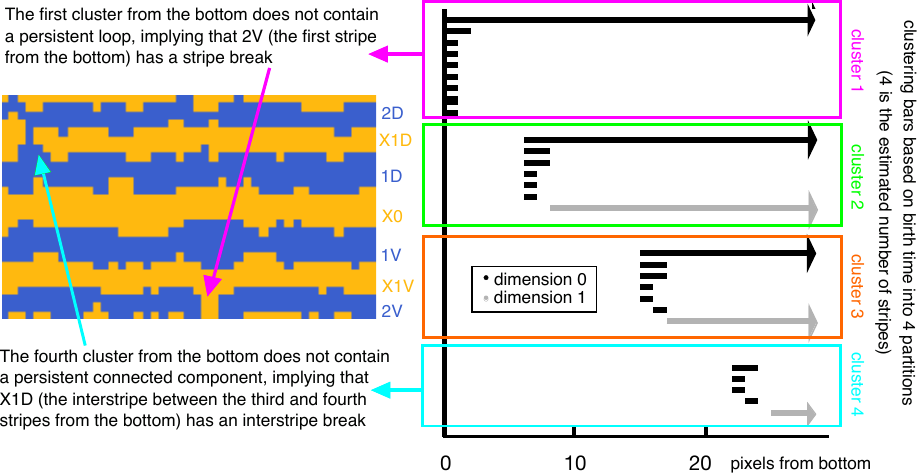}\vspace{0.5\baselineskip}
    \caption{Our approach to identifying the dorsal--ventral position of stripes and interstripes with interruptions. First, we cluster all of the connected components and loops in the BT barcode based on birth time, specifying that the number of clusters is the number of stripes estimated according to Eqn.~\eqref{eqn:stripe}. Thus, each cluster of features is associated with a stripe or stripe--interstripe pair. (Because we use the BT barcode, the clusters are $\{2\text{V}\}$, $\{1\text{V}, \text{X}1\text{V}\}$, $\{1\text{D}, \text{X}0\}$, and $\{2\text{D}, \text{X}1\text{D}\}$ in this example.) Second, we check each cluster for the presence of a persistent connected component and persistent loop. If no persistent loop is found, the corresponding stripe is broken; if no persistent connected component is found and an interstripe is associated with that cluster, that interstripe is broken.}
    \label{fig:clustering}
\end{figure}

Because the agent-based model \cite{volkening2018} does not incorporate differences in cell behavior or growth across the domain, we expect interruptions in patterns generated by this model to appear at random locations along the domain length. However, being able to identify where interruptions occur along the anterior--posterior axis opens the door to future work with empirical data; for example, the fish in Fig.~\ref{fig:biology}(b) \cite{fadeev2015} shows clear differences between its anterior and posterior patterns.
We thus define the position of each stripe or interstripe break along the horizontal length of the domain to be the center of the interruption in $x$. For breaks in stripes, we find this position by subtracting half of the estimated break width in Eqn.~\eqref{eq:gapwidth} from the birth time of the corresponding persistent nonzero-born connected component in the LR filtration. For breaks in interstripes, we define this position by adding half of the estimated break width in Eqn.~\eqref{eq:gapwidth2} to the death time of the corresponding non-persistent zero-born connected component in the LR filtration.

In comparison, because stripes and interstripes form sequentially outward from the center of the fish during development \cite{Jan,volkening2018}, we do expect differences in where stripe and interstripe interruptions occur along the dorsal--ventral axis.  
For our application, it is most meaningful to define the vertical position of each interruption in terms of whether it appears in interstripe X$0$, X$1$V, or X$1$D or stripe $1$V, $1$D, $2$V, or $2$D in Fig.~\ref{fig:biology}(c), as this is tied directly to developmental order. To do so, our approach relies on \textit{k}-means clustering in \texttt{MATLAB} based on the birth times $r_{\text{birth}}$ of the dimension-$0$ and dimension-$1$ topological features in the BT filtration. We partition all of the connected components and loops into a number of clusters equal to the stripe count according to Eqn.~\eqref{eqn:stripe2}; see 
Fig.~\ref{fig:clustering}. We then assign these clusters an order based on their average $r_\text{birth}$ values, meaning that the first cluster corresponds to the bottom blue stripe (X$2$V here) because we sweep from bottom to top.
For \textit{unbroken striped patterns}, each such cluster should contain one persistent connected component and one persistent loop, in addition to some short dimension-$0$ bars.

A cluster without a persistent loop implies a break in the corresponding blue stripe. For example, for the BT barcode in Fig.~\ref{fig:clustering}, when we cluster the dimension-$0$ and dimension-$1$ bars based on $r_\text{birth}$ into four clusters, notice that the first cluster does not contain a persistent loop. This implies that the first stripe from the bottom (stripe $2$V here) is broken.  
On the other hand, we interpret a cluster without a persistent connected component as indicating a break in the corresponding interstripe. In Fig.~\ref{fig:clustering}, the fourth $r_\text{birth}$-based cluster of features is missing a persistent connected component. This implies that the interstripe directly below the fourth stripe from the bottom is broken (interstripe X$1$D here). 
This method works well for patterns containing at most one stripe break and
at most one interstripe break, because it associates each break with its corresponding stripe or interstripe. However, it becomes more difficult to match individual breaks with their respective regions when multiple breaks occur, especially within the same stripe or interstripe. In such cases, we limit our analysis to reliably detecting which stripes or interstripes contain breaks, rather than matching each break with its corresponding stripe or interstripe.

\section{Results: Large-scale quantification of features and irregularities in zebrafish patterns}
\label{sec:results}

Considering a large set of $1000$ wild-type and $1000$ \textit{pfeffer} patterns \cite{mcguirl2020zebrafish} generated by the agent-based model \cite{volkening2018}, we now apply our TDA-based methodology in Sect.~\ref{sec:methodology} to automatically sort patterns by type and then generate detailed information about biologically meaningful features and irregularities. First, our classification pipeline in Sect.~\ref{sec:step3} blindly identifies all $1000$ \textit{pfeffer} images as spotted patterns, and $997$ of the $1000$ wild-type images as striped patterns (of various types, depending on whether breaks are present or not) with the remaining three wild-type patterns identified as spotted due to breaks in all stripes. For this reason, we often present our results in terms of \textit{pfeffer} and wild-type in the figures here. Because part of our motivation for this study is to encourage work combining multiple filtrations to provide complementary insight into biological systems, we also highlight how our results based on interpreting barcodes from the sweeping-plane filtration are related to prior studies \cite{Cleveland2023,McGuirl2020} of the same data with the Vietoris--Rips filtration. It is important to note, however, that these studies \cite{Cleveland2023,McGuirl2020} involved cropping the domain to focus on the pattern from X1D to X1V, while we consider the full domain; additionally, McGuirl \textit{et al.} \cite{McGuirl2020} considered \textit{pfeffer} patterns at a simulation time corresponding to $76$ days post fertilization, whereas we consider patterns at $66$ days post fertilization for both wild-type and \textit{pfeffer}.

\begin{figure}[t!]
    \centering 
    \includegraphics[width=1\linewidth]{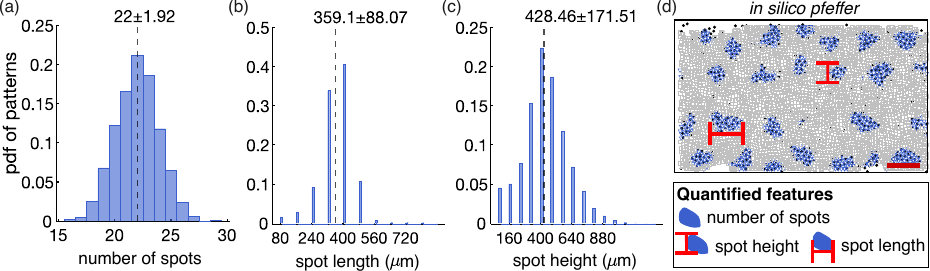}\vspace{0.5\baselineskip}
    \caption{Results of our large-scale quantitative study of \textit{pfeffer} mutant zebrafish patterns generated by the agent-based model \cite{volkening2018}. We show (a) the distribution of the number of spots across $1000$ \textit{pfeffer} patterns, as well as the distributions of (b) spot length and (c) spot height across all spots in these $1000$ patterns. (d) The \textit{pfeffer} pattern here provides a legend indicating our definitions of spot height and length; red scale bar is $500$~$\mu$m.}
    \label{fig:spot_distribution}
\end{figure}

We summarize our large-scale quantitative analysis of \textit{in silico} \textit{pfeffer} patterns in Fig.~\ref{fig:spot_distribution}, including distributions of our estimates for spot count, length, and height. These results highlight the population-scale variability that the model \cite{volkening2018} predicts is present in \textit{pfeffer}. For example, interpreting spots as persistent dimension-$0$ features when sweeping from top to bottom or from bottom to top with periodic boundary conditions in $x$, we show the distribution of our estimated number of spots in each pattern across $1000$ \textit{pfeffer} patterns 
in Fig.~\ref{fig:spot_distribution}(a). Interestingly, our methods suggest a mean of about $22$ spots with a standard deviation of about $2$~spots at $66$ days post fertilization, while McGuirl \textit{et al.} \cite{McGuirl2020} found a mean of about $23$ $\pm$ $2$ spots in \textit{pfeffer} patterns at $76$ days post fertilization based on the Vietoris--Rips filtration. We expect that the mean values are so similar because portions of the dorsal and ventral spots present at $66$ days post fertilization are likely still present even in the cropped domain at $76$ days post fertilization, given that the pattern forms from the center outward. 

\begin{table*}[!t]
\begin{center}
\small
\begin{tabular}{l c c c c }
  &  \textbf{\textit{Unbroken}} & \textbf{\textit{Broken stripe(s)}} & \textbf{\textit{Broken interstripe(s)}} & \textbf{\textit{Both broken}} \\
\toprule
Sweeping-plane: & 87.9\% & 3.8\% & 5.7\% &2.6\% \\ 
\midrule
Vietoris--Rips: & 86.1\% & 2.7\% & 9\% & 2.2\% \\
\bottomrule
\end{tabular}
\end{center}
\vspace{-0.5em}
\caption{Comparing different TDA perspectives on cropped striped patterns. We present our interpretations of the sweeping-plane filtration in comparison to the results of Cleveland \textit{et al.} \cite{Cleveland2023} based on the Vietoris--Rips filtration.  
The study \cite{Cleveland2023} involves cropping the top and bottom of the domain for $1000$ wild-type patterns generated by the agent-based model \cite{volkening2018}. This means that the focus in \cite{Cleveland2023} is on the central portion of each pattern spanned between X$1$D and X$1$V. For this table only, to allow for more direct comparison, we also apply our classification algorithm to cropped wild-type patterns. Neglecting stripes $2$D and $2$V, we find that the estimated percentages of \textit{unbroken}, \textit{broken stripe(s)}, \textit{broken interstripe(s)}, and \textit{broken stripe(s) and interstripe(s)} in $1000$ \textit{in silico} patterns are very similar across for sweeping-plane approach and the Vietoris--Rips perspective \cite{Cleveland2023}. 
(We note that the results in \cite{Cleveland2023} may be for a different random sample of $1000$ wild-type patterns than we quantify \cite{mcguirl2020zebrafish}.)}
\label{table:TDA_compare}
\end{table*}

We next turn to quantifying the $1000$ \textit{in silico} patterns in our dataset \cite{mcguirl2020zebrafish} representing juvenile wild-type zebrafish---$997$ of which we identify as striped (of some sort) based on our classification algorithm in Sect.~\ref{sec:step3}. First, as a means of validating our approach and drawing parallels between interpretations based on different filtration methods, we conduct a short study of cropped striped patterns with the sweeping-plane filtration that mirrors the work of Cleveland \textit{et al.} \cite{Cleveland2023} using the Vietoris--Rips filtration. Specifically, to focus on the portion of wild-type patterns spanned between X$1$D and X$1$V, the study \cite{Cleveland2023} involves cropping the top and bottom $10$\% of the domain. The result is that unbroken striped patterns have two stripes and three interstripes based on interpretations of the Vietoris--Rips filtration.  
Applying Steps~$1$--$2$ of our pipeline (Sections~\ref{sec:step1}--\ref{sec:step2}) on the full domain and then cropping the top and bottom $10$\% of our binary images before Steps~$3$--$4$ (Sections~\ref{sec:step3}--\ref{sec:step4}), produces a comparable situation. In this setting,
we interpret barcodes as indicating the presence of two stripes and three interstripes.

Focusing on $1000$ cropped striped patterns, Table \ref{table:TDA_compare} shows that the two TDA approaches---based on the Vietoris--Rips filtration \cite{Cleveland2023} or the sweeping-plane filtration here---yield largely similar results. We see the most disagreement in the percentage of patterns identified as \textit{broken interstripe(s)}, but we find the fraction of patterns identified as unbroken---either $87.9$\% or $86.1$\%---remarkably close. It is important to point out that the Vietoris--Rips-based method \cite{Cleveland2023} relies on \textit{a priori} knowledge of the target number of complete stripes and interstripes; broken patterns are then detected based on whether the number of complete stripes and interstripes is less than the target value. Our methodology based on the sweeping-plane filtration, on the other hand, bypasses this need for \textit{a priori} knowledge and directly counts breaks in stripes. We suggest that this is a particular benefit of the sweeping-plane filtration. With the intuition in Table~\ref{table:TDA_compare} in hand, we turn to quantifying full-size, uncropped patterns for the remainder of this manuscript.

Figure~\ref{fig:percentage_distributions} summarizes the results of our classification algorithm when we apply it to complete, uncropped wild-type patterns represented as binary images. Using our algorithm in Fig.~\ref{fig:algorihm_diagram} that is based fully on interpreting barcodes from the sweeping-plane filtration, we report the fractions of patterns identified as \textit{unbroken}, \textit{broken stripe(s)}, \textit{broken interstripe(s)}, or \textit{broken stripe(s) and interstripe(s)}. Notably, more than half of the striped patterns are classified as \textit{broken stripe(s)}, whereas only $30.8$\% are classified as \textit{unbroken}. Compared to our results on cropped patterns in Table~\ref{table:TDA_compare}, this alludes to more breaks occurring in the peripheral stripes, which we verify in Fig.~\ref{fig:break_percentage_position}. To validate our classification algorithm (Fig.~\ref{fig:algorihm_diagram}), we randomly chose $100$ wild-type patterns from our dataset and manually classified them; based on these qualitative observations, we estimate a $4$\% classification error. The most common misclassifications are in \textit{broken stripe(s) and interstripe(s)} patterns and \textit{broken stripe(s)} patterns. Such misclassifications can arise due to stray gold cells 
in stripes at the left or right domain boundaries. Alternatively, misclassifications may occur due when mismatches in multiple stripe and interstripe breaks cause our check that break widths are less than $40$\% of the image length 
to fail. See Fig.~\ref{fig:common_misclassifications} in the appendix for illustrative examples of the most common misclassifications.

\begin{figure}[t!]
    \centering 
    \includegraphics[width=1\linewidth]{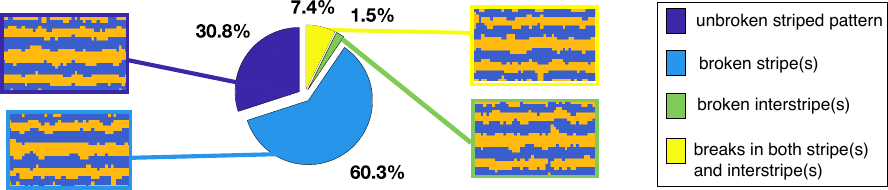}\vspace{0.5\baselineskip}
    \caption{Classification of $1000$ wild-type patterns generated by the agent-based model \cite{volkening2018}. Under wild-type conditions, the model \cite{volkening2018} aims to produce patterns with four blue stripes at $66$ days post fertilization.  Applying our pipeline in Fig.~\ref{fig:algorihm_diagram}, we distinguish between unbroken striped patterns, patterns with one or more broken stripes, patterns with one or more broken interstripes, and patterns with interruptions in both stripe(s) and interstripe(s). Importantly, we identify three of $1000$ wild-type patterns as having breaks in all stripes. Such patterns are classified as spotted by our algorithm, but, given the wild-type context, we report this $0.3$\% of patterns as part of the \textit{broken stripe(s) and interstripe(s)} group in this pie chart.}
    \label{fig:percentage_distributions}
\end{figure}

Focusing on \textit{unbroken striped patterns} (i.e., $30.8$\% of our wild-type dataset), we summarize distributions of minimum and maximum widths for stripes and interstripes in Fig.~\ref{fig:stripe_distributions}(a)--(b).
Our results show that interstripes tend, on average, to be slightly wider than stripes, with mean maximum stripe width of about $440$~$\mu$m and mean interstripe width of about $497$~$\mu$m. In comparison, Cleveland \textit{et al.} \cite{Cleveland2023} estimated the mean maximum stripe width as about $416$~$\mu$m and the mean maximum interstripe width as about $403$~$\mu$m. We consider these estimates very similar, particularly considering that the study \cite{Cleveland2023} factors in the separation between black and gold cells in stripes and interstripes, which is roughly $90$~$\mu$m \cite{McGuirl2020} in the model \cite{volkening2018}. Moreover, we expect our approach to slightly over-estimate stripe width because a blue pixel at the stripe boundary indicates that at least one melanophore is present somewhere in that pixel (e.g., on average, in its center). 
In terms of why we estimate stripes as narrower than interstripes while the Vietoris--Rips-based study \cite{Cleveland2023} has the opposite conclusion, this may be an artifact of our choice to build binary images based only on melanophores (i.e., cells in blue stripes). We also expect that this may be related to our analysis incorporating not just the widths of 1D, X0, and 1V, but also of 2D, X1D, X1V, and 2V. In particular, stripes 2D and 2V appear qualitatively narrower than interstripes X1D and X1V; see Fig.~\ref{fig:stripe_distributions}(d).
Interestingly, we observe that the standard deviations of measured stripe and interstripe widths are around $70$~$\mu$m, which is close to our voxel width $\varepsilon=80$~$\mu$m and to cell--cell distances in the agent-based model \cite{volkening2018}.

\begin{figure}[t!]
    \centering 
    \includegraphics[width=1\linewidth]{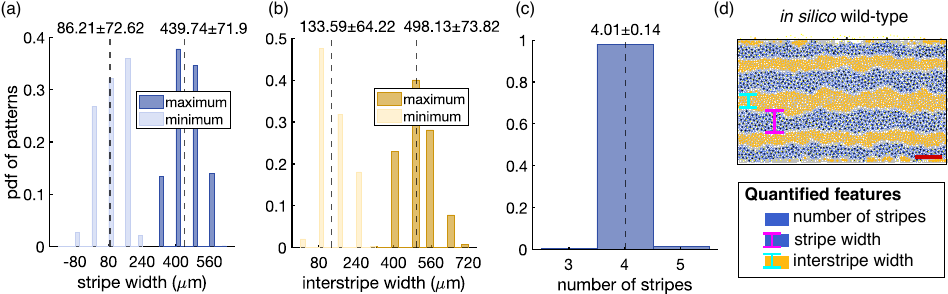}\vspace{0.5\baselineskip}
    \caption{Summary of striped-pattern features. Focusing on the subset of patterns in our dataset \cite{mcguirl2020zebrafish,volkening2018} that we classify as \textit{unbroken striped patterns}, we quantify (a) minimum and maximum stripe width, and (b) minimum and maximum interstripe width; also see Fig.~\ref{fig:stripe_width}. (Distributions in (a)--(b) are across all of the stripes in these patterns.) (c) Considering the $997$ patterns broadly identified as striped---whether broken or unbroken---in our dataset, our TDA-methods identify the vast majority as possessing four blue stripes, with a mean stripe number of about $4.01$. This shows that stripe number is a highly robust feature of the agent-based model \cite{volkening2018}, and it will be interesting to test this quantitative prediction with empirical data in the future. (d) We highlight stripe width (magenta) and interstripe width (cyan) in an example \textit{in silico} pattern as a reference; red scale bar is $500$~$\mu$m.}
    \label{fig:stripe_distributions}
\end{figure}

Our methodology directly counts stripes, regardless of if they are broken or unbroken, and we show the distribution of our estimated stripe counts across $997$ out of the $1000$ wild-type patterns that are classified as striped in Fig.~\ref{fig:stripe_distributions}(c). (If a stripe is broken into two pieces, for example, it is still counted as one stripe.) Because we consider wild-type patterns from the model \cite{volkening2018} at the juvenile developmental stage, there should be four stripes present in all of these patterns. Notably, we find that the average number of blue stripes is $4.01$, with just $2.2$\% of patterns having three or five stripes according to our interpretations of barcodes from the sweeping-plane filtration. The results in Fig.~\ref{fig:stripe_distributions}(c) serve both as additional support that our methodology is performing well at large-scale, and as additional insight into the nature of the agent-based model \cite{volkening2020}. In particular, while the stochastic cell interactions in the model \cite{volkening2020} lead to some variability in stripe width in Fig.~\ref{fig:stripe_distributions}(a) and in the presence of breaks in wild-type patterns in Fig.~\ref{fig:percentage_distributions}, the shear number of stripes formed by the juvenile developmental stage appears to be quite robust.

\begin{figure}[t!]
    \centering 
    \includegraphics[width=1\linewidth]{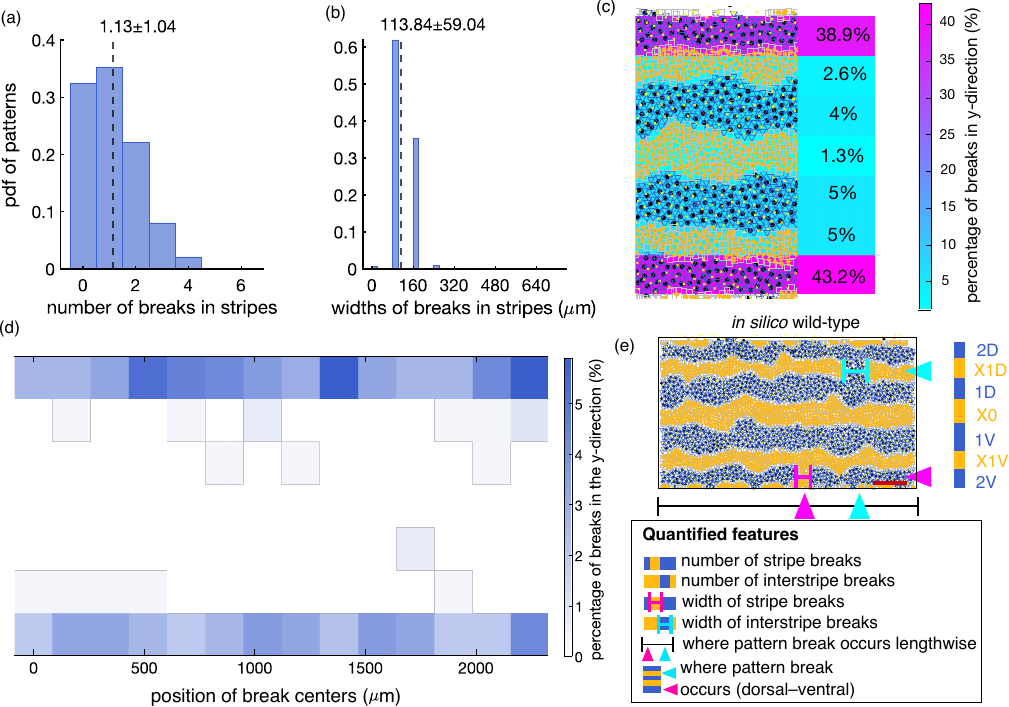}\vspace{0.5\baselineskip}
    \caption{Quantitative study of irregularities in \textit{in silico} zebrafish patterns. (a) Across the $997$ patterns in our dataset classified as striped---whether broken or unbroken---by our algorithm in Fig.~\ref{fig:algorihm_diagram}, we estimate $1.13$ stripe breaks per pattern on average. (This distribution does not include the three wild-type patterns with breaks in all of their stripes that our algorithm classifies as spotted.) (b) The vast majority of stripe breaks are one or two voxels wide, where the voxel width is $\varepsilon = 80$~$\mu$m. (c) Because stripes and interstripes form sequentially outward in wild-type zebrafish, starting from the center and progressing more dorsally (upward) and ventrally (downward) over the course of a few weeks \cite{Jan,Quigley2002}, the position of breaks along the dorsal--ventral axis can be thought of as related to developmental time. Notably, we estimate that about $82$\% of breaks occur in stripes $2$D or $2$V, which are the last to develop. (d) On the other hand, the model \cite{volkening2018} expects no pattern differences along the anterior--posterior axis. We find that percentages of stripe and interstripe breaks versus their position in the pattern domain suggests that the $x$-coordinate of break centers is uniform random. The distribution in (d) does not include patterns with multiple breaks of the same type---i.e., multiple stripe breaks (alternatively multiple interstripe breaks)---if the breaks occur in different stripes (respectively, interstripes), as it is challenging to reliably estimate break centers in this case. (e) As a guide we include a legend highlighting the features and irregularities quantified by our methods. Red scale bar is $500$~$\mu$m.}
    \label{fig:break_percentage_position}
\end{figure}

Lastly, we turn to a detailed study of irregularities in wild-type patterns. As we show in Fig.~\ref{fig:break_percentage_position}(a), the estimated average number of stripe breaks in striped patterns is about $1.1$ according to our methods, with about $30$\% of striped patterns having no breaks in stripes and about $35$\% having one stripe break. We estimate that the mean width of these interruptions is about $112$~$\mu$m, with the majority of breaks only one pixel (i.e., $80$~$\mu$m) wide. Importantly, this should be viewed as a minimum width, as we specify pixels are blue in wild-type patterns when they contain at least one melanophore. Because cells can appear anywhere along the length of a pixel, a break width of one pixel ($80$~$\mu$m) in a binary image can correspond to an agent-based pattern in which the distance between cells at opposite sides of the break is $80+2d$ $\mu$m, where $0 \le d \le 160$ $\mu$m and we expect $d = 80$~$\mu$m on average (assuming the average position of a cell is the voxel center).

Focusing on striped patterns with breaks of some kind, detailed heatmaps of the percentage distribution of the positions in which breaks in the stripes and interstripes occur in the ($x$,$y$)-plane are in Fig.~\ref{fig:break_percentage_position}(c)--(d). Our results show that stripes are broken much more frequently than interstripes, and that the vast majority of breaks occur in stripes $2$D and $2$V. This could be linked to the sequential development of stripes and interstripes over time and stochastic cell interactions. In particular, since the central stripes and interstripes are the first to develop in time, we expect that peripheral stripes that form later may experience more noise. For example, the formation of stripes $1$D and $1$V may be strongly guided by the model initial condition, which is meant to represent cells aligned along the horizontal myoseptum. (Notably, in the absence of the horizontal myoseptum, labyrinthine patterns without clear directionality form \cite{Frohnhofer,volkening2018}.) More broadly, by providing detailed, quantitative insight into interruptions and imperfections in messy striped patterns, our automated methods allow new predictions to be generated by the agent-based model \cite{volkening2018}, as we discuss in Sect.~\ref{sec:conclusion}.

\section{Discussion and conclusions}
\label{sec:conclusion}

Motivated by the presence of both characteristic features and variability in biological patterns, we developed a framework for automatically characterizing messy striped and spotted patterns based on topological data analysis.
We applied persistent homology to help select an appropriate resolution for transforming cell-based patterns into binary images, and we then computed persistent homology based on the sweeping-plane filtration applied to these binary images. Sweeping in four directions (from top to bottom, bottom to top, left to right, and right to left), we interpreted barcodes 
in terms of biologically meaningful characteristics. Throughout this study, we motivated and centered our methodology using a large dataset \cite{mcguirl2020zebrafish} of juvenile wild-type (striped) and \textit{pfeffer} mutant (spotted) zebrafish patterns generated by the agent-based model \cite{volkening2018}. We showed how to interpret the results of persistent homology with the sweeping-plane filtration to blindly classify these patterns by type and quantitatively characterize pattern features and irregularities in detail. For example, our methodology allowed us to estimate the number of stripes or spots present, stripe width, and spot size. Moreover, by interpreting barcodes from the sweeping-plane filtration, we identified the spatial position of interruptions which break stripes into separate pieces, and we characterized the width of these interruptions.

By conducting a large-scale study of $1000$ wild-type and $1000$ \textit{pfeffer} patterns \cite{mcguirl2020zebrafish} generated from the agent-based model \cite{volkening2018}, we provided new insight into robustness and variability that may emerge during development. We found that the model \cite{volkening2018} predicts that stripe number at the juvenile stage of wild-type development is highly robust, with less than $3$\% of patterns having any more or less than exactly four stripes. On the other hand, we automatically quantified (for the first time, to our knowledge) that the presence of breaks in stripes $2$D and $2$V in \textit{in silico} wild-type patterns \cite{volkening2018} is very common: about $68$\% of wild-type patterns feature broken stripes, and over $82$\% of these breaks occur in the last two stripes to develop (stripes $2$D and $2$V). In the future, it will be interesting to test these and other predictions based on our methods. For example, if a large-scale analysis of \textit{in vivo} zebrafish patterns shows that the fraction of breaks in stripes $2$D and $2$V is significantly different than $82$\%, then this could suggest that cells interact with the tissue environment to behave differently in specific regions of the fish body. As a strength of our approach, the sweeping-plane filtration works naturally on images. We thus expect our methodology to be directly applicable to empirical images of zebrafish after an added binarization step.

Choosing to focus our work on \textit{in silico} zebrafish patterns \cite{volkening2018} also set up an excellent opportunity for us to compare and contrast the insights provided by different topological perspectives. Prior studies \cite{McGuirl2020,Cleveland2023} of zebrafish based on the Vietoris--Rips filtration have interpreted persistent homology to estimate features including stripe width, stripe number, spot size, and spot number. (Notably, these studies involved cropping the pattern to remove the often messier, less fully formed stripes at the top and bottom of the domain.) We found the sweeping-plane filtration naturally amenable to messy patterns, particularly striped patterns. Sweeping from top to bottom or bottom to top allowed us to count horizontal stripes, while sweeping from left to right or right to left naturally encodes interruptions in stripes. Directly counting stripe interruptions with the Vietoris--Rips filtration may be more difficult (it is not considered in \cite{McGuirl2020,Cleveland2023}), and we would not expect the Vietoris--Rips filtration to provide information about the spatial position of stripe breaks. In comparison, we found the sweeping-plane filtration far less natural for spot patterns, and our measurements of spot size had high error. We conclude that the Vietoris–Rips filtration is more natural for spot patterns, whereas the sweeping-plane filtration is well-suited for irregular striped patterns. This may be because, in cases like wild-type zebrafish where stripes are clearly aligned, sweeping vertically and horizontally makes use of this directionality information. In spot patterns, on the other hand, the sweeping directions are less clearly related to pattern elements.

Although we addressed several challenges associated with automatically quantifying biological self-organization (specifically zebrafish skin patterns), our work has some limitations that suggest directions for future work.  
As one drawback, our pipeline is designed to detect horizontal stripes, so patterns with vertical stripes, for example, will not be correctly classified. Future work could involve creating a more flexible pipeline that also quantifies patterns with vertical stripes, but applying persistent homology to a dataset featuring stripes with a wider range of \textit{a priori}-unknown orientations may 
increase computational complexity. 
Another drawback is the sensitivity of our pipeline to stray cells. While we were able to address stray cells by introducing additional checks or hyperparameters in some cases, in other cases patterns were misclassified by our algorithm.
In the future, it would be interesting to incorporate an additional step to clean patterns before applying persistent homology as in \cite{Cleveland2023}, such as by swapping the color of any pixel surrounded by pixels of the opposite color. Additionally, we made choices about which cell types to base our binary images on, and it will be valuable to explore alternative choices in the future. Moreover, as we discuss above, our pipeline performs fairly poorly at estimating spot size. Combining Vietoris--Rips-based \cite{McGuirl2020} and sweeping-plane-based approaches in the future could circumvent this issue.

While we focused on zebrafish patterns at one stage of development, expanding our methodology to study pattern features and irregularities across the developmental timeline is an interesting direction for future work.
Likewise, we believe that a similar approach could be applied to quantify agent-based dynamics and collective behavior in other biological systems such as 
wound healing \cite{Cumming2010}, insect trails \cite{AMORIM2015160}, or other types of pigmentation patterns like those in lizards \cite{Manukyan2017}. Another valuable direction would be to adjust our methodology to
employ extended persistent homology \cite{Thorne,McDonald2025} in place of standard persistent homology, as this would allow us to consider two sweeping directions instead of four and could improve efficiency. More broadly, there are many filtrations available to choose from when computing persistent homology \cite{Otter2017mason,StolzThesis}, and applying them on the same data can provide complementary insight. This could open up additional opportunities for detailed quantitative studies of large datasets derived from both computational models and biological experiments, enabling model validations and increased understanding of the mechanisms underlying biological pattern formation.

\backmatter

\section*{Declarations}

\bmhead{Acknowledgments}
J.N.\ and A.V.\ gratefully acknowledge the SQuaRE program at the American Institute for Mathematics for providing an opportunity for them to discuss this project.

\bmhead{Data availability}
The simulated data that we quantified are publicly available on Figshare \cite{mcguirl2020zebrafish}.

\bmhead{Code availability}
Our code associated with this manuscript is publicly available on GitHub \cite{2025repo}.

\bmhead{Competing interests}
We have no competing interests to declare that are relevant to the content of this article.

\bibliography{ref}


\begin{thebibliography}{104}
\ifx \bisbn   \undefined \def \bisbn  #1{ISBN #1}\fi
\ifx \binits  \undefined \def \binits#1{#1}\fi
\ifx \bauthor  \undefined \def \bauthor#1{#1}\fi
\ifx \batitle  \undefined \def \batitle#1{#1}\fi
\ifx \bjtitle  \undefined \def \bjtitle#1{#1}\fi
\ifx \bvolume  \undefined \def \bvolume#1{\textbf{#1}}\fi
\ifx \byear  \undefined \def \byear#1{#1}\fi
\ifx \bissue  \undefined \def \bissue#1{#1}\fi
\ifx \bfpage  \undefined \def \bfpage#1{#1}\fi
\ifx \blpage  \undefined \def \blpage #1{#1}\fi
\ifx \burl  \undefined \def \burl#1{\textsf{#1}}\fi
\ifx \doiurl  \undefined \def \doiurl#1{\url{https://doi.org/#1}}\fi
\ifx \betal  \undefined \def \betal{\textit{et al.}}\fi
\ifx \binstitute  \undefined \def \binstitute#1{#1}\fi
\ifx \binstitutionaled  \undefined \def \binstitutionaled#1{#1}\fi
\ifx \bctitle  \undefined \def \bctitle#1{#1}\fi
\ifx \beditor  \undefined \def \beditor#1{#1}\fi
\ifx \bpublisher  \undefined \def \bpublisher#1{#1}\fi
\ifx \bbtitle  \undefined \def \bbtitle#1{#1}\fi
\ifx \bedition  \undefined \def \bedition#1{#1}\fi
\ifx \bseriesno  \undefined \def \bseriesno#1{#1}\fi
\ifx \blocation  \undefined \def \blocation#1{#1}\fi
\ifx \bsertitle  \undefined \def \bsertitle#1{#1}\fi
\ifx \bsnm \undefined \def \bsnm#1{#1}\fi
\ifx \bsuffix \undefined \def \bsuffix#1{#1}\fi
\ifx \bparticle \undefined \def \bparticle#1{#1}\fi
\ifx \barticle \undefined \def \barticle#1{#1}\fi
\bibcommenthead
\ifx \bconfdate \undefined \def \bconfdate #1{#1}\fi
\ifx \botherref \undefined \def \botherref #1{#1}\fi
\ifx \url \undefined \def \url#1{\textsf{#1}}\fi
\ifx \bchapter \undefined \def \bchapter#1{#1}\fi
\ifx \bbook \undefined \def \bbook#1{#1}\fi
\ifx \bcomment \undefined \def \bcomment#1{#1}\fi
\ifx \oauthor \undefined \def \oauthor#1{#1}\fi
\ifx \citeauthoryear \undefined \def \citeauthoryear#1{#1}\fi
\ifx \endbibitem  \undefined \def \endbibitem {}\fi
\ifx \bconflocation  \undefined \def \bconflocation#1{#1}\fi
\ifx \arxivurl  \undefined \def \arxivurl#1{\textsf{#1}}\fi
\csname PreBibitemsHook\endcsname

\bibitem[\protect\citeauthoryear{Giniunait\.{e} et~al.}{2020}]{Giniunaite2020}
\begin{barticle}
\bauthor{\bsnm{Giniunait\.{e}}, \binits{R.}},
\bauthor{\bsnm{Baker}, \binits{R.E.}},
\bauthor{\bsnm{Kulesa}, \binits{P.M.}},
\bauthor{\bsnm{Maini}, \binits{P.K.}}:
\batitle{Modelling collective cell migration: neural crest as a model
  paradigm}.
\bjtitle{J. Math. Biol.}
\bvolume{80},
\bfpage{481}--\blpage{504}
(\byear{2020})
\end{barticle}
\endbibitem

\bibitem[\protect\citeauthoryear{Buttensch\"{o}n and
  Edelstein-Keshet}{2020}]{Buttenschon2020}
\begin{botherref}
\oauthor{\bsnm{Buttensch\"{o}n}, \binits{A.}},
\oauthor{\bsnm{Edelstein-Keshet}, \binits{L.}}:
Bridging from single to collective cell migration: A review of models and links
  to experiments.
PLOS Comput. Biol.
\textbf{16}(12)
(2020)
\end{botherref}
\endbibitem

\bibitem[\protect\citeauthoryear{Volkening}{2020}]{VolkeningRev}
\begin{barticle}
\bauthor{\bsnm{Volkening}, \binits{A.}}:
\batitle{Linking genotype, cell behavior, and phenotype: multidisciplinary
  perspectives with a basis in zebrafish patterns}.
\bjtitle{Curr. Opin. Genet. Dev.}
\bvolume{63},
\bfpage{78}--\blpage{85}
(\byear{2020})
\end{barticle}
\endbibitem

\bibitem[\protect\citeauthoryear{Kondo et~al.}{2021}]{Kondo2021rev}
\begin{barticle}
\bauthor{\bsnm{Kondo}, \binits{S.}},
\bauthor{\bsnm{Watanabe}, \binits{M.}},
\bauthor{\bsnm{Miyazawa}, \binits{S.}}:
\batitle{Studies of {T}uring pattern formation in zebrafish skin}.
\bjtitle{Philos. Trans. Royal Soc. A}
\bvolume{379}(\bissue{2213}),
\bfpage{20200274}
(\byear{2021})
\end{barticle}
\endbibitem

\bibitem[\protect\citeauthoryear{Mogilner and
  Edelstein-Keshet}{1999}]{Mogilner1999}
\begin{barticle}
\bauthor{\bsnm{Mogilner}, \binits{A.}},
\bauthor{\bsnm{Edelstein-Keshet}, \binits{L.}}:
\batitle{A non-local model for a swarm}.
\bjtitle{J. Math. Biol.}
\bvolume{38},
\bfpage{534}--\blpage{570}
(\byear{1999})
\end{barticle}
\endbibitem

\bibitem[\protect\citeauthoryear{Bernoff and Topaz}{2011}]{Bernoff2011}
\begin{barticle}
\bauthor{\bsnm{Bernoff}, \binits{A.J.}},
\bauthor{\bsnm{Topaz}, \binits{C.M.}}:
\batitle{A primer of swarm equilibria}.
\bjtitle{SIAM J. Appl. Dyn. Sys.}
\bvolume{10}(\bissue{1}),
\bfpage{212}--\blpage{250}
(\byear{2011})
\end{barticle}
\endbibitem

\bibitem[\protect\citeauthoryear{Huepe and Aldana}{2008}]{Huepe2008}
\begin{barticle}
\bauthor{\bsnm{Huepe}, \binits{C.}},
\bauthor{\bsnm{Aldana}, \binits{M.}}:
\batitle{New tools for characterizing swarming systems: A comparison of minimal
  models}.
\bjtitle{Physica A}
\bvolume{387}(\bissue{12}),
\bfpage{2809}--\blpage{2822}
(\byear{2008})
\end{barticle}
\endbibitem

\bibitem[\protect\citeauthoryear{D'Orsogna et~al.}{2006}]{Dorsogna2006}
\begin{barticle}
\bauthor{\bsnm{D'Orsogna}, \binits{M.R.}},
\bauthor{\bsnm{Chuang}, \binits{Y.L.}},
\bauthor{\bsnm{Bertozzi}, \binits{A.L.}},
\bauthor{\bsnm{Chayes}, \binits{L.S.}}:
\batitle{Self-propelled particles with soft-core interactions: patterns,
  stability, and collapse}.
\bjtitle{Phys. Rev. Lett.}
\bvolume{96}(\bissue{10}),
\bfpage{104302}
(\byear{2006})
\end{barticle}
\endbibitem

\bibitem[\protect\citeauthoryear{Vicsek and Zafeiris}{2012}]{VicsekReview}
\begin{barticle}
\bauthor{\bsnm{Vicsek}, \binits{T.}},
\bauthor{\bsnm{Zafeiris}, \binits{A.}}:
\batitle{Collective motion}.
\bjtitle{Phys. Rep.}
\bvolume{517}(\bissue{3}),
\bfpage{71}--\blpage{140}
(\byear{2012})
\end{barticle}
\endbibitem

\bibitem[\protect\citeauthoryear{Katz et~al.}{2011}]{Katz2011fish}
\begin{barticle}
\bauthor{\bsnm{Katz}, \binits{Y.}},
\bauthor{\bsnm{Tunstr{\o}m}, \binits{K.}},
\bauthor{\bsnm{Ioannou}, \binits{C.C.}},
\bauthor{\bsnm{Huepe}, \binits{C.}},
\bauthor{\bsnm{Couzin}, \binits{I.D.}}:
\batitle{Inferring the structure and dynamics of interactions in schooling
  fish}.
\bjtitle{Proc. Natl. Acad. Sci. USA}
\bvolume{108}(\bissue{46}),
\bfpage{18720}--\blpage{18725}
(\byear{2011})
\end{barticle}
\endbibitem

\bibitem[\protect\citeauthoryear{Lukeman et~al.}{2010}]{Lukeman2010keshet}
\begin{barticle}
\bauthor{\bsnm{Lukeman}, \binits{R.}},
\bauthor{\bsnm{Li}, \binits{Y.-X.}},
\bauthor{\bsnm{Edelstein-Keshet}, \binits{L.}}:
\batitle{Inferring individual rules from collective behavior}.
\bjtitle{Proc. Natl. Acad. Sci. USA}
\bvolume{107}(\bissue{28}),
\bfpage{12576}--\blpage{12580}
(\byear{2010})
\end{barticle}
\endbibitem

\bibitem[\protect\citeauthoryear{Patterson and Parichy}{2019}]{ParichyRev2019}
\begin{barticle}
\bauthor{\bsnm{Patterson}, \binits{L.B.}},
\bauthor{\bsnm{Parichy}, \binits{D.M.}}:
\batitle{Zebrafish pigment pattern formation: Insights into the development and
  evolution of adult form}.
\bjtitle{Annu. Rev. Genet.}
\bvolume{53}(\bissue{1}),
\bfpage{505}--\blpage{530}
(\byear{2019})
\end{barticle}
\endbibitem

\bibitem[\protect\citeauthoryear{Irion and
  N\"{u}sslein-Volhard}{2019}]{IrionRev2019}
\begin{barticle}
\bauthor{\bsnm{Irion}, \binits{U.}},
\bauthor{\bsnm{N\"{u}sslein-Volhard}, \binits{C.}}:
\batitle{The identification of genes involved in the evolution of color
  patterns in fish}.
\bjtitle{Curr. Opin. Genet. Dev.}
\bvolume{57},
\bfpage{31}--\blpage{38}
(\byear{2019})
\end{barticle}
\endbibitem

\bibitem[\protect\citeauthoryear{Frohnh{\"o}fer et~al.}{2013}]{Frohnhofer}
\begin{barticle}
\bauthor{\bsnm{Frohnh{\"o}fer}, \binits{H.G.}},
\bauthor{\bsnm{Krauss}, \binits{J.}},
\bauthor{\bsnm{Maischein}, \binits{H.M.}},
\bauthor{\bsnm{N{\"u}sslein-Volhard}, \binits{C.}}:
\batitle{Iridophores and their interactions with other chromatophores are
  required for stripe formation in zebrafish}.
\bjtitle{Development}
\bvolume{140}(\bissue{14}),
\bfpage{2997}--\blpage{3007}
(\byear{2013})
\end{barticle}
\endbibitem

\bibitem[\protect\citeauthoryear{Volkening and Sandstede}{2018}]{volkening2018}
\begin{botherref}
\oauthor{\bsnm{Volkening}, \binits{A.}},
\oauthor{\bsnm{Sandstede}, \binits{B.}}:
Iridophores as a source of robustness in zebrafish stripes and variability in
  \emph{{D}anio} patterns.
Nat. Commun.
\textbf{9}(3231)
(2018)
\end{botherref}
\endbibitem

\bibitem[\protect\citeauthoryear{McCluskey et~al.}{2021}]{McCluskey2021}
\begin{botherref}
\oauthor{\bsnm{McCluskey}, \binits{B.M.}},
\oauthor{\bsnm{Liang}, \binits{Y.}},
\oauthor{\bsnm{Lewis}, \binits{V.M.}},
\oauthor{\bsnm{Patterson}, \binits{L.B.}},
\oauthor{\bsnm{Parichy}, \binits{D.M.}}:
{Pigment pattern morphospace of \textit{Danio} fishes: evolutionary
  diversification and mutational effects}.
Biol. Open
\textbf{10}(9)
(2021)
\end{botherref}
\endbibitem

\bibitem[\protect\citeauthoryear{Bendich et~al.}{2016}]{Bendich2016}
\begin{barticle}
\bauthor{\bsnm{Bendich}, \binits{P.}},
\bauthor{\bsnm{Marron}, \binits{J.S.}},
\bauthor{\bsnm{Miller}, \binits{E.}},
\bauthor{\bsnm{Pieloch}, \binits{A.}},
\bauthor{\bsnm{Skwerer}, \binits{S.}}:
\batitle{Persistent homology analysis of brain artery trees}.
\bjtitle{Ann. Appl. Stat.}
\bvolume{10}(\bissue{1}),
\bfpage{198}--\blpage{218}
(\byear{2016})
\end{barticle}
\endbibitem

\bibitem[\protect\citeauthoryear{Nardini et~al.}{2021}]{Nardini2021}
\begin{barticle}
\bauthor{\bsnm{Nardini}, \binits{J.T.}},
\bauthor{\bsnm{Stolz}, \binits{B.J.}},
\bauthor{\bsnm{Flores}, \binits{K.B.}},
\bauthor{\bsnm{Harrington}, \binits{H.A.}},
\bauthor{\bsnm{Byrne}, \binits{H.M.}}:
\batitle{{Topological data analysis distinguishes parameter regimes in the
  Anderson-Chaplain model of angiogenesis}}.
\bjtitle{PLOS Comput. Biol.}
\bvolume{17},
\bfpage{1009094}
(\byear{2021})
\end{barticle}
\endbibitem

\bibitem[\protect\citeauthoryear{Maderspacher and
  N{\"u}sslein-Volhard}{2003}]{Maderspacher2003}
\begin{barticle}
\bauthor{\bsnm{Maderspacher}, \binits{F.}},
\bauthor{\bsnm{N{\"u}sslein-Volhard}, \binits{C.}}:
\batitle{Formation of the adult pigment pattern in zebrafish requires
  \emph{leopard} and \emph{obelix} dependent cell interactions}.
\bjtitle{Development}
\bvolume{130}(\bissue{15}),
\bfpage{3447}--\blpage{3457}
(\byear{2003})
\end{barticle}
\endbibitem

\bibitem[\protect\citeauthoryear{Parichy and Turner}{2003}]{ParTur130}
\begin{barticle}
\bauthor{\bsnm{Parichy}, \binits{D.M.}},
\bauthor{\bsnm{Turner}, \binits{J.M.}}:
\batitle{Temporal and cellular requirements for {F}ms signaling during
  zebrafish adult pigment pattern development}.
\bjtitle{Development}
\bvolume{130}(\bissue{5}),
\bfpage{817}--\blpage{833}
(\byear{2003})
\end{barticle}
\endbibitem

\bibitem[\protect\citeauthoryear{Parichy et~al.}{2000}]{PatDev127}
\begin{barticle}
\bauthor{\bsnm{Parichy}, \binits{D.M.}},
\bauthor{\bsnm{Ransom}, \binits{D.G.}},
\bauthor{\bsnm{Paw}, \binits{B.}},
\bauthor{\bsnm{Zon}, \binits{L.I.}},
\bauthor{\bsnm{Johnson}, \binits{S.L.}}:
\batitle{{An orthologue of the kit-related gene fms is required for development
  of neural crest-derived xanthophores and a subpopulation of adult melanocytes
  in the zebrafish, \textit{Danio rerio}}}.
\bjtitle{Development}
\bvolume{127}(\bissue{14}),
\bfpage{3031}--\blpage{3044}
(\byear{2000})
\end{barticle}
\endbibitem

\bibitem[\protect\citeauthoryear{Singh and N{\"u}sslein-Volhard}{2015}]{Jan}
\begin{barticle}
\bauthor{\bsnm{Singh}, \binits{A.P.}},
\bauthor{\bsnm{N{\"u}sslein-Volhard}, \binits{C.}}:
\batitle{Zebrafish stripes as a model for vertebrate colour pattern formation}.
\bjtitle{Curr. Biol.}
\bvolume{25}(\bissue{2}),
\bfpage{81}--\blpage{92}
(\byear{2015})
\end{barticle}
\endbibitem

\bibitem[\protect\citeauthoryear{Quigley and Parichy}{2002}]{Quigley2002}
\begin{barticle}
\bauthor{\bsnm{Quigley}, \binits{I.K.}},
\bauthor{\bsnm{Parichy}, \binits{D.M.}}:
\batitle{Pigment pattern formation in zebrafish: A model for developmental
  genetics and the evolution of form}.
\bjtitle{Microsc. Res. Tech.}
\bvolume{58}(\bissue{6}),
\bfpage{442}--\blpage{455}
(\byear{2002})
\end{barticle}
\endbibitem

\bibitem[\protect\citeauthoryear{Yamaguchi et~al.}{2007}]{Yamaguchi}
\begin{barticle}
\bauthor{\bsnm{Yamaguchi}, \binits{M.}},
\bauthor{\bsnm{Yoshimoto}, \binits{E.}},
\bauthor{\bsnm{Kondo}, \binits{S.}}:
\batitle{Pattern regulation in the stripe of zebrafish suggests an underlying
  dynamic and autonomous mechanism}.
\bjtitle{Proc. Natl. Acad. Sci. USA}
\bvolume{104}(\bissue{12}),
\bfpage{4790}--\blpage{4793}
(\byear{2007})
\end{barticle}
\endbibitem

\bibitem[\protect\citeauthoryear{Cleveland et~al.}{2023}]{Cleveland2023}
\begin{botherref}
\oauthor{\bsnm{Cleveland}, \binits{E.}},
\oauthor{\bsnm{Zhu}, \binits{A.}},
\oauthor{\bsnm{Sandstede}, \binits{B.}},
\oauthor{\bsnm{Volkening}, \binits{A.}}:
Quantifying different modeling frameworks using topological data analysis: a
  case study with zebrafish patterns.
SIAM J. Appl. Dyn. Sys.
\textbf{22}(4)
(2023)
\end{botherref}
\endbibitem

\bibitem[\protect\citeauthoryear{Fadeev et~al.}{2015}]{fadeev2015}
\begin{barticle}
\bauthor{\bsnm{Fadeev}, \binits{A.}},
\bauthor{\bsnm{Krauss}, \binits{J.}},
\bauthor{\bsnm{Frohnhöfer}, \binits{H.G.}},
\bauthor{\bsnm{Irion}, \binits{U.}},
\bauthor{\bsnm{Nüsslein-Volhard}, \binits{C.}}:
\batitle{Tight junction protein 1a regulates pigment cell organisation during
  zebrafish colour patterning}.
\bjtitle{eLife}
\bvolume{4},
\bfpage{06545}
(\byear{2015})
\end{barticle}
\endbibitem

\bibitem[\protect\citeauthoryear{Eskova et~al.}{2017}]{eskova2017gain}
\begin{barticle}
\bauthor{\bsnm{Eskova}, \binits{A.}},
\bauthor{\bsnm{Chauvign{\'e}}, \binits{F.}},
\bauthor{\bsnm{Maischein}, \binits{H.-M.}},
\bauthor{\bsnm{Ammelburg}, \binits{M.}},
\bauthor{\bsnm{Cerd{\`a}}, \binits{J.}},
\bauthor{\bsnm{N{\"u}sslein-Volhard}, \binits{C.}},
\bauthor{\bsnm{Irion}, \binits{U.}}:
\batitle{Gain-of-function mutations in aqp3a influence zebrafish pigment
  pattern formation through the tissue environment}.
\bjtitle{Development}
\bvolume{144}(\bissue{11}),
\bfpage{2059}--\blpage{2069}
(\byear{2017})
\end{barticle}
\endbibitem

\bibitem[\protect\citeauthoryear{Patterson and Parichy}{2013}]{PatPLos}
\begin{barticle}
\bauthor{\bsnm{Patterson}, \binits{L.B.}},
\bauthor{\bsnm{Parichy}, \binits{D.M.}}:
\batitle{Interactions with iridophores and the tissue environment required for
  patterning melanophores and xanthophores during zebrafish adult pigment
  stripe formation}.
\bjtitle{PLOS Genet.}
\bvolume{9}(\bissue{5}),
\bfpage{1003561}
(\byear{2013})
\end{barticle}
\endbibitem

\bibitem[\protect\citeauthoryear{McGuirl et~al.}{2020}]{McGuirl2020}
\begin{botherref}
\oauthor{\bsnm{McGuirl}, \binits{M.R.}},
\oauthor{\bsnm{Volkening}, \binits{A.}},
\oauthor{\bsnm{Sandstede}, \binits{B.}}:
Topological data analysis of zebrafish patterns.
Proc. Natl. Acad. Sci. USA
\textbf{117}(10)
(2020)
\end{botherref}
\endbibitem

\bibitem[\protect\citeauthoryear{Watanabe et~al.}{2006}]{Watanabe2006}
\begin{barticle}
\bauthor{\bsnm{Watanabe}, \binits{M.}},
\bauthor{\bsnm{Iwashita}, \binits{M.}},
\bauthor{\bsnm{Ishii}, \binits{M.}},
\bauthor{\bsnm{Kurachi}, \binits{Y.}},
\bauthor{\bsnm{Kawakami}, \binits{A.}},
\bauthor{\bsnm{Kondo}, \binits{S.}},
\bauthor{\bsnm{Okada}, \binits{N.}}:
\batitle{Spot pattern of \emph{leopard {D}anio} is caused by mutation in the
  zebrafish \emph{connexin}$41.8$ gene}.
\bjtitle{EMBO Rep.}
\bvolume{7}(\bissue{9}),
\bfpage{893}--\blpage{897}
(\byear{2006})
\end{barticle}
\endbibitem

\bibitem[\protect\citeauthoryear{Watanabe and Kondo}{2012}]{2012changing}
\begin{barticle}
\bauthor{\bsnm{Watanabe}, \binits{M.}},
\bauthor{\bsnm{Kondo}, \binits{S.}}:
\batitle{Changing clothes easily: \textit{connexin41.8} regulates skin pattern
  variation}.
\bjtitle{Pigment Cell Melanoma Res.}
\bvolume{25}(\bissue{3}),
\bfpage{326}--\blpage{330}
(\byear{2012})
\end{barticle}
\endbibitem

\bibitem[\protect\citeauthoryear{Irion et~al.}{2014}]{IrionGap}
\begin{barticle}
\bauthor{\bsnm{Irion}, \binits{U.}},
\bauthor{\bsnm{Frohnh{\"o}fer}, \binits{H.G.}},
\bauthor{\bsnm{Krauss}, \binits{J.}},
\bauthor{\bsnm{Champollion}, \binits{T.C.}},
\bauthor{\bsnm{Maischein}, \binits{H.}},
\bauthor{\bsnm{Geiger-Rudolph}, \binits{S.}},
\bauthor{\bsnm{Weiler}, \binits{C.}},
\bauthor{\bsnm{N{\"u}sslein-Volhard}, \binits{C.}}:
\batitle{Gap junctions composed of connexins 41.8 and 39.4 are essential for
  colour pattern formation in zebrafish}.
\bjtitle{eLife}
\bvolume{3},
\bfpage{05125}
(\byear{2014})
\end{barticle}
\endbibitem

\bibitem[\protect\citeauthoryear{Iwashita et~al.}{2006}]{Iwashita}
\begin{barticle}
\bauthor{\bsnm{Iwashita}, \binits{M.}},
\bauthor{\bsnm{Watanabe}, \binits{M.}},
\bauthor{\bsnm{Ishii}, \binits{M.}},
\bauthor{\bsnm{Chen}, \binits{T.}},
\bauthor{\bsnm{Johnson}, \binits{S.L.}},
\bauthor{\bsnm{Kurachi}, \binits{Y.}},
\bauthor{\bsnm{Okada}, \binits{N.}},
\bauthor{\bsnm{Kondo}, \binits{S.}}:
\batitle{{Pigment pattern in \emph{jaguar/obelix} zebrafish is caused by a
  Kir7.1 mutation: implications for the regulation of melanosome movement}}.
\bjtitle{PLOS Genet.}
\bvolume{2}(\bissue{11}),
\bfpage{197}
(\byear{2006})
\end{barticle}
\endbibitem

\bibitem[\protect\citeauthoryear{Howe et~al.}{2013}]{genome}
\begin{barticle}
\bauthor{\bsnm{Howe}, \binits{K.}},
\bauthor{\bsnm{Clark}, \binits{M.D.}},
\bauthor{\bsnm{Torroja}, \binits{C.F.}},
\bauthor{\bsnm{Torrance}, \binits{J.}},
\bauthor{\bsnm{Berthelot}, \binits{C.}},
\bauthor{\bsnm{Muffato}, \binits{M.}},
\bauthor{\bsnm{Collins}, \binits{J.E.}},
\bauthor{\bsnm{Humphray}, \binits{S.}},
\bauthor{\bsnm{McLaren}, \binits{K.}},
\bauthor{\bsnm{Matthews}, \binits{L.}}, \betal:
\batitle{The zebrafish reference genome sequence and its relationship to the
  human genome}.
\bjtitle{Nature}
\bvolume{496}(\bissue{7446}),
\bfpage{498}--\blpage{503}
(\byear{2013})
\end{barticle}
\endbibitem

\bibitem[\protect\citeauthoryear{Bullara and De~Decker}{2015}]{Bullara}
\begin{botherref}
\oauthor{\bsnm{Bullara}, \binits{D.}},
\oauthor{\bsnm{De~Decker}, \binits{Y.}}:
Pigment cell movement is not required for generation of {T}uring patterns in
  zebrafish skin.
Nat. Commun.
\textbf{6}(6971)
(2015)
\end{botherref}
\endbibitem

\bibitem[\protect\citeauthoryear{Konow et~al.}{2021}]{Konow2021}
\begin{botherref}
\oauthor{\bsnm{Konow}, \binits{C.}},
\oauthor{\bsnm{Li}, \binits{Z.}},
\oauthor{\bsnm{Shepherd}, \binits{S.}},
\oauthor{\bsnm{Bullara}, \binits{D.}},
\oauthor{\bsnm{Epstein}, \binits{I.R.}}:
Influence of survival, promotion, and growth on pattern formation in zebrafish
  skin.
Sci. Rep.
\textbf{11}(9864)
(2021)
\end{botherref}
\endbibitem

\bibitem[\protect\citeauthoryear{Gaffney and Seirin~Lee}{2015}]{Gaffney}
\begin{barticle}
\bauthor{\bsnm{Gaffney}, \binits{E.A.}},
\bauthor{\bsnm{Seirin~Lee}, \binits{S.}}:
\batitle{{The sensitivity of Turing self-organization to biological feedback
  delays: 2D models of fish pigmentation}}.
\bjtitle{Math. Med. Biol.}
\bvolume{32},
\bfpage{57}--\blpage{79}
(\byear{2015})
\end{barticle}
\endbibitem

\bibitem[\protect\citeauthoryear{Nakamasu et~al.}{2009}]{Nakamasu}
\begin{barticle}
\bauthor{\bsnm{Nakamasu}, \binits{A.}},
\bauthor{\bsnm{Takahashi}, \binits{G.}},
\bauthor{\bsnm{Kanbe}, \binits{A.}},
\bauthor{\bsnm{Kondo}, \binits{S.}}:
\batitle{Interactions between zebrafish pigment cells responsible for the
  generation of {T}uring patterns}.
\bjtitle{Proc. Natl. Acad. Sci. USA}
\bvolume{106}(\bissue{21}),
\bfpage{8429}--\blpage{8434}
(\byear{2009})
\end{barticle}
\endbibitem

\bibitem[\protect\citeauthoryear{Painter et~al.}{2015}]{painter}
\begin{barticle}
\bauthor{\bsnm{Painter}, \binits{K.J.}},
\bauthor{\bsnm{Bloomfield}, \binits{J.M.}},
\bauthor{\bsnm{Sherratt}, \binits{J.A.}},
\bauthor{\bsnm{Gerisch}, \binits{A.}}:
\batitle{A nonlocal model for contact attraction and repulsion in heterogeneous
  cell populations}.
\bjtitle{Bull. Math. Biol.}
\bvolume{77}(\bissue{6}),
\bfpage{1132}--\blpage{1165}
(\byear{2015})
\end{barticle}
\endbibitem

\bibitem[\protect\citeauthoryear{Woolley}{2017}]{Woolley2017}
\begin{barticle}
\bauthor{\bsnm{Woolley}, \binits{T.E.}}:
\batitle{Pattern production through a chiral chasing mechanism}.
\bjtitle{Phys. Rev. E}
\bvolume{96}(\bissue{3}),
\bfpage{032401}
(\byear{2017})
\end{barticle}
\endbibitem

\bibitem[\protect\citeauthoryear{Volkening and Sandstede}{2015}]{volkening2015}
\begin{barticle}
\bauthor{\bsnm{Volkening}, \binits{A.}},
\bauthor{\bsnm{Sandstede}, \binits{B.}}:
\batitle{Modelling stripe formation in zebrafish: an agent-based approach}.
\bjtitle{J. R. Soc. Interface}
\bvolume{12}(\bissue{112}),
\bfpage{20150812}
(\byear{2015})
\end{barticle}
\endbibitem

\bibitem[\protect\citeauthoryear{Volkening et~al.}{2020}]{volkening2020}
\begin{botherref}
\oauthor{\bsnm{Volkening}, \binits{A.}},
\oauthor{\bsnm{Abbott}, \binits{M.R.}},
\oauthor{\bsnm{Chandra}, \binits{N.}},
\oauthor{\bsnm{Dubois}, \binits{B.}},
\oauthor{\bsnm{Lim}, \binits{F.}},
\oauthor{\bsnm{Sexton}, \binits{D.}},
\oauthor{\bsnm{Sandstede}, \binits{B.}}:
Modeling stripe formation on growing zebrafish tailfins.
Bull. Math. Biol.
\textbf{82}(56)
(2020)
\end{botherref}
\endbibitem

\bibitem[\protect\citeauthoryear{Owen et~al.}{2020}]{Owen2020}
\begin{barticle}
\bauthor{\bsnm{Owen}, \binits{J.P.}},
\bauthor{\bsnm{Kelsh}, \binits{R.N.}},
\bauthor{\bsnm{Yates}, \binits{C.A.}}:
\batitle{A quantitative modelling approach to zebrafish pigment pattern
  formation}.
\bjtitle{eLife}
\bvolume{9},
\bfpage{52998}
(\byear{2020})
\end{barticle}
\endbibitem

\bibitem[\protect\citeauthoryear{Moreira and Deutsch}{2005}]{MorDeutsch}
\begin{barticle}
\bauthor{\bsnm{Moreira}, \binits{J.}},
\bauthor{\bsnm{Deutsch}, \binits{A.}}:
\batitle{Pigment pattern formation in zebrafish during late larval stages: A
  model based on local interactions}.
\bjtitle{Dev. Dyn.}
\bvolume{232}(\bissue{1}),
\bfpage{33}--\blpage{42}
(\byear{2005})
\end{barticle}
\endbibitem

\bibitem[\protect\citeauthoryear{Woolley et~al.}{2014}]{Cops}
\begin{barticle}
\bauthor{\bsnm{Woolley}, \binits{T.E.}},
\bauthor{\bsnm{Maini}, \binits{P.K.}},
\bauthor{\bsnm{Gaffney}, \binits{E.A.}}:
\batitle{Is pigment cell pattern formation in zebrafish a game of cops and
  robbers?}
\bjtitle{Pigment Cell Melanoma Res.}
\bvolume{27}(\bissue{5}),
\bfpage{686}--\blpage{687}
(\byear{2014})
\end{barticle}
\endbibitem

\bibitem[\protect\citeauthoryear{Caicedo-Carvajal and
  Shinbrot}{2008}]{Shinbrot}
\begin{barticle}
\bauthor{\bsnm{Caicedo-Carvajal}, \binits{C.E.}},
\bauthor{\bsnm{Shinbrot}, \binits{T.}}:
\batitle{\emph{In silico} zebrafish pattern formation}.
\bjtitle{Dev. Biol.}
\bvolume{315}(\bissue{2}),
\bfpage{397}--\blpage{403}
(\byear{2008})
\end{barticle}
\endbibitem

\bibitem[\protect\citeauthoryear{Volkening}{2025}]{VolkeningQuant}
\begin{bchapter}
\bauthor{\bsnm{Volkening}, \binits{A.}}:
\bctitle{Methods for quantifying self-organization in biology: a
  forward-looking survey and tutorial}.
In: \beditor{\bsnm{Giuggioli}, \binits{L.}},
\beditor{\bsnm{Maini}, \binits{P.K.}} (eds.)
\bbtitle{The Mathematics of Movement: an Intderdisciplinary Approach to Mutual
  Challenges in Animal Ecology and Cell Biology}.
\bpublisher{Springer},
\blocation{Switzerland}
(\byear{2025})
\end{bchapter}
\endbibitem

\bibitem[\protect\citeauthoryear{Topaz et~al.}{2015}]{Topaz2015}
\begin{barticle}
\bauthor{\bsnm{Topaz}, \binits{C.M.}},
\bauthor{\bsnm{Ziegelmeier}, \binits{L.}},
\bauthor{\bsnm{Halverson}, \binits{T.}}:
\batitle{Topological data analysis of biological aggregation models}.
\bjtitle{PLOS ONE}
\bvolume{10}(\bissue{5}),
\bfpage{0126383}
(\byear{2015})
\end{barticle}
\endbibitem

\bibitem[\protect\citeauthoryear{Miyazawa
  et~al.}{2010}]{Miyazawa2010Simplicity}
\begin{barticle}
\bauthor{\bsnm{Miyazawa}, \binits{S.}},
\bauthor{\bsnm{Okamoto}, \binits{M.}},
\bauthor{\bsnm{Kondo}, \binits{S.}}:
\batitle{Blending of animal colour patterns by hybridization}.
\bjtitle{Nat. Commun.}
\bvolume{1},
\bfpage{66}
(\byear{2010})
\end{barticle}
\endbibitem

\bibitem[\protect\citeauthoryear{Djurdjevi\u{c} et~al.}{2019}]{Djur2019Trout}
\begin{botherref}
\oauthor{\bsnm{Djurdjevi\u{c}}, \binits{I.}},
\oauthor{\bsnm{Furmanek}, \binits{T.}},
\oauthor{\bsnm{Miyazawa}, \binits{S.}},
\oauthor{\bsnm{Bajec}, \binits{S.S.}}:
Comparative transcriptome analysis of trout skin pigment cells.
BMC Genom.
\textbf{20}(1)
(2019)
\end{botherref}
\endbibitem

\bibitem[\protect\citeauthoryear{Gavagnin et~al.}{2018}]{Gavagnin2018yates}
\begin{barticle}
\bauthor{\bsnm{Gavagnin}, \binits{E.}},
\bauthor{\bsnm{Owen}, \binits{J.P.}},
\bauthor{\bsnm{Yates}, \binits{C.A.}}:
\batitle{Pair correlation functions for identifying spatial correlation in
  discrete domains}.
\bjtitle{Phys. Rev. E}
\bvolume{97}(\bissue{6-1}),
\bfpage{062104}
(\byear{2018})
\end{barticle}
\endbibitem

\bibitem[\protect\citeauthoryear{Bull et~al.}{2024}]{Bull2024}
\begin{barticle}
\bauthor{\bsnm{Bull}, \binits{J.A.}},
\bauthor{\bsnm{Mulholland}, \binits{E.J.}},
\bauthor{\bsnm{Leedham}, \binits{S.J.}},
\bauthor{\bsnm{Byrne}, \binits{H.M.}}:
\batitle{Extended correlation functions for spatial analysis of multiplex
  imaging data}.
\bjtitle{Biol. Imaging}
\bvolume{4},
\bfpage{2}
(\byear{2024})
\end{barticle}
\endbibitem

\bibitem[\protect\citeauthoryear{Binder and Simpson}{2013}]{Binder2013}
\begin{barticle}
\bauthor{\bsnm{Binder}, \binits{B.J.}},
\bauthor{\bsnm{Simpson}, \binits{M.J.}}:
\batitle{Quantifying spatial structure in experimental observations and
  agent-based simulations using pair-correlation functions}.
\bjtitle{Phys. Rev. E}
\bvolume{88}(\bissue{2}),
\bfpage{022705}
(\byear{2013})
\end{barticle}
\endbibitem

\bibitem[\protect\citeauthoryear{Edelsbrunner and
  Harer}{2008}]{Edelsbrunner2008}
\begin{barticle}
\bauthor{\bsnm{Edelsbrunner}, \binits{H.}},
\bauthor{\bsnm{Harer}, \binits{J.}}:
\batitle{Persistent homology --- a survey}.
\bjtitle{Contemp. Math.}
\bvolume{453},
\bfpage{257}--\blpage{282}
(\byear{2008})
\end{barticle}
\endbibitem

\bibitem[\protect\citeauthoryear{Carlsson}{2009}]{Carlsson2009}
\begin{barticle}
\bauthor{\bsnm{Carlsson}, \binits{G.}}:
\batitle{Topology and data}.
\bjtitle{Bull. Am. Math. Soc.}
\bvolume{46}(\bissue{2}),
\bfpage{255}--\blpage{308}
(\byear{2009})
\end{barticle}
\endbibitem

\bibitem[\protect\citeauthoryear{Otter et~al.}{2017}]{Otter2017mason}
\begin{barticle}
\bauthor{\bsnm{Otter}, \binits{N.}},
\bauthor{\bsnm{Porter}, \binits{M.A.}},
\bauthor{\bsnm{Tillmann}, \binits{U.}},
\bauthor{\bsnm{Grindrod}, \binits{P.}},
\bauthor{\bsnm{Harrington}, \binits{H.A.}}:
\batitle{A roadmap for the computation of persistent homology}.
\bjtitle{EPJ Data Sci.}
\bvolume{6}(\bissue{1}),
\bfpage{1}--\blpage{38}
(\byear{2017})
\end{barticle}
\endbibitem

\bibitem[\protect\citeauthoryear{Ghrist}{2014}]{Ghrist2014}
\begin{bbook}
\bauthor{\bsnm{Ghrist}, \binits{R.}}:
\bbtitle{Elementary Applied Topology},
\bedition{1.0} edn.
\bpublisher{Createspace}, \blocation{}
(\byear{2014})
\end{bbook}
\endbibitem

\bibitem[\protect\citeauthoryear{Am\'{e}zquita et~al.}{2020}]{Munch2020Shape}
\begin{barticle}
\bauthor{\bsnm{Am\'{e}zquita}, \binits{E.J.}},
\bauthor{\bsnm{Quigley}, \binits{M.Y.}},
\bauthor{\bsnm{Ophelders}, \binits{T.A.}},
\bauthor{\bsnm{Munch}, \binits{E.}},
\bauthor{\bsnm{Chitwood}, \binits{D.H.}}:
\batitle{{The shape of things to come: Topological data analysis and biology,
  from molecules to organisms}}.
\bjtitle{Dev. Dyn.}
\bvolume{249}(\bissue{7}),
\bfpage{816}--\blpage{833}
(\byear{2020})
\end{barticle}
\endbibitem

\bibitem[\protect\citeauthoryear{Feng et~al.}{2022}]{Feng2021tda}
\begin{bchapter}
\bauthor{\bsnm{Feng}, \binits{M.}},
\bauthor{\bsnm{Hickok}, \binits{A.}},
\bauthor{\bsnm{Porter}, \binits{M.A.}}:
\bctitle{Topological data analysis of spatial systems}.
In: \beditor{\bsnm{Battiston}, \binits{F.}},
\beditor{\bsnm{Petri}, \binits{G.}} (eds.)
\bbtitle{Higher-Order Systems: Understanding Complex Systems},
pp. \bfpage{389}--\blpage{399}.
\bpublisher{Springer},
\blocation{Cham, Switzerland}
(\byear{2022})
\end{bchapter}
\endbibitem

\bibitem[\protect\citeauthoryear{Chazal and Michel}{2021}]{Chazal2021}
\begin{barticle}
\bauthor{\bsnm{Chazal}, \binits{F.}},
\bauthor{\bsnm{Michel}, \binits{B.}}:
\batitle{An introduction to topological data analysis: Fundamental and
  practical aspects for data scientists}.
\bjtitle{Front. Artif. Intell.}
\bvolume{4},
\bfpage{667963}
(\byear{2021})
\end{barticle}
\endbibitem

\bibitem[\protect\citeauthoryear{McDonald et~al.}{2023}]{Coral}
\begin{botherref}
\oauthor{\bsnm{McDonald}, \binits{R.A.}},
\oauthor{\bsnm{Neuhausler}, \binits{R.}},
\oauthor{\bsnm{Robinson}, \binits{M.}},
\oauthor{\bsnm{Larsen}, \binits{L.G.}},
\oauthor{\bsnm{Harrington}, \binits{H.A.}},
\oauthor{\bsnm{Bruna}, \binits{M.}}:
Zigzag persistence for coral reef resilience using a stochastic spatial model.
J. R. Soc. Interface
\textbf{20}(205)
(2023)
\end{botherref}
\endbibitem

\bibitem[\protect\citeauthoryear{Gharooni-Fard et~al.}{2024}]{PelegTDA2023}
\begin{botherref}
\oauthor{\bsnm{Gharooni-Fard}, \binits{G.}},
\oauthor{\bsnm{Byers}, \binits{M.}},
\oauthor{\bsnm{Deshmukh}, \binits{V.}},
\oauthor{\bsnm{Bradley}, \binits{E.}},
\oauthor{\bsnm{Mayo}, \binits{C.}},
\oauthor{\bsnm{Topaz}, \binits{C.M.}},
\oauthor{\bsnm{Peleg}, \binits{O.}}:
A computational topology-based spatiotemporal analysis technique for honeybee
  aggregation.
npj Complex.
\textbf{1}(3)
(2024)
\end{botherref}
\endbibitem

\bibitem[\protect\citeauthoryear{Ulmer et~al.}{2019}]{Ulmer2019topaz}
\begin{barticle}
\bauthor{\bsnm{Ulmer}, \binits{M.}},
\bauthor{\bsnm{Ziegelmeier}, \binits{L.}},
\bauthor{\bsnm{Topaz}, \binits{C.M.}}:
\batitle{A topological approach to selecting models of biological experiments}.
\bjtitle{PLOS ONE}
\bvolume{14}(\bissue{3}),
\bfpage{0213679}
(\byear{2019})
\end{barticle}
\endbibitem

\bibitem[\protect\citeauthoryear{Lawson et~al.}{2019}]{Lawson2019cancer}
\begin{barticle}
\bauthor{\bsnm{Lawson}, \binits{P.}},
\bauthor{\bsnm{Sholl}, \binits{A.B.}},
\bauthor{\bsnm{Brown}, \binits{J.Q.}},
\bauthor{\bsnm{Fasy}, \binits{B.T.}},
\bauthor{\bsnm{Wenk}, \binits{C.}}:
\batitle{Persistent homology for the quantitative evaluation of architectural
  features in prostate cancer histology}.
\bjtitle{Sci. Rep.}
\bvolume{9}(\bissue{1}),
\bfpage{1139}
(\byear{2019})
\end{barticle}
\endbibitem

\bibitem[\protect\citeauthoryear{Hartsock et~al.}{2025}]{Hartsock2025}
\begin{barticle}
\bauthor{\bsnm{Hartsock}, \binits{I.}},
\bauthor{\bsnm{Park}, \binits{E.}},
\bauthor{\bsnm{Toppen}, \binits{J.}},
\bauthor{\bsnm{Bubenik}, \binits{P.}},
\bauthor{\bsnm{Dimitrova}, \binits{E.S.}},
\bauthor{\bsnm{Kemp}, \binits{M.L.}},
\bauthor{\bsnm{Cruz}, \binits{D.A.}}:
\batitle{Topological data analysis of pattern formation of human induced
  pluripotent stem cell colonies}.
\bjtitle{Sci. Rep.}
\bvolume{15}(\bissue{1}),
\bfpage{11544}--\blpage{18}
(\byear{2025})
\end{barticle}
\endbibitem

\bibitem[\protect\citeauthoryear{Thorne et~al.}{2022}]{Thorne}
\begin{barticle}
\bauthor{\bsnm{Thorne}, \binits{T.}},
\bauthor{\bsnm{Kirk}, \binits{P.D.W.}},
\bauthor{\bsnm{Harrington}, \binits{H.A.}}:
\batitle{{Topological approximate {B}ayesian computation for parameter
  inference of an angiogenesis model}}.
\bjtitle{Bioinform.}
\bvolume{38}(\bissue{9}),
\bfpage{2529}--\blpage{2535}
(\byear{2022})
\end{barticle}
\endbibitem

\bibitem[\protect\citeauthoryear{Stolz et~al.}{2022}]{Stolz2022}
\begin{barticle}
\bauthor{\bsnm{Stolz}, \binits{B.J.}},
\bauthor{\bsnm{Kaeppler}, \binits{J.}},
\bauthor{\bsnm{Markelc}, \binits{B.}},
\bauthor{\bsnm{Braun}, \binits{F.}},
\bauthor{\bsnm{Lipsmeier}, \binits{F.}},
\bauthor{\bsnm{Muschel}, \binits{R.J.}},
\bauthor{\bsnm{Byrne}, \binits{H.M.}},
\bauthor{\bsnm{Harrington}, \binits{H.A.}}:
\batitle{Multiscale topology characterizes dynamic tumor vascular networks}.
\bjtitle{Sci. Adv.}
\bvolume{8}(\bissue{23}),
\bfpage{2456}
(\byear{2022})
\end{barticle}
\endbibitem

\bibitem[\protect\citeauthoryear{Bhaskar et~al.}{2019}]{Bhaskar2019}
\begin{barticle}
\bauthor{\bsnm{Bhaskar}, \binits{D.}},
\bauthor{\bsnm{Manhart}, \binits{A.}},
\bauthor{\bsnm{Milzman}, \binits{J.}},
\bauthor{\bsnm{Nardini}, \binits{J.T.}},
\bauthor{\bsnm{Storey}, \binits{K.M.}},
\bauthor{\bsnm{Topaz}, \binits{C.M.}},
\bauthor{\bsnm{Ziegelmeier}, \binits{L.}}:
\batitle{Analyzing collective motion with machine learning and topology}.
\bjtitle{Chaos}
\bvolume{29}(\bissue{12}),
\bfpage{123125}
(\byear{2019})
\end{barticle}
\endbibitem

\bibitem[\protect\citeauthoryear{Bhaskar et~al.}{2023}]{Bhaskar2023}
\begin{botherref}
\oauthor{\bsnm{Bhaskar}, \binits{D.}},
\oauthor{\bsnm{Zhang}, \binits{W.Y.}},
\oauthor{\bsnm{Volkening}, \binits{A.}},
\oauthor{\bsnm{Sandstede}, \binits{B.}},
\oauthor{\bsnm{Wong}, \binits{I.Y.}}:
Topological data analysis of spatial patterning in heterogeneous cell
  populations: I. clustering and sorting with varying cell-cell adhesion.
npj Sys. Biol. Appl.
\textbf{9}(43)
(2023)
\end{botherref}
\endbibitem

\bibitem[\protect\citeauthoryear{Ciocanel et~al.}{2021}]{Ciocanel2021}
\begin{botherref}
\oauthor{\bsnm{Ciocanel}, \binits{M.V.}},
\oauthor{\bsnm{Juenemann}, \binits{R.}},
\oauthor{\bsnm{Dawes}, \binits{A.T.}},
\oauthor{\bsnm{McKinley}, \binits{S.A.}}:
Topological data analysis approaches to uncovering the timing of ring structure
  onset in filamentous networks.
Bull. Math. Biol.
\textbf{83}(3)
(2021)
\end{botherref}
\endbibitem

\bibitem[\protect\citeauthoryear{Edelsbrunner et~al.}{2002}]{Edelsbrunner2002}
\begin{barticle}
\bauthor{\bsnm{Edelsbrunner}, \binits{H.}},
\bauthor{\bsnm{Letscher}, \binits{D.}},
\bauthor{\bsnm{Zomorodian}, \binits{A.}}:
\batitle{Topological persistence and simplification}.
\bjtitle{Discrete Comput. Geom.}
\bvolume{28}(\bissue{4}),
\bfpage{511}--\blpage{533}
(\byear{2002})
\end{barticle}
\endbibitem

\bibitem[\protect\citeauthoryear{Turkeš et~al.}{2021}]{Turke2021}
\begin{barticle}
\bauthor{\bsnm{Turkeš}, \binits{R.}},
\bauthor{\bsnm{Nys}, \binits{J.}},
\bauthor{\bsnm{Verdonck}, \binits{T.}},
\bauthor{\bsnm{Latré}, \binits{S.}}:
\batitle{Noise robustness of persistent homology on greyscale images, across
  filtrations and signatures}.
\bjtitle{PLOS ONE}
\bvolume{16}(\bissue{9}),
\bfpage{0257215}
(\byear{2021})
\end{barticle}
\endbibitem

\bibitem[\protect\citeauthoryear{Kramár et~al.}{2016}]{Kramar2016}
\begin{barticle}
\bauthor{\bsnm{Kramár}, \binits{M.}},
\bauthor{\bsnm{Levanger}, \binits{R.}},
\bauthor{\bsnm{Tithof}, \binits{J.}},
\bauthor{\bsnm{Suri}, \binits{B.}},
\bauthor{\bsnm{Xu}, \binits{M.}},
\bauthor{\bsnm{Paul}, \binits{M.}},
\bauthor{\bsnm{Schatz}, \binits{M.F.}},
\bauthor{\bsnm{Mischaikow}, \binits{K.}}:
\batitle{{Analysis of Kolmogorov flow and Rayleigh--B\'{e}nard convection using
  persistent homology}}.
\bjtitle{Physica D}
\bvolume{334},
\bfpage{82}--\blpage{98}
(\byear{2016})
\end{barticle}
\endbibitem

\bibitem[\protect\citeauthoryear{Stolz-Pretzer}{2019}]{StolzThesis}
\begin{botherref}
\oauthor{\bsnm{Stolz-Pretzer}, \binits{B.}}:
Global and local persistent homology for the shape and classification of
  biological data.
PhD thesis,
University of Oxford
(2019)
\end{botherref}
\endbibitem

\bibitem[\protect\citeauthoryear{Khoudari et~al.}{2025}]{2025repo}
\begin{botherref}
\oauthor{\bsnm{Khoudari}, \binits{N.}},
\oauthor{\bsnm{Nardini}, \binits{J.}},
\oauthor{\bsnm{Volkening}, \binits{A.}}:
Quantifying zebrafish skin patterns using sweeping-plane TDA.
\url{https://github.com/nourkhoudari/quantifying-zebrafish-skin-patterns-using-sweeping-plane-TDA}
(2025)
\end{botherref}
\endbibitem

\bibitem[\protect\citeauthoryear{Hamada et~al.}{2014}]{delta}
\begin{barticle}
\bauthor{\bsnm{Hamada}, \binits{H.}},
\bauthor{\bsnm{Watanabe}, \binits{M.}},
\bauthor{\bsnm{Lau}, \binits{H.E.}},
\bauthor{\bsnm{Nishida}, \binits{T.}},
\bauthor{\bsnm{Hasegawa}, \binits{T.}},
\bauthor{\bsnm{Parichy}, \binits{D.M.}},
\bauthor{\bsnm{Kondo}, \binits{S.}}:
\batitle{{Involvement of Delta/Notch signaling in zebrafish adult pigment
  stripe patterning}}.
\bjtitle{Development}
\bvolume{141}(\bissue{2}),
\bfpage{318}--\blpage{324}
(\byear{2014})
\end{barticle}
\endbibitem

\bibitem[\protect\citeauthoryear{Inaba et~al.}{2012}]{Inaba}
\begin{barticle}
\bauthor{\bsnm{Inaba}, \binits{M.}},
\bauthor{\bsnm{Yamanaka}, \binits{H.}},
\bauthor{\bsnm{Kondo}, \binits{S.}}:
\batitle{Pigment pattern formation by contact-dependent depolarization}.
\bjtitle{Science}
\bvolume{335}(\bissue{6069}),
\bfpage{677}--\blpage{677}
(\byear{2012})
\end{barticle}
\endbibitem

\bibitem[\protect\citeauthoryear{Patterson et~al.}{2014}]{Patterson2014}
\begin{botherref}
\oauthor{\bsnm{Patterson}, \binits{L.B.}},
\oauthor{\bsnm{Bain}, \binits{E.J.}},
\oauthor{\bsnm{Parichy}, \binits{D.M.}}:
Pigment cell interactions and differential xanthophore recruitment underlying
  zebrafish stripe reiteration and \emph{{D}anio} pattern evolution.
Nat. Commun.
\textbf{5}(5299)
(2014)
\end{botherref}
\endbibitem

\bibitem[\protect\citeauthoryear{Mahalwar et~al.}{2016}]{2016heterotypic}
\begin{barticle}
\bauthor{\bsnm{Mahalwar}, \binits{P.}},
\bauthor{\bsnm{Singh}, \binits{A.P.}},
\bauthor{\bsnm{Fadeev}, \binits{A.}},
\bauthor{\bsnm{N{\"u}sslein-Volhard}, \binits{C.}},
\bauthor{\bsnm{Irion}, \binits{U.}}:
\batitle{Heterotypic interactions regulate cell shape and density during color
  pattern formation in zebrafish}.
\bjtitle{Biol. Open}
\bvolume{5}(\bissue{11}),
\bfpage{1680}--\blpage{1690}
(\byear{2016})
\end{barticle}
\endbibitem

\bibitem[\protect\citeauthoryear{Parichy and Turner}{2003}]{ParTur256}
\begin{barticle}
\bauthor{\bsnm{Parichy}, \binits{D.M.}},
\bauthor{\bsnm{Turner}, \binits{J.M.}}:
\batitle{Zebrafish \emph{puma} mutant decouples pigment pattern and somatic
  metamorphosis}.
\bjtitle{Dev. Biol.}
\bvolume{256}(\bissue{2}),
\bfpage{242}--\blpage{257}
(\byear{2003})
\end{barticle}
\endbibitem

\bibitem[\protect\citeauthoryear{Takahashi and
  Kondo}{2008}]{TakahashiMelDisperse}
\begin{barticle}
\bauthor{\bsnm{Takahashi}, \binits{G.}},
\bauthor{\bsnm{Kondo}, \binits{S.}}:
\batitle{Melanophores in the stripes of adult zebrafish do not have the nature
  to gather, but disperse when they have the space to move}.
\bjtitle{Pigment Cell Melanoma Res.}
\bvolume{21}(\bissue{6}),
\bfpage{677}--\blpage{686}
(\byear{2008})
\end{barticle}
\endbibitem

\bibitem[\protect\citeauthoryear{Fadeev et~al.}{2016}]{fadeev2016}
\begin{barticle}
\bauthor{\bsnm{Fadeev}, \binits{A.}},
\bauthor{\bsnm{Krauss}, \binits{J.}},
\bauthor{\bsnm{Singh}, \binits{A.P.}},
\bauthor{\bsnm{N{\"u}sslein-Volhard}, \binits{C.}}:
\batitle{Zebrafish leucocyte tyrosine kinase controls iridophore establishment,
  proliferation and survival}.
\bjtitle{Pigment Cell Melanoma Res.}
\bvolume{29}(\bissue{3}),
\bfpage{284}--\blpage{296}
(\byear{2016})
\end{barticle}
\endbibitem

\bibitem[\protect\citeauthoryear{Mahalwar et~al.}{2014}]{Mahalwar}
\begin{barticle}
\bauthor{\bsnm{Mahalwar}, \binits{P.}},
\bauthor{\bsnm{Walderich}, \binits{B.}},
\bauthor{\bsnm{Singh}, \binits{A.P.}},
\bauthor{\bsnm{N{\"u}sslein-Volhard}, \binits{C.}}:
\batitle{Local reorganization of xanthophores fine-tunes and colors the striped
  pattern of zebrafish}.
\bjtitle{Science}
\bvolume{345}(\bissue{6202}),
\bfpage{1362}--\blpage{1364}
(\byear{2014})
\end{barticle}
\endbibitem

\bibitem[\protect\citeauthoryear{Dooley et~al.}{2013}]{Dooley}
\begin{barticle}
\bauthor{\bsnm{Dooley}, \binits{C.M.}},
\bauthor{\bsnm{Mongera}, \binits{A.}},
\bauthor{\bsnm{Walderich}, \binits{B.}},
\bauthor{\bsnm{N{\"u}sslein-Volhard}, \binits{C.}}:
\batitle{On the embryonic origin of adult melanophores: the role of {ErbB} and
  {K}it signalling in establishing melanophore stem cells in zebrafish}.
\bjtitle{Development}
\bvolume{140}(\bissue{5}),
\bfpage{1003}--\blpage{1013}
(\byear{2013})
\end{barticle}
\endbibitem

\bibitem[\protect\citeauthoryear{McMenamin et~al.}{2014}]{Mcmen2014}
\begin{barticle}
\bauthor{\bsnm{McMenamin}, \binits{S.K.}},
\bauthor{\bsnm{Bain}, \binits{E.J.}},
\bauthor{\bsnm{McCann}, \binits{A.E.}},
\bauthor{\bsnm{Patterson}, \binits{L.B.}},
\bauthor{\bsnm{Eom}, \binits{D.S.}},
\bauthor{\bsnm{Waller}, \binits{Z.P.}},
\bauthor{\bsnm{Hamill}, \binits{J.C.}},
\bauthor{\bsnm{Kuhlman}, \binits{J.A.}},
\bauthor{\bsnm{Eisen}, \binits{J.S.}},
\bauthor{\bsnm{Parichy}, \binits{D.M.}}:
\batitle{Thyroid hormone--dependent adult pigment cell lineage and pattern in
  zebrafish}.
\bjtitle{Science}
\bvolume{345}(\bissue{6202}),
\bfpage{1358}--\blpage{1361}
(\byear{2014})
\end{barticle}
\endbibitem

\bibitem[\protect\citeauthoryear{Budi et~al.}{2011}]{Budi}
\begin{barticle}
\bauthor{\bsnm{Budi}, \binits{E.H.}},
\bauthor{\bsnm{Patterson}, \binits{L.B.}},
\bauthor{\bsnm{Parichy}, \binits{D.M.}}:
\batitle{Post-embryonic nerve-associated precursors to adult pigment cells:
  genetic requirements and dynamics of morphogenesis and differentiation}.
\bjtitle{PLOS Genet.}
\bvolume{7}(\bissue{5}),
\bfpage{1002044}
(\byear{2011})
\end{barticle}
\endbibitem

\bibitem[\protect\citeauthoryear{Gur et~al.}{2020}]{Gur2020}
\begin{barticle}
\bauthor{\bsnm{Gur}, \binits{D.}},
\bauthor{\bsnm{Bain}, \binits{E.J.}},
\bauthor{\bsnm{Johnson}, \binits{K.R.}},
\bauthor{\bsnm{Aman}, \binits{A.J.}},
\bauthor{\bsnm{Amalia~Pasolli}, \binits{H.}},
\bauthor{\bsnm{Flynn}, \binits{J.D.}},
\bauthor{\bsnm{Allen}, \binits{M.C.}},
\bauthor{\bsnm{Deheyn}, \binits{D.D.}},
\bauthor{\bsnm{Lee}, \binits{J.C.}},
\bauthor{\bsnm{Lippincott-Schwartz}, \binits{J.}},
\bauthor{\bsnm{Parichy}, \binits{D.M.}}:
\batitle{In situ differentiation of iridophore crystallotypes underlies
  zebrafish stripe patterning}.
\bjtitle{Nat. Commun.}
\bvolume{11}(\bissue{1}),
\bfpage{6391}
(\byear{2020})
\end{barticle}
\endbibitem

\bibitem[\protect\citeauthoryear{Eom and Parichy}{2017}]{eom2017macrophage}
\begin{barticle}
\bauthor{\bsnm{Eom}, \binits{D.S.}},
\bauthor{\bsnm{Parichy}, \binits{D.M.}}:
\batitle{A macrophage relay for long-distance signaling during postembryonic
  tissue remodeling}.
\bjtitle{Science}
\bvolume{355}(\bissue{6331}),
\bfpage{1317}--\blpage{1320}
(\byear{2017})
\end{barticle}
\endbibitem

\bibitem[\protect\citeauthoryear{Singh et~al.}{2014}]{Singh}
\begin{barticle}
\bauthor{\bsnm{Singh}, \binits{A.P.}},
\bauthor{\bsnm{Schach}, \binits{U.}},
\bauthor{\bsnm{N{\"u}sslein-Volhard}, \binits{C.}}:
\batitle{Proliferation, dispersal and patterned aggregation of iridophores in
  the skin prefigure striped colouration of zebrafish}.
\bjtitle{Nat. Cell Biol.}
\bvolume{16}(\bissue{6}),
\bfpage{604}--\blpage{611}
(\byear{2014})
\end{barticle}
\endbibitem

\bibitem[\protect\citeauthoryear{Walderich
  et~al.}{2016}]{walderich2016homotypic}
\begin{botherref}
\oauthor{\bsnm{Walderich}, \binits{B.}},
\oauthor{\bsnm{Singh}, \binits{A.P.}},
\oauthor{\bsnm{Mahalwar}, \binits{P.}},
\oauthor{\bsnm{N{\"u}sslein-Volhard}, \binits{C.}}:
Homotypic cell competition regulates proliferation and tiling of zebrafish
  pigment cells during colour pattern formation.
Nat. Commun.
\textbf{7}
(2016)
\end{botherref}
\endbibitem

\bibitem[\protect\citeauthoryear{McMenamin et~al.}{2016}]{McMen2016}
\begin{barticle}
\bauthor{\bsnm{McMenamin}, \binits{S.K.}},
\bauthor{\bsnm{Chandless}, \binits{M.N.}},
\bauthor{\bsnm{Parichy}, \binits{D.M.}}:
\batitle{Working with zebrafish at postembryonic stages}.
\bjtitle{Methods Cell Biol.}
\bvolume{134},
\bfpage{587}--\blpage{607}
(\byear{2016})
\end{barticle}
\endbibitem

\bibitem[\protect\citeauthoryear{Parichy et~al.}{2009}]{Parichy}
\begin{barticle}
\bauthor{\bsnm{Parichy}, \binits{D.M.}},
\bauthor{\bsnm{Elizondo}, \binits{M.R.}},
\bauthor{\bsnm{Mills}, \binits{M.G.}},
\bauthor{\bsnm{Gordon}, \binits{T.N.}},
\bauthor{\bsnm{Engeszer}, \binits{R.E.}}:
\batitle{Normal table of postembryonic zebrafish development: staging by
  externally visible anatomy of the living fish}.
\bjtitle{Dev. Dyn.}
\bvolume{238}(\bissue{12}),
\bfpage{2975}--\blpage{3015}
(\byear{2009})
\end{barticle}
\endbibitem

\bibitem[\protect\citeauthoryear{McGuirl et~al.}{2020}]{mcguirl2020zebrafish}
\begin{botherref}
\oauthor{\bsnm{McGuirl}, \binits{M.}},
\oauthor{\bsnm{Volkening}, \binits{A.}},
\oauthor{\bsnm{Sandstede}, \binits{B.}}:
Zebrafish Simulation Data
(2020)
\end{botherref}
\endbibitem

\bibitem[\protect\citeauthoryear{Chazal et~al.}{2016}]{chazal}
\begin{bbook}
\bauthor{\bsnm{Chazal}, \binits{F.}},
\bauthor{\bsnm{Silva}, \binits{V.}},
\bauthor{\bsnm{Glisse}, \binits{M.}},
\bauthor{\bsnm{Oudot}, \binits{S.}}:
\bbtitle{The Structure and Stability of Persistence Modules},
\bedition{1st} edn.
\bpublisher{Springer},
\blocation{Switzerland}
(\byear{2016})
\end{bbook}
\endbibitem

\bibitem[\protect\citeauthoryear{Heiss et~al.}{2021}]{Heiss2021}
\begin{bchapter}
\bauthor{\bsnm{Heiss}, \binits{T.}},
\bauthor{\bsnm{Tymochko}, \binits{S.}},
\bauthor{\bsnm{Story}, \binits{B.}},
\bauthor{\bsnm{Garin}, \binits{A.}},
\bauthor{\bsnm{Bui}, \binits{H.}},
\bauthor{\bsnm{Bleile}, \binits{B.}},
\bauthor{\bsnm{Robins}, \binits{V.}}:
\bctitle{The impact of changes in resolution on the persistent homology of
  images}.
In: \bbtitle{2021 IEEE International Conference on Big Data (Big Data)},
pp. \bfpage{3824}--\blpage{3834}
(\byear{2021})
\end{bchapter}
\endbibitem

\bibitem[\protect\citeauthoryear{Hu et~al.}{2021}]{Hu2021}
\begin{bchapter}
\bauthor{\bsnm{Hu}, \binits{C.-S.}},
\bauthor{\bsnm{Lawson}, \binits{A.}},
\bauthor{\bsnm{Chung}, \binits{Y.-M.}},
\bauthor{\bsnm{Keegan}, \binits{K.}}:
\bctitle{Two-parameter persistence for images via distance transform}.
In: \bbtitle{2021 IEEE/CVF International Conference on Computer Vision
  Workshops (ICCVW)},
pp. \bfpage{4159}--\blpage{4167}
(\byear{2021})
\end{bchapter}
\endbibitem

\bibitem[\protect\citeauthoryear{Carlsson and Zomorodian}{2009}]{CarlssonMulti}
\begin{barticle}
\bauthor{\bsnm{Carlsson}, \binits{G.}},
\bauthor{\bsnm{Zomorodian}, \binits{A.}}:
\batitle{The theory of multidimensional persistence}.
\bjtitle{Discrete Comput. Geom.}
\bvolume{42}(\bissue{1}),
\bfpage{71}--\blpage{93}
(\byear{2009})
\end{barticle}
\endbibitem

\bibitem[\protect\citeauthoryear{Carlsson et~al.}{2009}]{CarlssonMulti2}
\begin{bchapter}
\bauthor{\bsnm{Carlsson}, \binits{G.}},
\bauthor{\bsnm{Singh}, \binits{G.}},
\bauthor{\bsnm{Zomorodian}, \binits{A.}}:
\bctitle{Computing multidimensional persistence}.
In: \beditor{\bsnm{Dong}, \binits{Y.}},
\beditor{\bsnm{Du}, \binits{D.-Z.}},
\beditor{\bsnm{Ibarra}, \binits{O.}} (eds.)
\bbtitle{Algorithms and Computation},
pp. \bfpage{730}--\blpage{739}.
\bpublisher{Springer},
\blocation{Berlin, Heidelberg}
(\byear{2009})
\end{bchapter}
\endbibitem

\bibitem[\protect\citeauthoryear{Harrington et~al.}{2019}]{Harrington2019}
\begin{barticle}
\bauthor{\bsnm{Harrington}, \binits{H.A.}},
\bauthor{\bsnm{Otter}, \binits{N.}},
\bauthor{\bsnm{Schenck}, \binits{H.}},
\bauthor{\bsnm{Tillmann}, \binits{U.}}:
\batitle{Stratifying multiparameter persistent homology}.
\bjtitle{SIAM J. Appl. Algebra Geom.}
\bvolume{3}(\bissue{3}),
\bfpage{439}--\blpage{471}
(\byear{2019})
\end{barticle}
\endbibitem

\bibitem[\protect\citeauthoryear{Nardini et~al.}{2023}]{Nardini2023}
\begin{botherref}
\oauthor{\bsnm{Nardini}, \binits{J.}},
\oauthor{\bsnm{Pugh}, \binits{C.}},
\oauthor{\bsnm{Byrne}, \binits{H.}}:
Statistical and topological summaries aid disease detection for segmented
  retinal vascular images.
Microcirculation
\textbf{30}(4)
(2023)
\end{botherref}
\endbibitem

\bibitem[\protect\citeauthoryear{Cumming et~al.}{2010}]{Cumming2010}
\begin{barticle}
\bauthor{\bsnm{Cumming}, \binits{B.D.}},
\bauthor{\bsnm{McElwain}, \binits{D.L.S.}},
\bauthor{\bsnm{Upton}, \binits{Z.}}:
\batitle{{A mathematical model of wound healing and subsequent scarring}}.
\bjtitle{J. R. Soc. Interface}
\bvolume{7}(\bissue{42}),
\bfpage{19}--\blpage{34}
(\byear{2010})
\end{barticle}
\endbibitem

\bibitem[\protect\citeauthoryear{Amorim}{2015}]{AMORIM2015160}
\begin{barticle}
\bauthor{\bsnm{Amorim}, \binits{P.}}:
\batitle{Modeling ant foraging: A chemotaxis approach with pheromones and trail
  formation}.
\bjtitle{J. Theor. Biol.}
\bvolume{385},
\bfpage{160}--\blpage{173}
(\byear{2015})
\end{barticle}
\endbibitem

\bibitem[\protect\citeauthoryear{Manukyan et~al.}{2017}]{Manukyan2017}
\begin{botherref}
\oauthor{\bsnm{Manukyan}, \binits{L.}},
\oauthor{\bsnm{Montandon}, \binits{S.A.}},
\oauthor{\bsnm{Fofonjka}, \binits{A.}},
\oauthor{\bsnm{Smirnov}, \binits{S.}},
\oauthor{\bsnm{Milinkovitch}, \binits{M.C.}}:
A living mesoscopic cellular automaton made of skin scales.
Nature
\textbf{544}(7649)
(2017)
\end{botherref}
\endbibitem

\bibitem[\protect\citeauthoryear{McDonald et~al.}{2025}]{McDonald2025}
\begin{botherref}
\oauthor{\bsnm{McDonald}, \binits{R.A.}},
\oauthor{\bsnm{Byrne}, \binits{H.M.}},
\oauthor{\bsnm{Harrington}, \binits{H.A.}},
\oauthor{\bsnm{Thorne}, \binits{T.}},
\oauthor{\bsnm{Stolz}, \binits{B.J.}}:
{Topological model selection: a case-study in tumour-induced angiogenesis}
(2025).
\doiurl{10.48550/arXiv.2504.15442}
\end{botherref}
\endbibitem

\end{thebibliography}

\appendix

\FloatBarrier
\section*{Appendix}

While we developed our approach for interpreting the sweeping-plane filtration in terms of pattern features and irregularities with zebrafish in mind, we expect our methodology to be more widely applicable. To help encourage future studies of other biological systems based on the sweeping-plane filtration, we thus illustrate a few patterns that present challenges for our methodology here. We note that some of these examples are rare patterns generated by the agent-based model \cite{volkening2018}, but others are synthetic patterns---i.e., not found in our dataset \cite{mcguirl2020zebrafish}---that we created by hand to highlight places where we introduce additional steps or where our methods may fail.

First, in Fig.~\ref{fig:misclassified_patterns}, we show examples of patterns that are misclassified by our three-step algorithm (Fig.~\ref{fig:algorihm_diagram}), and we provide more information on how we use measurements of the widths of supposed breaks based on Equations~\eqref{eq:gapwidth}--\eqref{eq:gapwidth2} and hyperparameters to refine classifications. Second, in Fig.~\ref{fig:common_misclassifications}, we show illustrative examples of the most common pattern misclassifications that arise in our methodology, even after applying the additional checks and hyperparameter thresholds that we discuss in the caption of Fig.~\ref{fig:misclassified_patterns}. Lastly, in Fig.~\ref{fig:multiple_breaks_patterns}, we show how patterns with multiple breaks are handled by Equations~\eqref{eq:gapwidth}--\eqref{eq:gapwidth2} for estimating break width. Namely, we show how to detect the presence of a stripe break at the right or left domain boundary in Fig.~\ref{fig:multiple_breaks_patterns}(a), as well as how to use Eqn.~\eqref{eq:gapwidth} to estimate break width when there are multiple stripe breaks in the interior of the domain. We also highlight, in Fig.~\ref{fig:multiple_breaks_patterns}(b), how Eqn.~\eqref{eq:gapwidth} fails to estimate break width correctly when breaks of different widths occur at the same position across multiple stripes.

\begin{figure}[h]
    \centering 
    \includegraphics[width=1\linewidth]{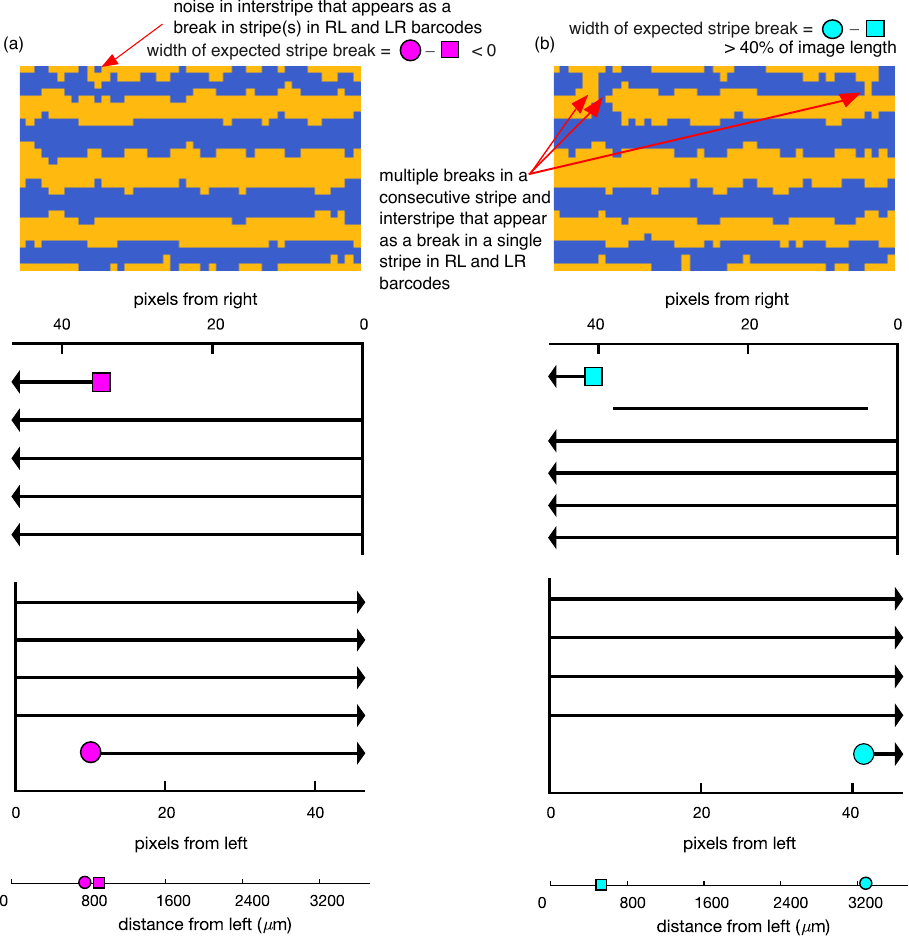}\vspace{0.5\baselineskip}
    \caption{Examples of patterns that are initially misclassified by our three-step algorithm (Fig.~\ref{fig:algorihm_diagram}) and later reclassified using hyperparameter-based conditions that we enforce on interruption width. (a) A stray blue pixel in an interstripe is interpreted by our three-step classification algorithm as a broken stripe, leading us to initially classify this unbroken striped pattern as \textit{broken stripe(s)}. However, applying Eqn.~\eqref{eq:gapwidth} to this pattern outputs a negative width for the stripe break. We thus add an additional check to our classification algorithm that no interruptions have negative width. If we do detect a negative width later in our pipeline, we redefine the number of breaks as our original number of breaks minus the number of negative-width occurrences, 
    and we reclassify the pattern as an \textit{unbroken striped pattern} if the count of breaks in stripes is now zero. (b) Striped patterns occasionally feature multiple breaks in a consecutive stripe and interstripe (here two stripe breaks in $2$D and one interstripe break in X$1$D) that cause difficulties for our three-step classification algorithm; these patterns are very messy and rare. Our main algorithm in Fig.~\ref{fig:algorihm_diagram} classifies this pattern---which should be labeled \textit{broken stripe(s) and interstripe(s)}---as \textit{broken stripe(s)}. Under this misclassification, we find a break wider than $40\%$ of the image length. To address this, we introduce a hyperparameter-based condition on maximum interruption width admissible. Specifically, if a break width is greater than $40\%$, we reclassify the pattern as \textit{broken stripe(s) and interstripe(s)}, but exclude it from our distribution of break width in Fig.~\ref{fig:break_percentage_position}(b), since we cannot reliable interpret the barcodes in this setting.}
    \label{fig:misclassified_patterns}
\end{figure}

\begin{figure}[h]
    \centering 
    \includegraphics[width=1\linewidth]{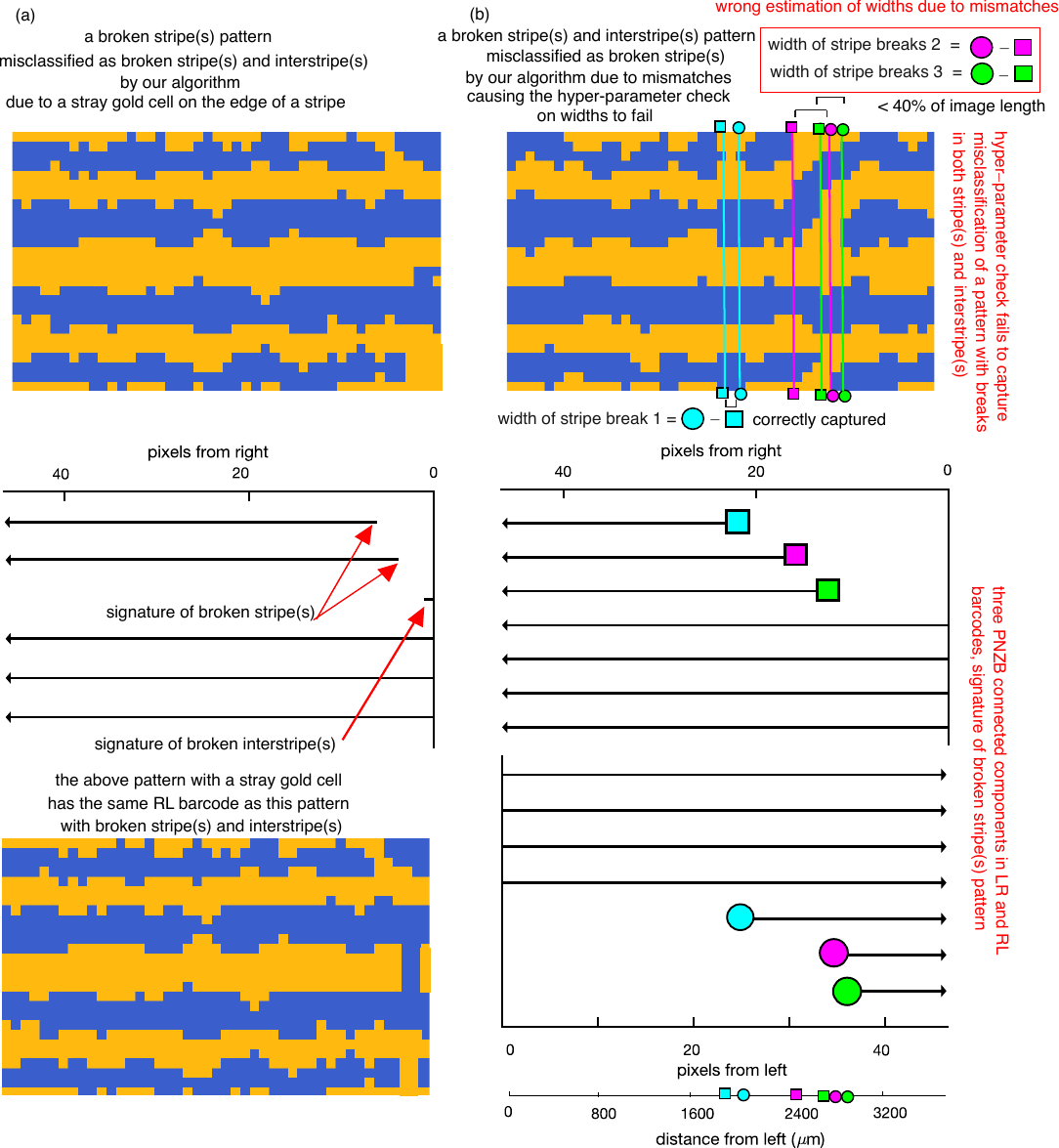}\vspace{0.5\baselineskip}
    \caption{Examples of very rare, messy patterns illustrating the most common places that our classification algorithm fails even after applying the additional checks that we discuss in Sect.~\ref{sec:step3} and Fig.~\ref{fig:misclassified_patterns}. (a) A stray gold cell in a blue stripe at the domain boundary causes this pattern---which should be classified as \textit{broken stripe(s)}---to be misclassified as \textit{broken stripe(s) and interstripe(s)}. This pattern has a zero-born connected component of very low persistence (i.e., one pixel), and its RL barcode looks similar to that of a \textit{broken stripe(s) and interstripe(s)} pattern (bottom row in (a)) with an interstripe break located one pixel away from the boundary. Because it is challenging to distinguish these two cases based on their barcodes, we highlight this as a limitation of our methodology. (b) Very messy patterns which should be classified as \textit{broken stripe(s) and interstripe(s)} are occasionally misclassified as \textit{broken stripe(s)} due to multiple breaks as well as mismatches between breaks and their corresponding PNZB connected components in the RL and LR barcodes. In these rare cases, mismatches lead us to incorrectly estimate interruption width as below $40$\% of the image width, so that our hyperparameter check in Fig.~\ref{fig:misclassified_patterns} fails to correct the misclassification.}
    \label{fig:common_misclassifications}
\end{figure}

\begin{figure}[h]
    \centering 
    \includegraphics[width=1\linewidth]{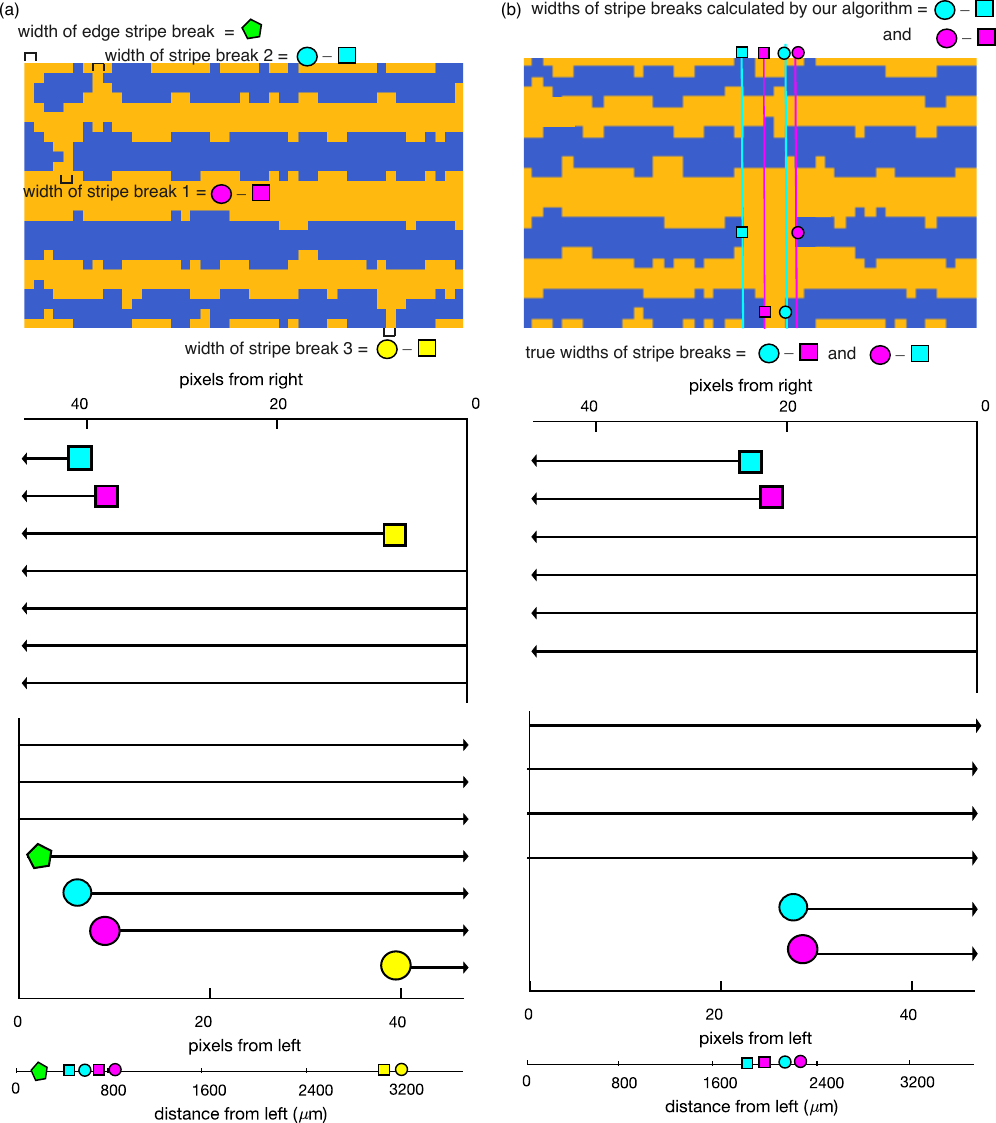}\vspace{0.5\baselineskip}
    \caption{Rare or pathological examples related to estimating stripe width. To make our pipeline more readily applicable to other data, we highlight some limitations and challenges. (a) As a rare example, this pattern features a stripe break at the domain boundary (the left boundary of $2$D) and multiple interior breaks in stripes. It has a different number of PNZB connected components in the RL and LR barcodes, rendering Eqn.~\eqref{eq:gapwidth} challenging to use. We observe that there are four PNZB features when sweeping left to right, and three PNZB features when sweeping right to left; notably, the difference in these these two values is precisely the number of breaks on domain edges ($4-3=1$ here). Differences in the number of PNZB connected components in the RL and LR barcodes typically arise due to the presence of breaks on domain edges. To address this, we associate the earliest born PNZB connected components with the edge breaks, and match the remaining death times to estimate the width of each interior break according to Eqn.~\eqref{eq:gapwidth}. 
    (b) This  synthetic example (not from our dataset) demonstrates that if breaks of the same type align, it is challenging to match death times of connected components associated with the RL and LR filtrations to their corresponding stripe break. 
    Since it is not possible to automatically detect such cases by only analyzing barcodes, they represent a limitation of our approach to estimating break width.}
    \label{fig:multiple_breaks_patterns}
\end{figure}

\end{document}